\documentclass[aps,prd,nofootinbib,twocolumn,superscriptaddress,preprintnumbers,letterpaper,natbib,nofootinbib]{revtex4-1}

\pdfoutput=1
\usepackage{amssymb,amsmath,latexsym,mathrsfs}
\usepackage[normalem]{ulem}
\usepackage{url}
\usepackage{enumitem}
\usepackage{graphicx}
\usepackage[usenames,dvipsnames]{color}
\usepackage[breaklinks,colorlinks,urlcolor=blue,citecolor=blue,linkcolor=magenta]{hyperref}
\usepackage{comment}
\usepackage{bm}
\usepackage{fancybox}
\usepackage[table]{xcolor}
\usepackage{physics}
\usepackage{tikz}
\usetikzlibrary{positioning,shapes}
\usepackage{relsize} 
\usepackage[latin1]{inputenc}
\usepackage{tikz}
\usetikzlibrary{trees}
\usetikzlibrary{decorations.pathmorphing}
\usetikzlibrary{decorations.markings}
\usetikzlibrary{positioning,arrows}
\usetikzlibrary{decorations.pathmorphing}
\usetikzlibrary{decorations.markings}
\usetikzlibrary{decorations.pathreplacing,calc}
\usetikzlibrary{decorations.pathmorphing,decorations.markings,trees,positioning,arrows}   
\newif\ifmirrorsemicircle

\usepackage{hhline}
\usepackage{multirow}

\newcommand{\bea}{\begin{eqnarray}}
\newcommand{\eea}{\end{eqnarray}}

\def\slashD{D \!\!\! \slash}

\def\slashq{q \!\!\! \slash}
\def\diff{\mathrm{d}}

\makeatletter

\begin{document}

\title{The Strange Physics of Dark Baryons}

\preprint{TUM-HEP-1373/21}
\preprint{MITP-21-060}
\author{Gonzalo~Alonso-\'Alvarez}
\affiliation{McGill University, Department of Physics, 3600 Rue University,
Montr\'eal, QC H3A\,2T8 Canada}
\author{Gilly~Elor}
\affiliation{PRISMA$^+$ Cluster of Excellence \& Mainz Institute for Theoretical Physics\\
Johannes Gutenberg University, 55099 Mainz, Germany}
\author{Miguel~Escudero}
\affiliation{Physik-Department, Technische Universit{\"{a}}t, M{\"{u}}nchen, James-Franck-Stra{\ss}e, 85748 Garching, Germany}
\author{Bartosz~Fornal}
\affiliation{Department of Chemistry and Physics, Barry University,\\ Miami Shores, Florida 33161, USA}
\author{Benjam\'{i}n~Grinstein}
\affiliation{Department of Physics, University of California, San Diego,\\ 9500 Gilman Drive, La Jolla, CA 92093, USA} 
\author{Jorge~Martin~Camalich}
\affiliation{Instituto de Astrof\'isica de Canarias, C/ V\'ia L\'actea, s/n
E38205 - La Laguna, Tenerife,  Spain}
\affiliation{Universidad de La Laguna, Departamento de Astrof\'isica - La Laguna, Tenerife, Spain\vspace{1mm}}

\setcounter{footnote}{0}
\def\thefootnote{\arabic{footnote}}

\begin{abstract}
Dark sector particles at the GeV scale carrying baryon number provide an attractive framework for understanding the origin of dark matter and the matter-antimatter asymmetry of the universe.
We demonstrate that dark decays of hadronic states containing strange quarks ---hyperons--- offer excellent prospects for discovering such dark baryons.
Building up on novel calculations of the matrix elements relevant for hyperon dark decays, and in view of various collider, flavor, and astrophysical constraints, we determine the expected rates at hyperon factories like BESIII and LHCb.
We also highlight the interesting theoretical connections of hyperon dark decays to the neutron lifetime anomaly and Mesogenesis.
\end{abstract}

\maketitle
{
  \hypersetup{linkcolor=black}
 \setlength\parskip{-0.2pt}
  \setlength\parindent{0.0pt}
\tableofcontents
}

\section{Introduction}\label{intro}

The standard model of inflationary cosmology predicts a Universe born with equal amounts of matter and anti-matter, see, e.g., \cite{Canetti:2012zc}. 
In order to explain the existence of the complex structures encountered today, it is necessary to introduce a  dynamical mechanism that can generate the primordial asymmetry between matter and antimatter. 
To add to this conundrum, a plethora of observations~\cite{Bertone:2016nfn} show that the visible baryonic matter only makes up about 15\% of the matter in the Universe.
The remaining 85\% must therefore consist of some unknown constituent dubbed \emph{dark matter}.
While there is plenty of evidence supporting the existence of dark matter through its gravitational interaction, to date experiments have failed to detect it in the laboratory (see, e.g., \cite{Bertone:2018krk}). 

Questions about the nature of dark matter and the origin of the baryon asymmetry of the universe are among the most relevant outstanding open problems in fundamental physics. Although many scenarios have been developed to tackle these problems, most of them can only solve a single one of them. On the one hand, popular mechanisms explaining the observed matter-antimatter asymmetry of the Universe include leptogenesis~\cite{Fukugita:1986hr,Bodeker:2020ghk}, GUT baryogenesis~\cite{Ellis:1978xg,Riotto:1999yt}, electroweak baryogenesis~\cite{Kuzmin:1985mm,Cohen:1993nk}, and Affleck-Dine baryogenesis~\cite{Affleck:1984fy,Dine:2003ax}. On the other, some of the  many well-motivated dark matter  candidates proposed are axions~\cite{Sikivie:2006ni,Marsh:2015xka}, WIMPs~\cite{Bertone:2004pz,Feng:2010gw}, sterile neutrinos~\cite{Dodelson:1993je,Drewes:2016upu}, and primordial black holes~\cite{Carr:2016drx,Villanueva-Domingo:2021spv}.
These scenarios predict dark matter and baryon abundances which are, in principle, not linked to each other. However, observations show that the baryon and dark matter energy densities are fairly similar, $\Omega_{\rm DM}/\Omega_{b}\simeq 5.36\pm0.06$~\cite{Planck:2018vyg}, leading to a \emph{coincidence problem} that in most cases has to be addressed \emph{ad hoc}.

\vspace{0.05in}
\noindent \textbf{Dark Sector Baryons ---}
The similar magnitudes of the baryon and dark matter abundances suggest that there might exist a connection between  the dark sector and baryogenesis. Indeed, theories of asymmetric dark matter  have been proposed to explain this connection~\cite{Petraki:2013wwa,Zurek:2013wia,Davoudiasl:2010am,Davoudiasl:2012uw}. These scenarios predict similar particle/antiparticle asymmetries in the visible and dark sectors, $n_b-n_{\bar{b}} \sim n_{\overline{\rm DM}}-n_{\rm DM}$, which explains  $\Omega_{\rm DM} \sim \Omega_b$ provided that the dark matter particle mass is at the GeV scale. Importantly, in these scenarios the dark sector typically contains several particles  and at least some of these states interact with the visible sector.

The simplest effective interactions connecting the dark and baryonic sectors arise from dimension six operators
\bea
\mathcal{O}_{abc} = u_a d_b d_c \chi \,,
\label{eq:generalOp}
\eea 
where, schematically, $u$ ($d$) is an \textit{up}- (\textit{down}-)type quark of flavor labeled by the subindex, and $\chi$ is a fermionic dark sector state (but not necessarily the dark matter particle itself). The existence of such operators implies that the dark sector particle $\chi$ interacts with hadrons, which can give rise to several interesting effects beyond the Standard Model (SM). 

\vspace{0.05in}
\noindent \textbf{Neutron Lifetime Anomaly and Mesogenesis ---} In  recent years, several proposals have triggered interest in dark baryonic sectors interacting with the SM via operators  in Eq.~\eqref{eq:generalOp}. Firstly, it has been pointed out that a dark decay of the neutron can resolve the long-standing \emph{neutron lifetime anomaly}~\cite{Fornal:2018eol} (see also \cite{Fornal:2020gto} for related works). 
Secondly, a new set of low-scale baryogenesis models has been recently put forward~\cite{McKeen:2015cuz,Aitken:2017wie,Elor:2018twp,Nelson:2019fln, Alonso-Alvarez:2019fym,Alonso-Alvarez:2021qfd,Elor:2020tkc,Elahi:2021jia}, in which a crucial ingredient are decays of hadrons into dark sector baryons triggered by operators in Eq.~\eqref{eq:generalOp}.
These low-scale baryogenesis scenarios are collectively referred to as \emph{Mesogenesis}. One may argue that, given that the dark decay of the neutron requires one specific variation of the operators in Eq.~\eqref{eq:generalOp} involving only first-generation quarks, while Mesogenesis models need different ones involving heavy flavors, both paradigms are in principle unconnected.
However, in the absence of any concrete symmetry argument, all flavor combinations of the operators in Eq.~\eqref{eq:generalOp} are expected to be present simultaneously.
This leads to interesting connections between the aforementioned scenarios and phenomenological predictions regarding \emph{apparent} baryon number violating signatures in hadron decays~\cite{Barducci:2018rlx,Fajfer:2020tqf,Heeck:2020nbq,Fajfer:2021woc}.

Contrary to other dark matter candidates, GeV-scale dark baryons remain largely unconstrained. For example, studies of the neutron lifetime anomaly have broadly showcased that the neutron may have a branching ratio to dark sector states as large as $1\%$~\cite{Fornal:2018eol}, whereas studies of the $B$-Mesogenesis paradigm \cite{Alonso-Alvarez:2021qfd} highlighted that $B$ mesons can have a branching ratio as large as $0.5\%$ to GeV-scale dark sector baryons (in stark contrast to other modes, such as $B_s \to \mu^+\mu^-$, see, e.g.,~\cite{Geng:2021nhg}, for which sensitivities at the level  $10^{-9}$  have been reached). These theoretical developments have  triggered dedicated searches for new neutron decay channels such as $n\to \chi\gamma$~\cite{PhysRevLett.121.022505}, $n\to\chi\, e^+e^-$\cite{Sun:2018yaw} and nuclear dark decays \cite{Pfutzner:2018ieu,Ayyad:2019kna}, as well as searches for the decays of non-standard $b$-flavored hadrons at experiments like BaBar, Belle and Belle-II~\cite{Belle:2021gmc}, and LHCb~\cite{Borsato:2021aum,Rodriguez:2021urv}. 

\begin{figure}[t]
\centering
\includegraphics[width=0.48\textwidth]{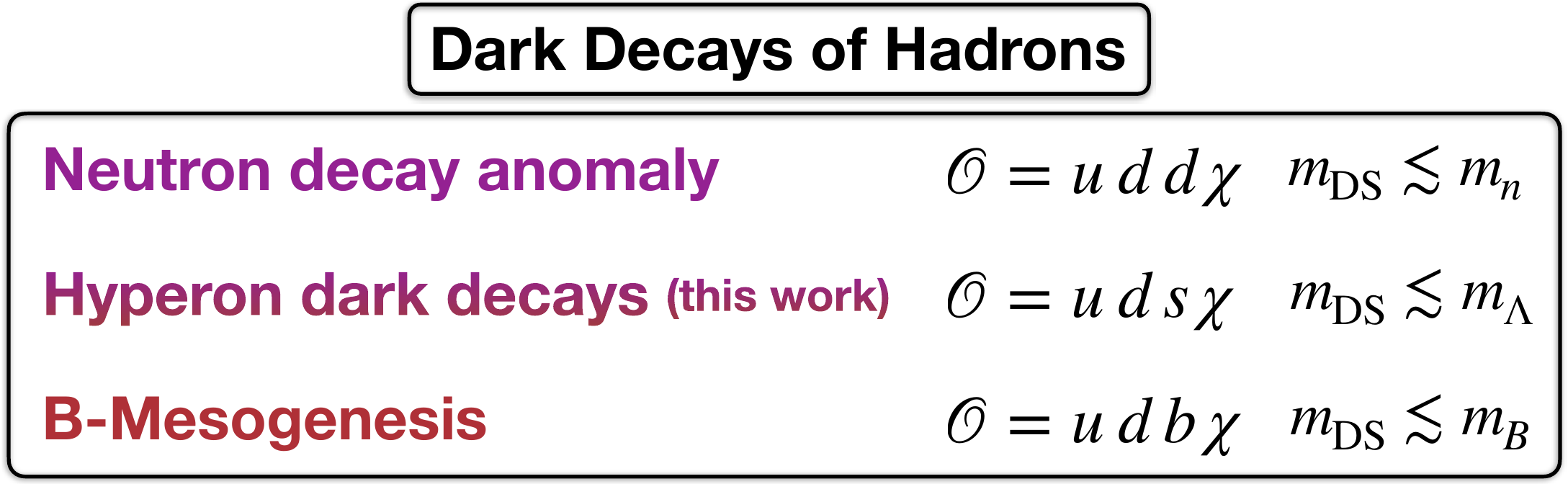}
\caption{Phenomenology of various flavor variations of the  dark baryon $\chi$ -- SM interactions. Couplings to first generation quarks lead to a dark decay solution to the neutron lifetime anomaly, while interactions with third generation quarks enable $B$-Mesogenesis. Interactions with second generation quarks lead to the \emph{apparent} baryon-number-violating signatures in hyperon decays studied in this work. The maximum mass allowed for dark sector particles ($m_{\rm DS}$) produced in the respective hadron decay is also indicated.
}
\label{fig:mindthegap}
\end{figure}

\vspace{0.05in}
\noindent \textbf{Strangeness and Dark Baryon Sectors ---} The discussion above highlights the incipient developments in the study of baryonic dark sectors interacting with first ($u$, $d$) and third ($b$) generation quarks.
Couplings to the second generation have received much less attention~\cite{Heeck:2020nbq,Fajfer:2020tqf,Fajfer:2021woc}, in part due to the lack of direct phenomenological applications.
However, from a theoretical perspective one would expect that baryonic dark sectors interact with all SM quark flavors.
As illustrated in Fig.~\ref{fig:mindthegap}, in this work we attempt to fill this gap by considering the possible presence of GeV-scale dark baryons in hyperon decays.
This is a very timely endeavour, since a large number of hyperons is currently being produced and analyzed at BESIII~\cite{BESIII:2018cnd,Li:2016tlt}, in $e^+e^-$ collisions with energy at the invariant mass of the $J/\psi$ resonance, and at LHCb~\cite{Alves:2018npj}. In fact, the BESIII collaboration has very recently reported a search for totally invisible decays of $\Lambda$ hyperons yielding $\text{BR}(\Lambda \to \text{invisible}) < 7.4\times10^{-5}$ at 90\%~\cite{BESIII:2021slv}. Moreover, there are plans to build a Super Charm-Tau Factory which would considerably increase the hyperon data sets with richer physical information provided by polarized $e^+e^-$ beams~\cite{Barnyakov:2020vob,Goudzovski:2022vbt}.

In this paper, we investigate all relevant aspects of hyperon decays into GeV-scale dark sector particles. In particular, we:
\begin{enumerate}[label=\roman*)]
\item Characterize all experimentally relevant decay modes using the framework of chiral effective field theory to reliably predict the relevant hadronic form factors.
\item Derive astrophysical bounds on these new decay channels from the duration of the neutrino signal from supernova SN~1987A. 
\item Obtain indirect bounds from LHC searches and neutral meson oscillations on the colored bosons needed to mediate the exotic hyperon decays.
\end{enumerate}
In light of the combined bounds from the LHC and SN 1987A, we discuss the required sensitivities that the searches at BES III and LHCb need to achieve in order to test uncharted regions of parameter space. In addition, we discuss the relevance of our results for Mesogenesis and the neutron lifetime anomaly. 

\vspace{0.05in}
\noindent \textbf{Structure of this paper ---} Our work and results are divided into the following sections:

\textit{Section~\ref{sec:Models}: Models and Effective Operators}. We start by compiling an exhaustive list of operators that can trigger the decay of SM hadrons into dark sector baryons and their possible UV completions. 
   
\textit{Section~\ref{sec:DarkDecays}: Hyperon Dark Decays}. In this section, we develop the chiral perturbation theory framework needed to calculate the rates of hyperon decays into dark sector baryons (Sec.~\ref{sec:Matching}). We consider invisible decays in Sec.~\ref{sec:Baryontoinv}, pionic decays in Sec.~\ref{sec:Baryontopion}, and radiative decays in Sec.~\ref{sec:Baryontogamma}.
Although Sec.~\ref{sec:DarkDecays} contains the foundational elements of our present work, a reader interested primarily in  phenomenology may wish to skip the technical details and go directly to the summary provided in Sec.~\ref{sec:decaysummary}, including the list of the most promising channels in Table.~\ref{tab:decayChannels}.

\textit{Section~\ref{sec:supernova}: Supernova constraints}. Here we derive an upper limit on the dark hyperon decay rates using a stellar cooling argument based on the observed duration of the neutrino pulse from SN~1987A. We present detailed calculations of the luminosity in form of dark-sector particles and explain the specific numerical implementation using state-of-the-art simulations of core-collapse supernovae.
    
\textit{Section~\ref{sec:phenomenology}: Phenomenology}. This section is devoted to exploring the relevant phenomenology associated with dark hyperon decays. In particular, we determine  the LHC constraints (Sec.~\ref{sec:LHC}) and meson mixing bounds (Sec.~\ref{sec:meson_mixing}) on the mediators needed to generate the effective operators under consideration.

\textit{Section~\ref{sec:theory}: Results and Implications}.  
We present and discuss our results in Sec.~\ref{sec:decaysummary}. This section contains a summary of our findings (see Table.~\ref{tab:decayChannels}), including a list of  strange hadron dark decay channels offering the most promising discovery  prospects in upcoming hyperon facilities. Armed with the machinery of Sec.~\ref{sec:DarkDecays} and the phenomenological analysis of Sec.~\ref{sec:phenomenology}, in this section we examine the extent to which the upcoming searches at hyperon factories can test the regions of parameter space relevant for the neutron decay anomaly and $B$-Mesogenesis. 

\section{Models and effective operators}
\label{sec:Models}

The simplest way to construct UV-complete models producing the effective operators in Eq.~\eqref{eq:generalOp} is to introduce a new heavy colored bosonic field.
This field should have a coupling to two quarks, and a coupling to a quark and the dark baryon $\chi$. There are three possible gauge-invariant realizations of such a scenario, which correspond to the following Lagrangians:
\begin{eqnarray}
\label{eq:Models}
\mathcal{L}_{1}&&\,\,\, \supset \,\,\,-\,y_{d_ad_b} \epsilon_{ijk} \Psi^i d_{Ra}^j d_{Rb}^{k}  -\,y_{\chi u_c}\Psi^{*}_i {\chi_R} u_{Rc}^i  + {\rm h.c.}, \nonumber \\
\mathcal{L}_{2}&& \,\,\, \supset \,\,\, -\,y_{u_ad_b} \epsilon_{ijk} \Phi^i u_{Ra}^j d_{Rb}^k -\,y_{\chi d_c} \Phi^{*}_i {\chi_R}d_{Rc}^i\nonumber\\
&&\qquad  -\,y_{Q_aQ_b} \epsilon_{ijk}\epsilon_{\alpha\beta} \Phi^i Q_{La}^{j\alpha} Q_{Lb}^{k\beta} +{\rm h.c.},
\nonumber \\   
\mathcal{L}_{3}&& \,\,\, \supset \,\,\,-\,y_{Q_ad_b} \epsilon_{ijk}\epsilon_{\alpha\beta} X_{\mu}^{i\alpha} Q_{La}^{j\beta}\sigma^\mu d_{Rb}^k \nonumber\\
&&\qquad \,-\,y_{\chi Q_c} X_{\mu}^{\dagger i\alpha} Q_{Lc}^{i\alpha}\sigma^\mu {\chi_R}   + {\rm h.c.},
\end{eqnarray}
where $u_R$, $d_R$ and $Q_L$ denote the SM quark fields (in the 2-spinor notation of \cite{Abbott:1980zj}), $i$, $j$ and $k$ are color indices, $a$, $b$ and $c$ are generation indices and $\alpha$ and $\beta$ are $SU(2)_L$ indices. 
The quantum numbers of the new heavy colored mediator particles in Eq.~\eqref{eq:Models} are $\Psi = (3,1)_{\frac23}$ and $\Phi = (3,1)_{-\frac13}$ for the color-triplet scalars, and $X_\mu = (3,2)_{\frac16}$ for the color-triplet vector. Note that in $\mathcal{L}_{1}$ the two $d_{R}$ quarks must belong to different generations due to the antisymmetry of the color indices.

The field $\chi$ is a SM singlet Dirac spinor with baryon number $B_\chi=1$. Note that we impose baryon number conservation\footnote{This way the coupling of $\Phi$ and $X_\mu$, which carry quantum numbers of leptoquarks, to a quark-lepton pair is automatically forbidden. Such a coupling would generically lead to proton decay.} on the models in Eq.~\eqref{eq:Models}. To be consistent with experiments, such a dark baryon must be heavy enough to kinematically block the proton and all known stable nuclei from decaying. This requires \cite{Fornal:2018eol,Pfutzner:2018ieu}
\begin{align}
\label{eq:ndecay}
    m_\chi > 937.993 \ {\rm MeV} \ .
\end{align}
In particular, this bound assures that neutrons inside the $^9{\rm Be}$ nucleus do not undergo the decay $n \to \chi \,\gamma$. 
This process would otherwise occur at unacceptable rates even in the absence of tree-level $\chi$--first generation quark couplings through loop effects. 
Enforcing Eq.~\eqref{eq:ndecay} along with $m_\chi<m_n = 939.565\,\text{MeV}$, nuclei are stable but a free neutron can  undergo dark decays as will be discussed in Sec.\,\ref{sec:theory}.
If $m_\chi>m_n$, the neutron remains stable with respect to dark decays, but heavier baryons and mesons can still experience exotic desintegrations~\cite{Barducci:2018rlx,Heeck:2020nbq,Fajfer:2021woc}.

As will be shown in Sec.~\ref{sec:LHC}, direct searches~\cite{ATLAS:2017jnp,CMS:2018mts,Sirunyan:2019ctn,Aad:2020aze} for new colored states like those in Eq.~\eqref{eq:Models} set an absolute lower bound on the masses of $\Psi$, $\Phi$ and $X_\mu$ of
\begin{align}\label{LHCm}
    M > 0.52\, \text{TeV} \,.~ 
\end{align}
The heavy colored states can therefore be integrated out at the lower energies relevant for hadron decays.
This procedure leads to the effective Lagrangian
\begin{eqnarray}
\label{eq:SMEFTchi}
\mathcal L_{\rm eff}&& \,\,=\,\,C_{ab,c}\,\mathcal O_{ab,c}+C^\prime_{ab,c}\mathcal O^\prime_{ab,c}, 
\end{eqnarray}
with contains the four-fermion operators
\begin{eqnarray}
\mathcal O_{ab,c}&&\,\,=\,\,\epsilon_{ijk}(u^i_{Ra}d^j_{Rb})(\chi_R d^k_{Rc}),\nonumber\\
\mathcal O^\prime_{ab,c}&&\,\,=\,\,\epsilon_{ijk}\epsilon_{\alpha\beta}(Q^{i\alpha}_{La}Q^{j\beta}_{Lb})(\chi_R d^k_{Rc}).
\label{eq:SMEFTchi_operators}
\end{eqnarray}
Here, $C_{ab,c}^{(\prime)}\sim y^2/M^2$ are the Wilson coefficients and $C_{ab,c}^{\prime}$ is symmetric under the permutation of the flavor indices $a$ and $b$. 
In deriving the effective Lagrangian in Eq.~\eqref{eq:SMEFTchi} we have used Fierz relations to express the current-current operators produced by the exchange of $X_\mu$ (Model 3) in terms of the scalar operator $\mathcal O^\prime_{ab,c}$~\cite{Weinberg:1979sa}. An additional operator
\begin{equation}
\mathcal O_{ab,c}^{\prime\prime}=\epsilon_{ijk}(d_{Ra}^i d_{Rb}^j)(\chi_R u_{Rc}^k)
\end{equation}
is produced directly by the exchange of $\Psi$ in  Model 1; this can however also be related to the previous ones by means of the Fierz relation~\cite{Abbott:1980zj}
\begin{equation}
\label{eq:FierzModel1}
\mathcal O^{\prime\prime}_{ab,c}=\mathcal O_{cb,a}-\mathcal O_{ca,b} \,. 
\end{equation}
The contributions of the three models in Eq.~\eqref{eq:Models} to each of the Wilson coefficients in Eq.~\eqref{eq:SMEFTchi} are shown in Table~\ref{tab:WCs}. The operators in $\mathcal L_{\rm eff}$ are the only dimension-six operators which are $SU(2)_L\times U(1)$ invariant and include SM quark fields coupled to the singlet field $\chi$ linearly. 

In addition to the effective couplings in Eq.\,(\ref{eq:SMEFTchi}), the dark particle $\chi$ may have interactions with other dark sector states\footnote{In fact, these are essential to allow for a \emph{pure} dark decay channel \cite{Fornal:2018eol} of the neutron, and are also needed in the $B$-Mesogenesis framework to avoid the washout of the generated baryon asymmetry~\cite{Elor:2018twp}.} 
A minimal model for such an interaction is provided by the Lagrangian term
\begin{align} \label{eq:Lag_dark}
    \mathcal{L} \ &\supset \  y_{\xi\phi}\, \bar\chi\, \xi\, \phi \ + \ \text{h.c.}\,,
\end{align}
where $\xi$ and $\chi$ denote four-component Dirac spinors and $\phi$ is a complex scalar, all of them singlets under the SM gauge group.
Conservation of baryon number implies that $B_\xi + B_\phi = 1$.
A direct coupling of $\xi$ and $\phi$ to quarks can be excluded by imposing a $\mathbb{Z}_2$ symmetry under which $\xi$ and $\phi$ are odd and all other particles are even.
Such a discrete symmetry ensures that the lightest among $\xi$ and $\phi$ is absolutely stable: this state thus becomes a candidate for dark matter. 
Stability of nuclei against decays to $\xi+\phi$ is guaranteed by the requirement $m_\xi+ m_\phi>937.993\,\text{MeV}$, with similar implications to those discussed below Eq.\,(\ref{eq:ndecay}). 

\begin{table}[t!]
\renewcommand{\arraystretch}{1.8}
  \setlength{\arrayrulewidth}{.25mm}
\centering
\begin{tabular}{ c | c | c | c  }
    \hline \hline
   Model & Mediator & $M^2\,C_{ab,c}$ & $M^2\,C^\prime_{ab,c}$\\ 
   \hline \hline
1& $\Psi = $ $(3,1)_{\frac23}$  & $(y_{d_bd_c}-y_{d_cd_b})y_{\chi u_a}$&0\\ \hline
2& \,$\Phi =  $ $(3,1)_{-\frac13}$ & $-y_{u_ad_b}y_{\chi d_c}$& $-y_{Q_aQ_b}y_{\chi d_c}$\\ \hline
3& $ X_\mu =$ $(3,2)_{\frac16}$  & 0 &$2y_{Q_ad_c} y_{\chi Q_b}$\\
 \hline \hline
\end{tabular}
\caption{Contributions of the three different models in Eq.~\eqref{eq:Models} to the Wilson coefficients of the operators $\mathcal O_{ab,c}^{(\prime)}$. The flavor indices $a$ and $b$ are understood to be symmetryzed for the Wilson coefficient $C^{\prime}_{ab,c}$.
\label{tab:WCs}}
\end{table}

\section{Hyperon dark decays}\label{sec:DarkDecays}

\begin{figure*}[t!]
		\includegraphics[width=0.85\textwidth]{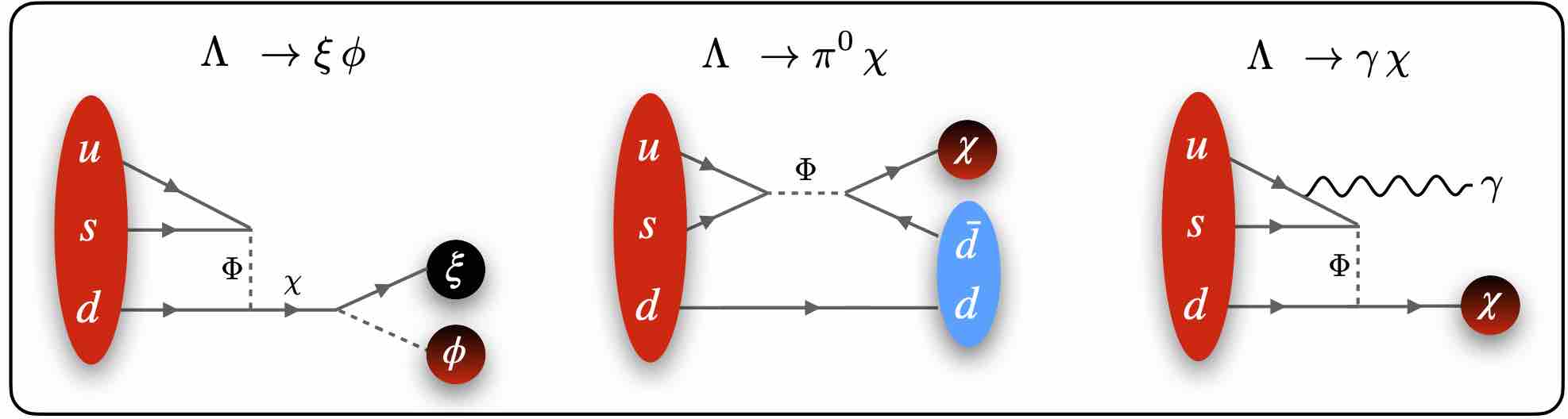}
\caption{Three possible decay channels of the $\Lambda$ hyperon into final states containing dark baryonic particles. The processes displayed can be induced by the effective operators $\mathcal{O}_{ud,s}$ and $\mathcal{O}'_{ud,s}$ in Eq.~\eqref{eq:SMEFTchi_operators}. For concreteness and illustration purposes, we choose to show the scalar mediator $\Phi$ corresponding to Model 2 in Table~\ref{tab:WCs}, but $\Psi$ and $X_\mu$ can mediate analogous decays. 
}
\label{fig:cartoon} 
\end{figure*}

The effective Lagrangian in Eq.~\eqref{eq:SMEFTchi} induces interactions between $\chi$ and baryons.
In particular, if these interactions involve a strange quark, various new decay channels for hyperons are enabled. 
For each one of the initial states $\Lambda \, (uds)$, $\Sigma^0 \, (uds)$, $\Sigma^+ \, (uus)$, $\Sigma^+ \, (dds)$, $\Xi^0 \, (uss)$ and $\Xi^{-} \, (dss)$, 
in this section we compute the exclusive branching fractions for the following channels:
\begin{itemize}
\item \, Fully invisible decay.
\item \, Decay to $\pi^{0\,,\pm}$ and a dark baryon. 
\item \, Decay to a photon and a dark baryon. 
\end{itemize}
These processes are shown in Fig.~\ref{fig:cartoon} for the exemplary case of the $\Lambda$ hyperon and the scalar mediator $\Phi$.

Our goal is to produce theoretical predictions that can be used in present and upcoming hyperon factories to search for dark baryon sectors.
In the absence of a positive signal, our calculations would allow to set constraints on the Wilson coefficients of the operators $\mathcal O_{ab,c}^{(\prime)}$, and therefore on the couplings of the mediators listed in Table~\ref{tab:WCs}. 
Deriving the predictions for the hyperon exclusive branching fractions induced by the operators in Eq.~\eqref{eq:SMEFTchi} requires the knowledge of relevant baryonic form factors as input.
Those can be computed within the framework of chiral effective field theory, also known as chiral perturbation theory (ChPT). We do this by following the formalism introduced for proton decay in the context of GUTs in \cite{Claudson:1981gh} (see also~\cite{Aoki:2008ku,Aoki:2017puj}).

\subsection{Matching to the chiral EFT} \label{sec:Matching}

In order to connect the effective Lagrangian in Eq.~\eqref{eq:SMEFTchi} to the operators triggering hyperon decays to dark baryons, one needs to break up the doublets and rotate the quark fields from the gauge to the mass eigenstate basis. For simplicity, we assume that the right-handed fields and $d_L$ are defined in their mass eigenstate basis. 
Focusing on the couplings to light quarks and neglecting contributions suppressed by $\lambda\sim V_{us}\simeq0.22$, one obtains
\begin{align}
\label{eq:LEEFTchi}
&\mathcal L_{\rm eff} \,\, \supset \,\, C^{R}_{ud,d}\mathcal O^{R}_{ud,d}+ C^{L}_{ud,d}\mathcal O^{L}_{ud,d}\nonumber\\
&+C^{R}_{ud,s}\mathcal O^{R}_{ud,s}+C^{R}_{us,d}\mathcal O^{R}_{us,d}+C^{L}_{ud,s}\mathcal O^{L}_{ud,s}+C^{L}_{us,d}\mathcal O^{L}_{us,d}\nonumber\\
&\qquad +\,\, C^{R}_{us,s}O^{R}_{us,s}+C^{L}_{us,s}\mathcal O^{L}_{us,s}\,,
\end{align}
where the operators in the first, second and third line correspond to the change of strangeness by $\Delta S=0$, 1 and 2, respectively. We have also  relabeled the operators,
\begin{align}
\mathcal O_{ud_a,d_b}^{R}&=\epsilon_{ijk}(u_R^i d_{Ra}^j)(\chi_R d_{Rb}^k)\,,\nonumber\\
\mathcal O_{ud_a,d_b}^{L}&=\epsilon_{ijk}(u_L^i d_{La}^j)(\chi_R d_{Rb}^k) \,,
\end{align}
and the Wilson coefficients, $C^{R}_{ud_a,d_b}=C_{1a,b}$,  $C^{L}_{ud_a,d_b}\simeq2 C^\prime_{1a,b}$, using those in Table~\ref{tab:WCs} under our flavor assumptions and ignoring renormalization group effects when connecting the heavy scale $M$ to the scale characteristic for hyperon decays, $\sim1$ GeV. In the discussion  below we use the short-hand notation $(ud_a)d_b$ for the operator $\mathcal O_{ud_a,d_b}^{L/R}$. In addition,  $(u[d),s]$ represents the combination $\mathcal O^R_{ud,s}-\mathcal O^R_{us,d}$ relevant for Model 1 (see Eqs.~\eqref{eq:Models} and \eqref{eq:FierzModel1}).

At the low scale relevant for hyperon decays, we match the particle-level Lagrangian to the chiral Lagrangian, where hadrons, not quarks and gluons, are the explicit degrees of freedom. Chiral perturbation theory~\cite{Weinberg:1968de,Coleman:1969sm,Callan:1969sn,Weinberg:1978kz,Gasser:1983yg,Gasser:1984gg,Gasser:1987rb,Pich:1995bw,Scherer:2012zzd}  implements the spontaneously broken $SU(3)_L\times SU(3)_R$ chiral symmetry of QCD supplemented by a ``power counting''  based on a momentum expansion $p\ll\Lambda_{\rm ChPT}\sim1$ GeV. The pseudo-Goldstone bosons are described in terms of a unitary $3\times3$ matrix $U(x)$,
\begin{equation}
 U(x)=\exp\left(i\,\frac{\phi(x)}{f}\right),
\end{equation}
where $f$ is the pion decay constant and
\begin{eqnarray}
\phi=\left(\begin{array}{ccc}\pi^0+\frac{1}{\sqrt{3}}\eta_8&\sqrt{2}\pi^+&\sqrt{2}K^+\\
           \sqrt{2}\pi^-&-\pi^0+\frac{1}{\sqrt{3}}\eta_8&\sqrt{2}K^0\\
           \sqrt{2}K^-&\sqrt{2}\bar{K}^0&-\frac{2}{\sqrt{3}}\eta_8\end{array}\right).
\end{eqnarray}
Under chiral transformations $(L,R)\in SU(3)_L\times SU(3)_R$, $U$ transforms like
\begin{equation}
 U\rightarrow L \,U R^{\dagger}. \label{Eq:Chiraltransfmesons}
\end{equation}
The lowest-lying $J^P=1/2^+$ baryon octet fields, on the other hand, are described by a traceless $3\times3$ complex matrix
\begin{equation}
B=\left(\begin{array}{ccc}\frac{1}{\sqrt{2}}\Sigma^0+\frac{1}{\sqrt{6}}\Lambda&\Sigma^+&p\\
           \Sigma^-&-\frac{1}{\sqrt{2}}\Sigma^0+\frac{1}{\sqrt{6}}\Lambda&n\\
           \Xi^-&\Xi^0&-\frac{2}{\sqrt{6}}\Lambda\end{array}\right).
\end{equation}
The $B$ fields and the variable $u=\sqrt{U}$ transform as
\begin{eqnarray}
u\rightarrow L u K^\dagger=KuR^\dagger,\hspace{0.7cm}B\rightarrow KBK^{-1}\,,\label{Eq:TrnsfCompensator}
\end{eqnarray}
where
\begin{equation}
K\equiv K(L,R,U)=\left(\sqrt{UL^\dagger}\right)^{-1}\!\left(\sqrt{RU}\right) \label{Eq:EqCompensator}
\end{equation}
is the so-called {\it compensator field}. For a vector transformation $L=R=V$ one can easily see 
from Eqs. \eqref{Eq:Chiraltransfmesons}  and \eqref{Eq:EqCompensator} that $K=V$, transforming the 
baryons $B$ linearly as an octet under the vectorial $SU(3)$ subgroup of the chiral group. Next, one 
introduces the $vielbein$
\begin{align}
u^\mu&=i\left(u^\dagger\left(\partial^\mu-il^\mu\right) u-u\left(\partial^\mu-ir^\mu\right)u^\dagger\right),\nonumber\\
u^\mu &\rightarrow K u^\mu K^\dagger, \label{Eq:vielbein}
\end{align}
and a covariant derivative (with the $connection$ $\Gamma^\mu$)
\begin{eqnarray}
&&\Gamma_\mu=\frac{1}{2}\left(u^\dagger\left(\partial_\mu-il_\mu\right)u+u\left(\partial_\mu-ir_\mu\right)u^\dagger\right),\nonumber\\
&&D_\mu X=\partial_\mu X+[\Gamma_\mu,X],~~~D_\mu X\rightarrow K D_\mu X\,, 
\label{Eq:connection}
\end{eqnarray}
where $l^\mu$ and $r^\mu$ are external left- and right-handed vector sources. 
One assigns the following power counting:
\begin{eqnarray}
&D_\mu B,\,\bar{B},\, B \sim 
\mathcal{O}(1),\nonumber\\
&u_\mu,~D_\mu u_\nu,~D_\mu 
u\sim \mathcal{O}(p)\,. 
\end{eqnarray}
With these ingredients,  the leading order chiral Lagrangian containing octet baryons is
\begin{align}
\mathcal{L}_{\phi B}^{(1)}&=\langle 
\bar{B}\left(i\slashD-M_{0}\right)B\rangle\nonumber\\
&+\frac{D}{2}\langle\bar{B}\gamma^\mu\gamma_5\{u_\mu,B\}\rangle 
+\frac{F}{2}\langle\bar{B}\gamma^\mu\gamma_5[u_\mu,B]\rangle, \label{Eq:LeadingLagOctet}
\end{align}
where the angle brackets indicate that a trace is taken, $D$ and $F$ are nonperturbative couplings, $\left(i\slashD-M_{0}\right)B\sim\mathcal{O}(p)$ and $M_{0}$ is a common mass of the octet baryons obtained in the limit of massless quarks (chiral limit).

The operators in Eq.~\eqref{eq:LEEFTchi}  have well-defined  $SU(3)_L\times SU(3)_R$ transformations and this allows one to find their counterparts in the chiral EFT~\cite{Claudson:1981gh}. For instance, $\mathcal O^{R}_{ud_a,d_b}$ and $\mathcal O^{L}_{ud_a,d_b}$ transform as $(1,8)$ and $(\bar 3,3)$ under $SU(3)_L\times SU(3)_R$, respectively.\footnote{Note that $3\otimes3=\bar3\oplus6$, but only the antisymmetric irreducible representation  $\bar3$ is possible due to the simultaneous action of the antisymmetric color tensor. Similarly, the purely singlet representation (1,1) is identically zero.} To construct the chiral representation of the effective Lagrangian we treat the coefficients $C^{R}$ and $C^{L}$ as spurions transforming as components of a $(1,8)$ and a $(3,\bar 3)$ under $SU(3)_L\times SU(3)_R$, respectively.  Defining $\hat C^{L/R}_{nm}\equiv \tfrac12\sum_{i,j} \epsilon_{mij}C^{L/R}_{ij,n}$, so that $\hat C^{R}\to R \hat C^{R} R^\dagger$ and $\hat C^{L}\to R \hat C^{L} L^\dagger$, the chiral representation of the effective Lagrangian is, at leading order, 
\begin{equation}
\label{eq:EFTChiral}
\mathcal L^{(0)}_{\rm eff,ChPT}= \alpha\big\langle  \hat C^{L} u^\dagger  B_R\chi_R u^\dagger \big\rangle
+\beta \big\langle\hat C^{R} u^\dagger B_R\chi_R u\big\rangle,
\end{equation}
where $\alpha$ and $\beta$ are two new nonperturbative couplings. Note that our formalism is equivalent to  introducing flavor projection matrices, as described  in~\cite{Claudson:1981gh}.

Given the Lagrangian in Eq.~\eqref{eq:EFTChiral}, one can compute all the relevant matrix elements involving the decays of hyperons, including any possible number of pions in the final state, at  leading order in the chiral expansion. For the calculations, we take the numerical inputs\footnote{The normalization of the pion field $\phi$ differs from that in \cite{Claudson:1981gh} by a factor of $\sqrt2$. Hence,  the value of the decay constant in \cite{Claudson:1981gh} is correspondingly larger, 131~MeV.} $f=f_\pi=92.4$ MeV from pion decay, $D=0.80$ and $F=0.46$ from the leading order fit applying Eq.~\eqref{Eq:LeadingLagOctet} to semileptonic hyperon decays~\cite{Ledwig:2014rfa}, and $\alpha\simeq-\beta=-0.014(2)$ GeV$^3$ obtained from lattice QCD calculations~\cite{Aoki:2017puj}. 

In our computation of the matrix elements, we also implement some  contributions which are  higher order in the chiral expansion. On one hand, we use the physical masses of the octet baryons (for which we take the PDG values~\cite{pdg}) instead of the commonly used $SU(3)$-symmetric value $M_0$ in $\mathcal L^{(1)}_{\phi B}$.  This is equivalent to introducing $SU(3)$-breaking corrections starting at $\mathcal{O}(p^2)$, which already provide an accurate description of the data. On the other hand, the radiative decays $\mathfrak B\to \gamma\,\chi$ are induced by the magnetic moment of an octet baryon $\mathfrak B$, which formally enters in the $SU(3)$-symmetric  limit at $\mathcal O(p^2)$. In our calculation we use the experimental values  for these quantities, described by
\begin{align}
\label{eq:TransMagMom}
\mathcal L_{\rm ph}=\frac{e\kappa_{\mathfrak B^\prime\mathfrak B}}{4m_p} \bar{\mathfrak B}^\prime \sigma^{\mu\nu}\mathfrak B F_{\mu\nu}\,,
\end{align}
where $\kappa_{\mathfrak B^\prime\mathfrak B}$ is the experimental value of the magnetic moment expressed  in nuclear magnetons, $\mu_{\rm N}=e/2m_p$.\footnote{The value of the $\Sigma^0$ magnetic moment and the sign of the $\Lambda\Sigma^0$-transition one are not known experimentally. We use the values predicted by ChPT: $\mu_{\Sigma^0}=0.66~\mu_{\rm N}$ and  $\mu_{\Lambda\Sigma^0}=+1.58~\mu_{\rm N}$~\cite{Geng:2008mf}  (the experimental absolute value agrees with the latter prediction at a few-percent level).}

Finally, for  the dark sector particles at the GeV scale  considered in this work, the final photons and pions produced in  hyperon decays carry energies $\sim 100$ MeV. This is smaller than the cutoff of the EFT, $\Lambda_{\rm ChPT}$, and the ChPT predictions are expected to be reliable. This differs from the typical kinematics encountered in GUT-induced proton decay, such as $p\to \pi^0 e^+$, where the SM particles recoil with a higher energy and the predictions are less reliable (see~\cite{Aoki:2017puj} for a detailed discussion).

\subsection{Matrix elements for $\mathfrak B\to\text{Dark Sector}$}\label{sec:Baryontoinv}

We first define the matrix elements
\begin{align}
\label{eq:Btovac}
&\langle 0|\epsilon_{ijk}(u^{iT}CP_L d_{a}^j) P_R d_{b}^k|\mathfrak B(p)\rangle=\gamma^L_{\mathfrak B} P_R u(p)\,,\nonumber
\\
&\langle 0|\epsilon_{ijk}(u^{iT}C P_R d_{a}^j) P_Rd_{b}^k|\mathfrak B(p)\rangle=\gamma^R_{\mathfrak B} P_R u(p)\,,
\end{align}
in terms of the 4-component spinors as in \cite{Aoki:2008ku,Aoki:2017puj}. Here, $C$ is the charge-conjugation matrix, $\mathfrak B$ is an octet baryon with the valence quark composition $(u d_ad_b)$, and $u(p)$ is the corresponding spinor. 
Expanding Eq.~\eqref{eq:EFTChiral} at zeroth order in the meson fields, one obtains predictions for the baryon decay constants $\gamma_{\mathfrak B}^{L/R}$. These are proportional to $\alpha$/$\beta$ with the coefficients shown in Table~\ref{tab:Btovac}.
\begin{table}[t!]
\renewcommand{\arraystretch}{1.3}
\begin{tabular}{c|c|c|c|c|c|c}
\hline\hline
Baryon & $n$ & \multicolumn{2}{c|}{$\Lambda$} & \multicolumn{2}{c|}{$\Sigma^0$} & $\Xi^0$\\
\hline \hline
Operator & $(ud)d$ & $(ud)s$ & $(us)d$ & $(ud)s$ & $(us)d$ &  $(us)s$\\
\hline
$a_{\mathfrak B}$ & 1 & $-\sqrt{2/3}$ &$-1/\sqrt{6}$ & 0 &$1/\sqrt{2}$&$-1$\\ 
\hline\hline
\end{tabular}
\caption{
{Baryon decay constants relevant for the ${\mathfrak B}\to \xi \,\phi$ decays}. They are mapped to the coefficients  in Eq.~\eqref{eq:Btovac} via $\gamma^{L}_{\mathfrak B}=a_{\mathfrak B}\alpha$ and $\gamma^R_{\mathfrak B}=a_{\mathfrak B}\beta$, where $\alpha\simeq-\beta=-0.014(2)$ GeV$^3$~\cite{Aoki:2017puj}. We display the values of $a_{\mathfrak B}$ for each of the  flavor configurations $(ud_a)d_b$ in the operators $\mathcal O^{L/R}_{ud_a,d_b}$.\label{tab:Btovac}}
\end{table}

The corresponding decay rate for $\mathfrak{B}\rightarrow\xi\phi$, using the dark sector interaction Lagrangian in Eq.~\eqref{eq:Lag_dark}, is thus given by
\begin{equation}\label{eq:DecayRate}
    \Gamma_{\mathfrak{B}\rightarrow\xi\phi} = \frac{|\vec{k}|}{8\pi m_{\mathfrak{B}}} \, |y_{\xi\phi}|^2 \,|C_{ud_a,d_b}^{L/R}|^2\,\mathcal{H}_{\mathfrak{B}}^{L/R},
\end{equation}
where $\vec k$ is the 3-momentum of the decay products in the $\mathfrak B$ rest frame and
\begin{equation}\label{eq:DSmatrixele}
    \mathcal{H}_{\mathfrak{B}}^{L/R} = |\gamma_\mathfrak{B}^{L/R}|^2 \frac{E_\xi(m_\mathfrak{B}^2+m_\chi^2) + 2m_\mathfrak{B}m_\chi m_\xi}{(m_\mathfrak{B}^2-m_\chi^2)^2}\,,
\end{equation}
given in terms of the various masses and momenta of the $\xi$ particle in the rest frame of the decaying baryon: $E_\xi = (m_\mathfrak{B}^2+m_\xi^2-m_\phi^2)/(2m_\mathfrak{B})$ and $|\vec{k}| = \sqrt{E_\xi^2-m_\xi^2}$. We note that the result in Eq.~\eqref{eq:DSmatrixele} depends on the interactions assumed for the dark sector in Eq.~\eqref{eq:Lag_dark}. 

\begin{figure}[t]
		\includegraphics[width=0.48\textwidth]{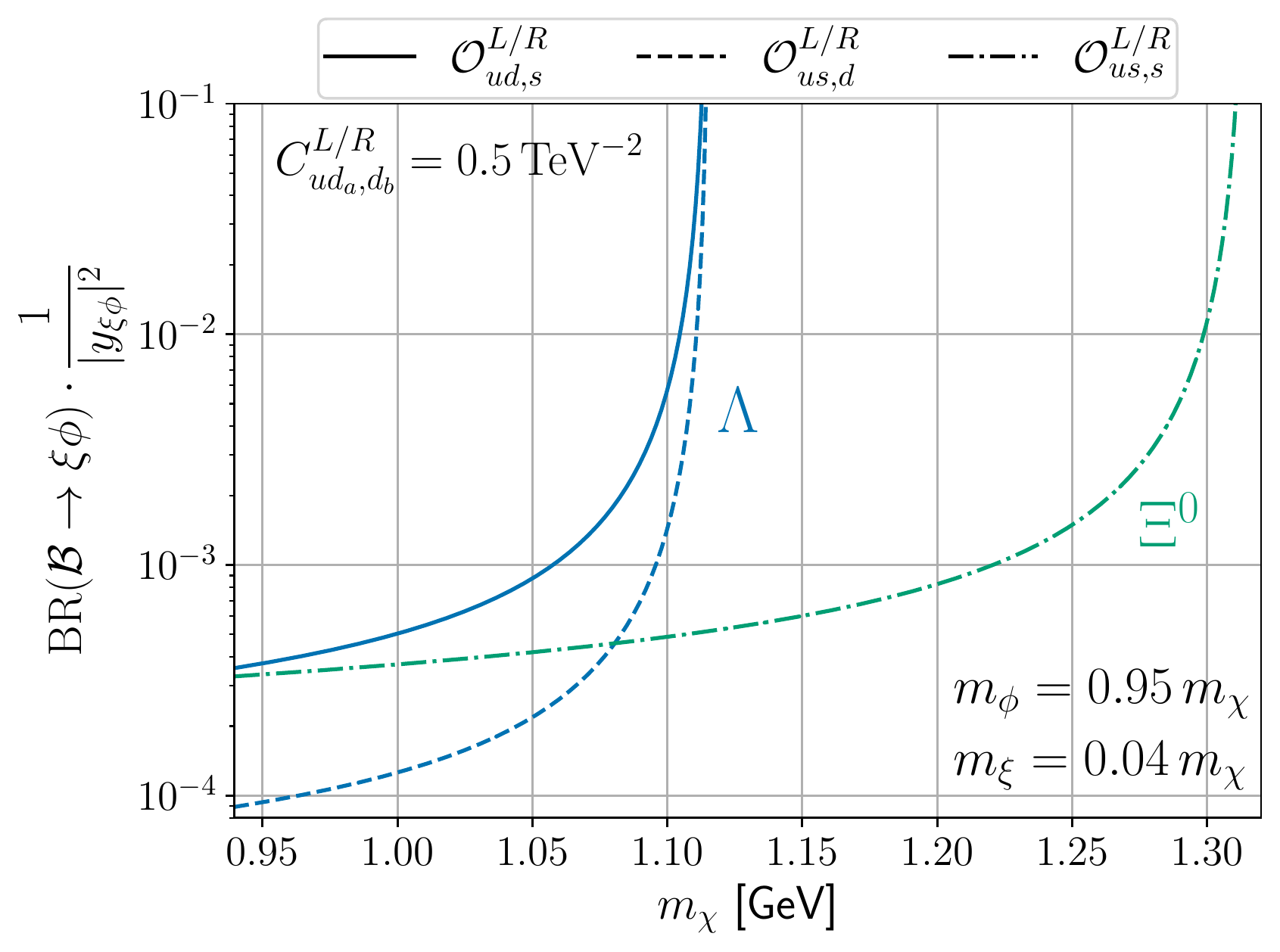}
\caption{
Branching ratios of neutral hyperons into a purely invisible final state, with the dark sector particle masses and couplings fixed to $m_\phi=0.95\,m_\chi$, $m_\xi=0.04\,m_\chi$, $y_{\xi\phi}=1$. Different colors correspond to different hyperons undergoing the dark decay, whereas different line styles represent different combinations or orderings of the down-type quarks in Eq.~\eqref{eq:LEEFTchi}.
We fix the Wilson coefficient to  roughly saturate the LHC constraints on the mediators, as discussed in the text (see Sec.~\ref{sec:LHC}). The branching ratios for other values of the Wilson coefficients can be obtained simply by noting that ${\rm BR} \propto |C^{L/R}_{ud_a,d_b}|^2$.
\label{fig:ChPT_xi_phi}}
\end{figure}

In Fig.~\ref{fig:ChPT_xi_phi} we plot the branching ratios for  neutral hyperon invisible decays  as a function of $m_\chi$, assuming  $m_\phi=0.95\,m_\chi$ and $m_\xi=0.04\,m_\chi$.  The Wilson coefficients have been adjusted to saturate the LHC constraints (see Sec.~\ref{sec:LHC}). Fig.~\ref{fig:ChPT_xi_phi} showcases that the purely invisible branching ratios can, in principle, be as large as $\sim10^{-4} - 10^{-2}$.

\subsection{Matrix elements for $\mathfrak B\to\pi\chi$}
\label{sec:Baryontopion}

\begin{figure}[t!]
\centering
		\includegraphics[width=0.4\textwidth]{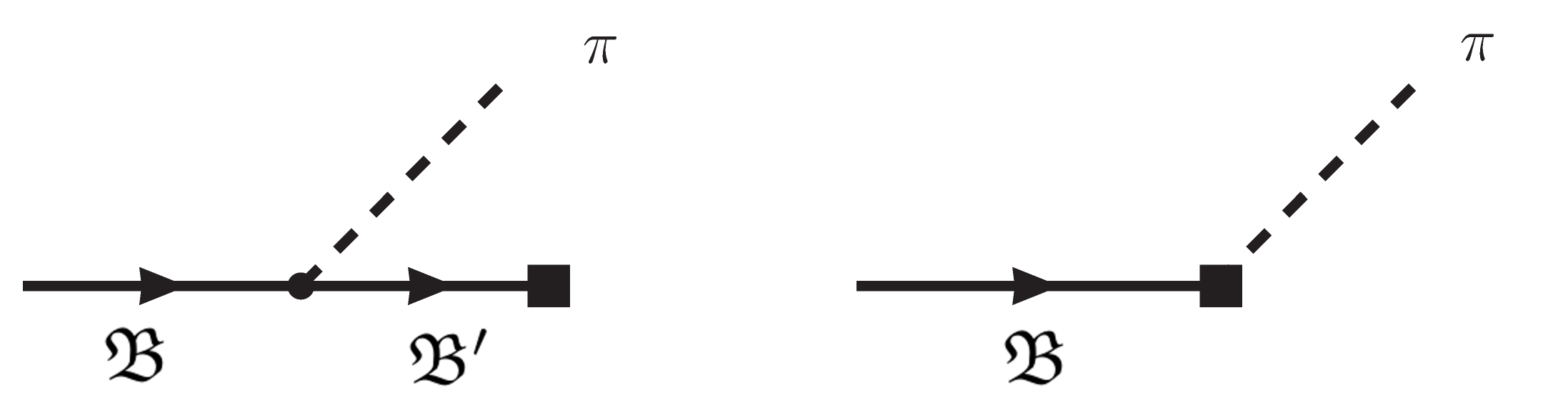}
\caption{Contributions to $\mathfrak B\to \pi$ form factors at leading order $\mathcal O(p^0)$ induced by the chiral representation of the operators coupled to $\chi$. The arrow indicates the flow of baryon number, the box is an insertion of a coupling from $\mathcal L^{(0)}_{\rm eff,ChPT}$ and the dot is the leading-order strong pion-baryon coupling.
}
\label{fig:Baryontopi}
\end{figure}

The hadronic matrix elements for the decays $\mathfrak B\to\pi\chi$ are parameterized in terms of two form factors,
\begin{multline}
\langle \pi(k)|\epsilon_{ijk}(u^{iT}CP_L d_{a}^j) P_R d_{b}^k|\mathfrak B(p)\rangle\\
=iP_R\left[W^{L}_{\mathfrak B 0}(q^2)
-\slashq W^{L}_{\mathfrak B1}(q^2)\right]u(p)\,,
\end{multline}
\begin{multline}
\langle \pi(k)|\epsilon_{ijk}(u^{iT}CP_R d_{a}^j) P_R d_{b}^k|\mathfrak B(p)\rangle\\
=iP_R\left[W^{R}_{\mathfrak B 0}(q^2)
-\slashq W^{R}_{\mathfrak B1}(q^2)\right]u(p)\,,
\end{multline}
where $q=p-k$ and $q^2=m_\chi^2$ if the final state $\chi$ is on shell. There are two possible contributions to these matrix elements at leading order $\mathcal O(p^0)$ in the chiral EFT, as shown in Fig.~\ref{fig:Baryontopi}.

The diagram on the left in Fig.~\ref{fig:Baryontopi} is the ``baryon pole'' diagram and is given by the $\mathfrak B^\prime$ decay constant. The second diagram on the right of Fig.~\ref{fig:Baryontopi} is a ``contact term'' induced by the expansion of $\mathcal L^{(0)}_{\rm eff,ChPT}$ in Eq.~\eqref{eq:EFTChiral} to first order in the pion field. The pole terms contribute to both form factors, giving
\begin{align}
\label{eq:W01q2}
W_{\mathfrak B0}^{L,{\rm pole}}(q^2)&=\frac{\alpha b_{\mathfrak B}}{f}\frac{m_{\mathfrak B}m_{\mathfrak B^\prime}+q^2}{m_{\mathfrak B^\prime}^2-q^2}\,,\nonumber\\
W_{\mathfrak B1}^{L,{\rm pole}}(q^2)&=\frac{\alpha b_{\mathfrak B}}{f}\frac{m_{\mathfrak B}+ m_{\mathfrak B^\prime}}{m_{\mathfrak B^\prime}^2-q^2}\,,
\end{align}
where the coefficients $b_{\mathfrak B}$ are listed in Table~\ref{tab:Btopipole} of   Appendix~\ref{sec:Appendix} for other decay channels. The results for the pole contributions to $W_{\mathfrak B0,1}^R(q^2)$ are the same as those in Eq.~\eqref{eq:W01q2} upon replacing $\alpha$ with $\beta$. The contact term diagram contributes only to $W_{0\mathfrak B}^{L/R}$, 
\begin{equation}
\label{eq:W0ct}
    W_{0\mathfrak B}^{L,\rm ct}= \frac{\alpha c^{L}_\mathfrak B}{f} ~~\text{and}~~  W_{0\mathfrak B}^{R,\rm ct}=\frac{\beta c^{R}_\mathfrak B}{f}, 
\end{equation}
where the contributions $c^{L/R}_\mathfrak B$ to each channel and operator are listed in  Appendix~\ref{sec:Appendix}, Table~\ref{tab:BtopiCT}.

\begin{figure*}[t!]
\centering
\begin{tabular}{cc}
		\label{fig:Decays1}
		\includegraphics[width=0.48\textwidth]{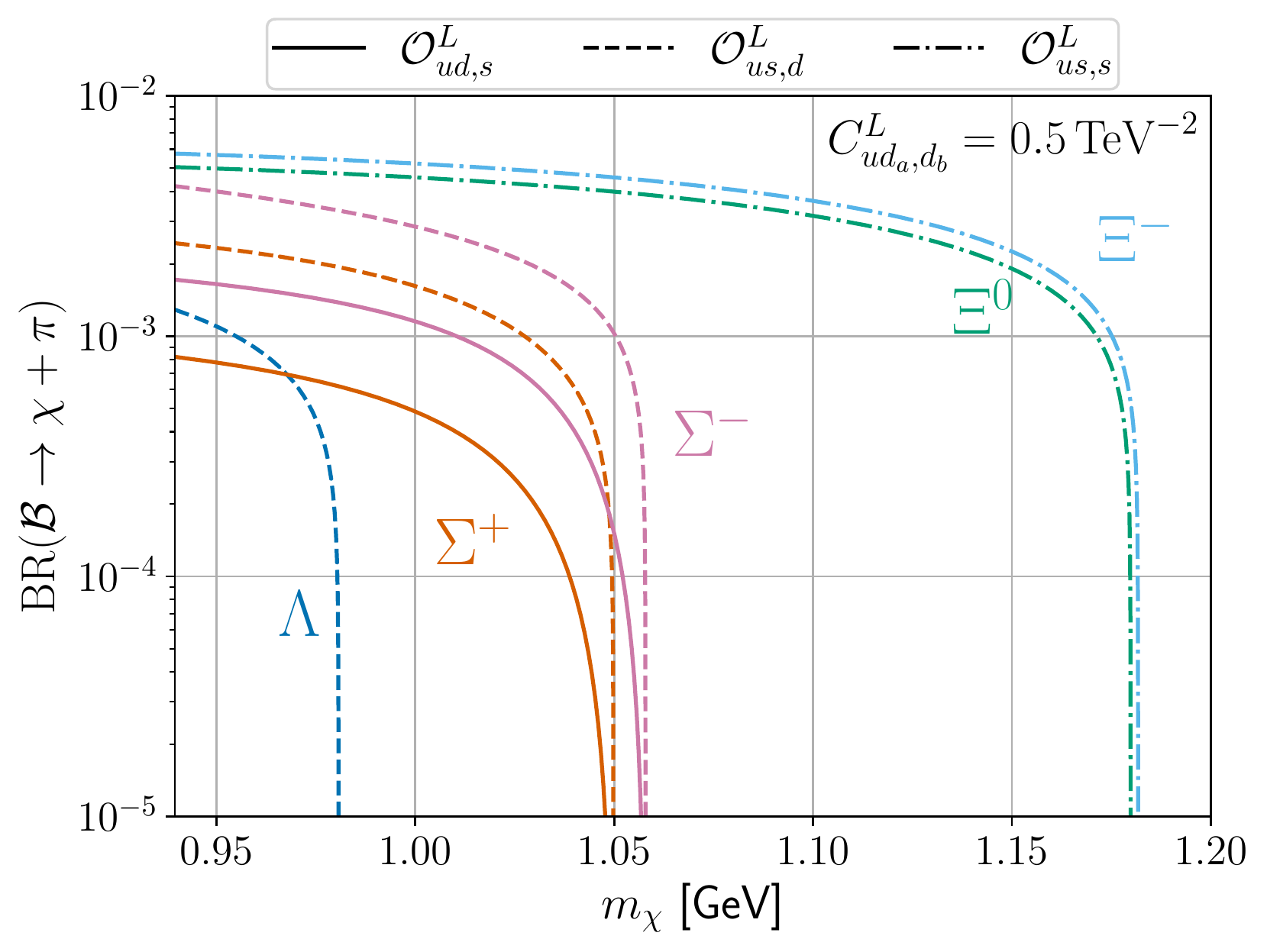}
&
		\label{fig:Decays2}
		\includegraphics[width=0.48\textwidth]{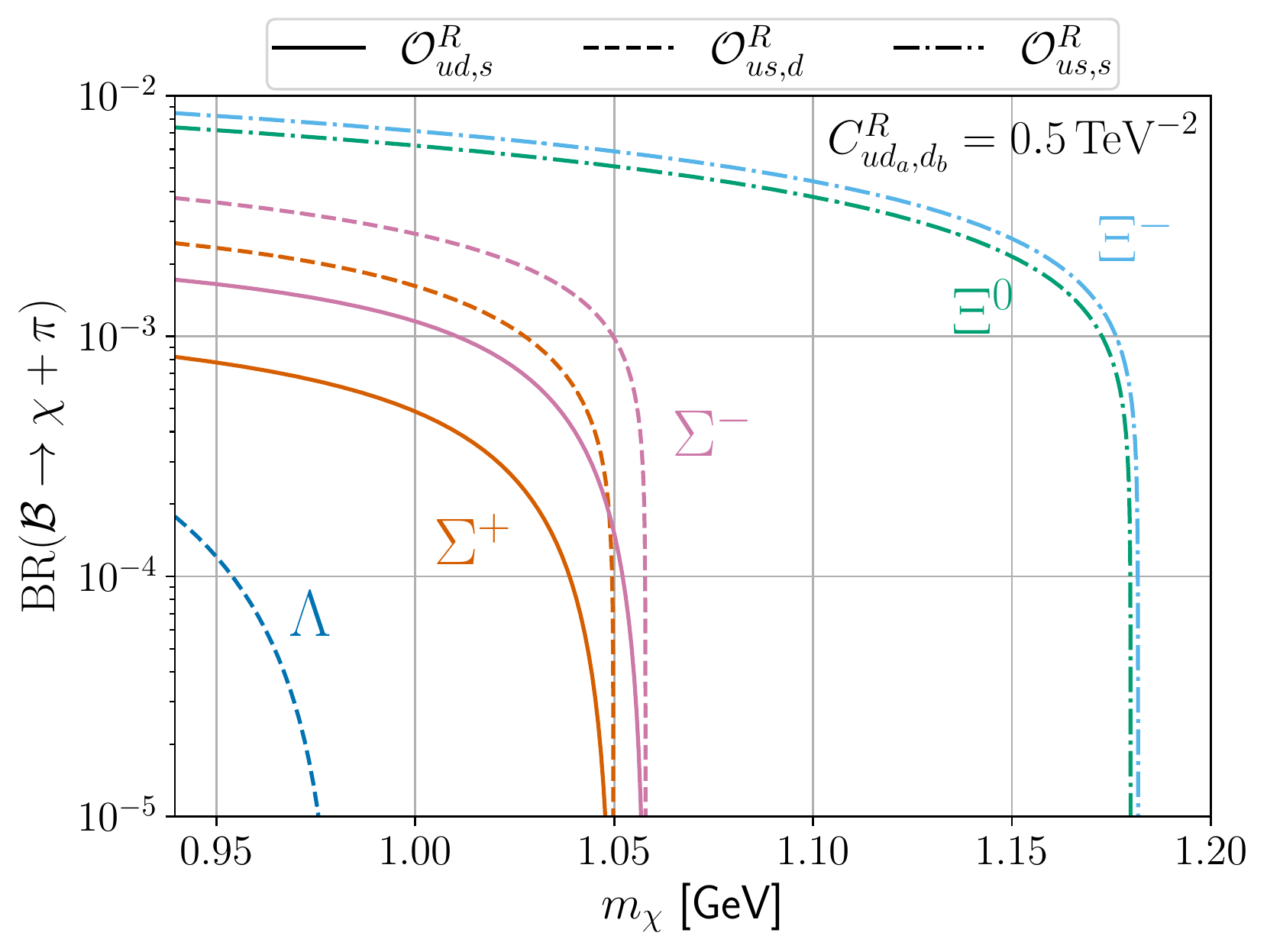}
\end{tabular}
\vspace{-0.2cm}
\caption{Branching ratios for hyperons decaying to a dark baryon and a pion, as a function of the dark baryon mass.
The left panel shows the branching ratios for the decays induced by the left-handed type operators, while the right panel corresponds to those induced by right-handed ones, as defined in Eq.~\eqref{eq:LEEFTchi}.
Different colors correspond to different hyperon decays, while different line styles represent different combinations or orderings of the down-type quarks in Eq.~\eqref{eq:LEEFTchi}.
For all of them, we fix the Wilson coefficient to roughly saturate the LHC constraints on the mediators, as discussed in the text (see Section~\ref{sec:LHC}). The branching ratios for other values of the Wilson coefficients can be obtained simply by noting that ${\rm BR} \propto |C^{L/R}_{ud_a,d_b}|^2$.
\label{fig:DecaysBNV1}}
\end{figure*}

Given the contribution described by the Wilson coefficient $C^{L/R}_{ud_a,d_b}$ and the corresponding form factors, one can calculate the rate for $\mathfrak B\to\pi\chi$,
\begin{align}
\label{eq:RateBpi}
\Gamma_{\mathfrak B\to\pi\chi}=\frac{|\vec k|}{8\pi m_{\mathfrak B}}|C_{ud_a,d_b}^{L/R}|^2\mathcal F^{L/R}_{\mathfrak B},
\end{align}
where 
\begin{align}
\label{eq:Ffunction}
\mathcal F^{L/R}_{\mathfrak B}=&E_\chi\left[\left(W^{L/R}_{0\mathfrak B}\right)^2+m_\chi^2\left(W^{L/R}_{1\mathfrak B}\right)^2\right]\nonumber\\
&-2m_\chi^2W^{L/R}_{0\mathfrak B}W^{L/R}_{1\mathfrak B},
\end{align}
with $E_\chi=(m_\chi^2+m_{\mathfrak B}^2-m_\pi^2)/(2 m_{\mathfrak B})$ being the energy of $\chi$ in the $\mathfrak B$ rest frame, $|\vec{k}|$ its momentum, and the form factors are evaluated at $q^2=m_\chi^2$. In the limit where the dark baryon $\chi$ is nonrelativistic in the $\mathfrak B$ rest frame, this expression is approximated by
\begin{align}
\mathcal F_{\mathfrak B}^{L/R}\simeq m_\chi\left(W^{L/R}_{0\mathfrak B}-m_\chi W^{L/R}_{1\mathfrak B}\right)^2.
\end{align}
Furthermore, if the $\mathfrak B$-$\chi$ mass splitting $\delta$ is small compared to $m_\mathfrak B \sim m_\chi$, then the baryon-pole contributions to the two form factors in Eq.~\eqref{eq:W01q2} cancel each other\footnote{We  assume $m_{\mathfrak B}=m_{\mathfrak B^\prime}$ and use $\alpha\simeq-\beta$.}, and one finds
\begin{align}
\label{eq:Fapp}
\mathcal F_{\mathfrak B}^{L/R}\simeq\frac{\alpha^2 m_\chi}{f^2}\left(c_{\mathfrak B}^{L/R}+b_{\mathfrak B}\frac{\delta}{2 m_\chi}\right)^2+\mathcal O(\delta^2).
\end{align}

The rich strong dynamics encoded in the form factors is reflected by the numerical results for the functions $\mathcal F^{L/R}_\mathfrak B$ shown in Table~\ref{tab:Ffunctions}. The sensitivity to the rates for various channels and contributions can differ by more than an order of magnitude in some cases. In particular, the channels which only receive a contribution from the pole diagrams (see Tab.~\ref{tab:BtopiCT} in Appendix~\ref{sec:Appendix}) can suffer from the suppression described by Eq.~\eqref{eq:Fapp}, which is visible for higher masses (lower velocities) of the $\chi$. This is also observed in Fig.~\ref{fig:DecaysBNV1}, where we show the predicted branching ratios for different channels and operators, taking as a reference the upper limits set by  LHC searches for the mediators (see Sec.~\ref{sec:LHC}). 

Our results in Fig.~\ref{fig:DecaysBNV1} indicate that branching fractions can be as large as $\sim10^{-3}$ in case of the pionic decays of the $\Xi$ baryons. The decay $\Xi^-\to\pi^-\chi$ may be particularly preferred experimentally, given that the charged pion in the final state may be easier to trigger on. The decay with the best theoretical sensitivity to both $(ud)s$- and $(us)d$-type operators is $\Sigma^-\to\pi^-\chi$. It has up to a factor of $\sim10$ more sensitivity compared to the $\Lambda\to\pi^0\chi$ decay and it is approximately a factor of $2$ more sensitive than $\Sigma^+\to\pi^+\chi$, since the lifetime of  $\Sigma^-$ is approximately twice that of $\Sigma^+$~\cite{pdg}. Note also that the branching ratio of the $\Sigma^0$ decay is very suppressed because it mainly decays through the electromagnetic channel $\Sigma^0\to\Lambda\gamma$. 

Finally, we have not considered the decays of the negatively-charged decuplet baryon $\Omega^-(sss)$. This is  because the dominant decay channel of interest would be $\Omega^-\to K^-\chi$. This probes operators with flavor $uss$ and $m_\chi^{\rm max}=1.179$ GeV, which is the same as for $\Xi$ decays. The $\Omega^-$ are also more difficult to produce in experiments~\cite{Li:2016tlt,Alves:2018npj} and  their decay rates are harder to predict theoretically. 

\begin{table}[t]
\renewcommand{\arraystretch}{1.5}
\setlength{\arrayrulewidth}{.1mm}
\begin{tabular}{cccccc}
\hline\hline
Channel & Operator &\multicolumn{2}{c}{$10^3\times\mathcal F_{\mathfrak B}^{L}$ [GeV$^5$]}& \multicolumn{2}{c}{$10^3\times\mathcal F_{\mathfrak B}^{R}$ [GeV$^5$]}\\
\hline
$\Lambda\to\pi^0$ &$(us)d$ & \multicolumn{2}{c}{$3.5$}&\multicolumn{2}{c}{$0.48$}\\
\hline
\multirow{5}{*}{$\Sigma^0\to\pi^0$}&&\multicolumn{4}{c}{$m_\chi$ [GeV]}\\
\cline{3-6}
&&0.94&1.05&0.94&1.05\\
\cline{3-6}
&$(us)d$&$11$&$12$&$1.2$&$0.42$\\
&$(ud)s$&$4.6$&$1.7$&$4.6$&$1.7$\\
&$(u[d),s]$&--&--&$1.2$&$0.42$\\
\hline
\multirow{5}{*}{$\Sigma^-\to\pi^-$}&&\multicolumn{4}{c}{$m_\chi$ [GeV]}\\
\cline{3-6}
&&0.94&1.05&0.94&1.05\\
\cline{3-6}
&$(us)d$&$12$&$12$&$10$&$12$\\
&$(ud)s$&$4.8$&$1.8$&$4.8$&$1.8$\\
&$(u[d),s]$&--&--&$13$&$12$\\
\hline
\multirow{3}{*}{$\Xi^-\to\pi^-$}&&\multicolumn{4}{c}{$m_\chi$ [GeV]}\\
\cline{3-6}
&&0.94 &1.15 &0.94 &1.15\\
\cline{3-6}
&$(us)s$ & $10$&$13$&$15$&$15$\\
\hline
\hline
\end{tabular}
\caption{\label{tab:Ffunctions}Predictions for the $\mathcal F^{L/R}_{\mathfrak B}$ factor entering the decay $\mathfrak B\to \pi\chi$, shown for different channels and $\chi$ masses. In the case of $\Lambda\to\pi\chi$,  $m_\chi=0.94$ GeV was assumed. The notation $(u[d),s]$ represents the combination  $C^R_{ud,s}-C^R_{us,d}$ provided by the Model 1, see Eqs.~\eqref{eq:Models} and \eqref{eq:FierzModel1}. }
\end{table}

\subsection{Matrix elements for $\mathfrak B\to\gamma\chi$}\label{sec:Baryontogamma}

\begin{figure}[t]
\centering
		\includegraphics[width=0.2\textwidth]{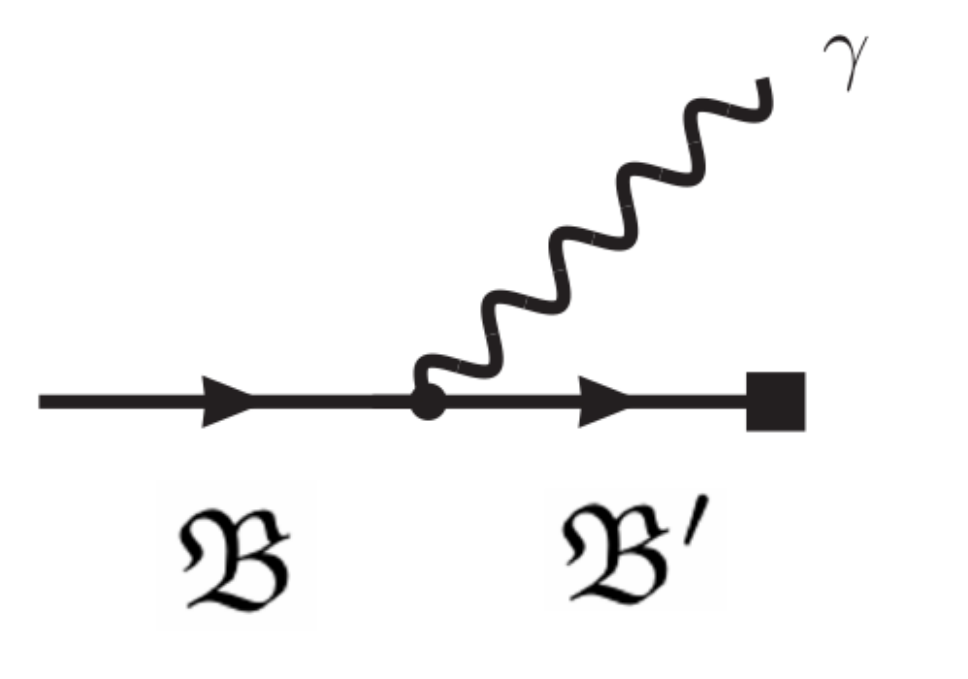}
\caption{Contributions to $\mathfrak B\to \gamma$ form factors at leading order $\mathcal O(p)$ induced by the chiral representation of the operators coupled to $\chi$. The arrow indicates the flow of baryon number, the box is an insertion of a coupling from $\mathcal L^{(0)}_{\rm eff,ChPT}$ and the dot is the magnetic moment of the hyperons, which is a vertex starting at $\mathcal O(p^2)$ in the chiral Lagrangian. }
\label{fig:Baryontogamma}
\end{figure}

We can parametrize the matrix elements of the radiative $\mathfrak B\to\gamma\,\chi$ decays induced by the hyperon magnetic moments as
\begin{multline}
\label{eq:radiativeFFs}
\langle \gamma(k)|
\epsilon_{ijk}(u^{iT}CP_L d_{a}^j) P_R d_{b}^k
|\mathfrak B(p)\rangle = \\
i\mu_{\rm N}[V^L_{\mathfrak B0}(q^2)\sigma^{\mu\nu}P_R
+V^L_{\mathfrak B1}(q^2)\slashq \sigma^{\mu\nu}P_L]u(p)\varepsilon_\mu^* k_\nu,
\end{multline}
where $\varepsilon^\mu(k)$ is the polarization vector of the photon, $\mu_{\rm N}=e/2m_p$ is the nuclear magneton, and $V^L_{\mathfrak B0,1}(q^2)$ are the corresponding form factors. A similar definition $V^R_{\mathfrak B0,1}(q^2)$ follows for the matrix elements of
$\epsilon_{ijk}(u^{iT}CP_R d_{a}^j) P_R d_{b}^k$.
In the chiral EFT, the leading-order contribution enters at $\mathcal O(p)$ (an order higher than in the pionic decays) and is given exclusively by baryon-pole contributions, as shown in Fig.~\ref{fig:Baryontogamma}.
Higher-order chiral operators linearly coupled to $\chi$ including the photon field and with unknown coefficients are possible.
They can be constructed by including $F_{\mu\nu}$ or covariant derivatives acting on the $u$ and $u^\dagger$ fields; however, none give contact-term contributions to the form factors at $\mathcal O(p)$. 

The diagram in Fig.~\ref{fig:Baryontogamma} yields
\begin{align}
\label{eq:V01q2}
V_{\mathcal B0}^{L/R}=m_{\mathfrak B^\prime}V_{\mathcal B1}^{L/R}=\frac{m_{\mathfrak B^\prime}\gamma^{L/R}_{\mathfrak B^\prime} \kappa_{\mathfrak B\mathfrak B^\prime}}{q^2-m_{\mathfrak B^\prime}^2},
\end{align}
where $\kappa_{\mathfrak B\mathfrak B^\prime}$, defined in Eq.~\eqref{eq:TransMagMom}, is the value of the   magnetic moment of the $\mathfrak B\to\mathfrak B^\prime$ transition (simply $\kappa_{\mathfrak B}$ in case of diagonal couplings) in units of the nuclear magneton and $\gamma^{L/R}_{\mathfrak B^\prime}$ are the coefficients defined in Eq.~\eqref{eq:Btovac}, which can be read from Table~\ref{tab:Btovac} for each of the baryons. In Table~\ref{tab:Btogamma} of Appendix~\ref{sec:Appendix}, we show the value of the coefficients for the different contributions and channels.

The decay rate for $\mathfrak B\to\chi\gamma$ in the $\mathfrak B$ rest frame can then be computed as
\begin{align}
\label{eq:RateBgamma}
\Gamma_{\mathfrak B\to \chi\gamma}=\frac{\alpha_{\rm em}}{8m_p^2} \frac{|\vec{k}|}{m_{\mathfrak{B}}}|C_{ud_a,d_b}^{L/R}|^2\mathcal G^{L/R}_{\mathfrak B},
\end{align}
where $\alpha_{\rm em}=e^2/4\pi$ denotes the fine structure constant, $\vec k$ is the three-momentum of the decay products in the $\mathfrak B$ rest frame, and
\begin{align}
\label{eq:GBLRfunc}
\mathcal G^{L/R}_{\mathfrak B}&=\frac{(m_{\mathfrak B}^2-m_\chi^2)^2}{m_{\mathfrak B}}\left[m_\chi^2 \left(V_{\mathfrak B1}^{L/R}\right)^2+\left(V_{\mathfrak B0}^{L/R}\right)^2\right]\nonumber\\
&=(\gamma^{L/R}_{\mathfrak B^\prime})^2\kappa_{\mathfrak B\mathfrak B^\prime}^2\frac{m_{\mathfrak B^\prime}^2+m_\chi^2}{m_{\mathfrak B}}\frac{(m_{\mathfrak B}^2-m_\chi^2)^2}{(m_{\mathfrak B^\prime}^2-m_\chi^2)^2}.
\end{align}
In the second line, we have replaced the form factors by the chiral predictions in Eq.~\eqref{eq:V01q2} evaluated at $q^2=m_\chi^2$.
In decays where only the diagram with $\mathfrak B=\mathfrak B^\prime$ contributes (see Table~\ref{tab:Btogamma}), the poles cancel exactly. On the other hand, in cases where $\mathfrak B\neq\mathfrak B^\prime$ the pole enhancement can give rise to a large resonant contribution to the rate when $m_\chi\simeq m_{\mathfrak B^\prime}$. In practice, this can only occur for the decays of $\Sigma^0$ when $m_\chi\simeq m_\Lambda$. 
However, as the $\Sigma^0$ lifetime is dominated by the standard radiative decays $\Sigma^0\to\Lambda \gamma$, a contribution to the total rate from the dark radiative decay would only be significant for a very finely-tuned $m_\chi$.\footnote{Note that in the extreme case $m_\chi=m_\Lambda$ one resolves the (weak) width of the $\Lambda$ in the propagator of Fig.~\ref{fig:Baryontogamma}, and there is a scaling of the branching ratio, $\text{BR}(\Sigma^0\to\chi\gamma)\sim|C^{L(R)}_{ud_a,d_b}|^2/(m_\Lambda^2 G_F^4)$. We will ignore this extremely fine-tuned scenario in our paper.} 

\begin{figure}[t]
		\includegraphics[width=0.48\textwidth]{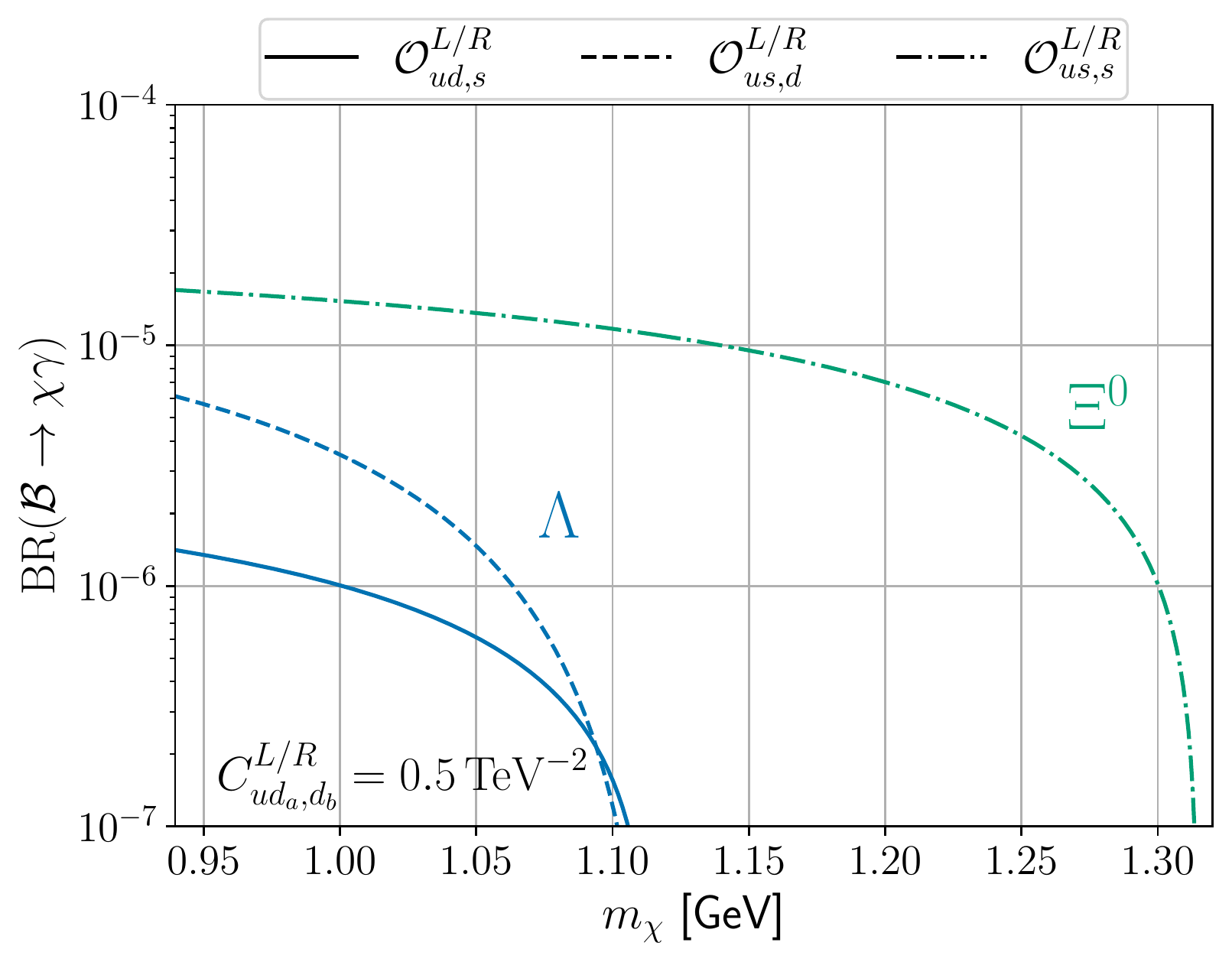}
\caption{
Branching ratio for neutral hyperon decays into a dark baryon accompanied by a photon. The Wilson coefficients are fixed to saturate the LHC constraints on the mediators, as discussed in the text. The branching ratios for other values of the Wilson coefficients can be obtained simply by noting that ${\rm BR} \propto |C^{L/R}_{ud_a,d_b}|^2$. 
\label{fig:DecaysBNV2}}
\end{figure}

In Table~\ref{tab:Gfunctions} we show the predictions for the $\mathcal G^{L/R}_{\mathfrak B}$ functions evaluated at various $\chi$ masses and for different operators. We observe that the predictions for the dark radiative hyperon decay rates for the different possible channels roughly span one order of magnitude (with the exception of the $\Sigma^0$, which is suppressed by the large $\Sigma^0$ width as discussed above). In Fig.~\ref{fig:DecaysBNV2} we show the largest possible branching ratios for different channels and operators in light of the LHC constraints derived in Sec.~\ref{sec:LHC}. Note that even though the radiative decay rates are suppressed by $\alpha_{\rm em}$ compared to the pionic decays, they probe a much broader range of $\chi$ masses.

\begin{table}[t]
\renewcommand{\arraystretch}{1.5}
\setlength{\arrayrulewidth}{.1mm}
\begin{tabular}{cccc}
\hline\hline
Channel & Operator &\multicolumn{2}{c}{$10^3\times\mathcal G_{\mathfrak B}^{L/R}$ [GeV$^7$]}\\
\hline
$n\to\gamma$& $(ud)d$ & \multicolumn{2}{c}{0.13}\\
\hline
\multirow{5}{*}{$\Lambda\to\gamma$}&&\multicolumn{2}{c}{$m_\chi$ [GeV]}\\
\cline{3-4}
& &0.94&1.05 \\
\cline{3-4}
& $(us)d$ &0.40&0.24\\
& $(ud)s$ &0.09&0.10\\
& $(u[d)s]$ &0.11&0.03\\
\hline
\multirow{5}{*}{$\Sigma^0\to\gamma$}&&\multicolumn{2}{c}{$m_\chi$ [GeV]}\\
\cline{3-4}
& &0.94&1.1 \\
\cline{3-4}
& $(us)d$ &0.08&4.6\\
& $(ud)s$ &1.3&24\\
& $(u[d)s]$ &0.73&7.6\\
\hline
$\Xi^0\to\gamma$ & $(us)s$ &0.61 &0.68\\
\hline
\hline
\end{tabular}
\caption{\label{tab:Gfunctions}Predictions for the $\mathcal G^{L/R}_{\mathfrak B}$ factor entering the $\mathcal{B}\rightarrow\chi\gamma$ decays, evaluated at two different $\chi$ masses. In the case of $n\to\chi\gamma$,  $m_\chi=0.938$ GeV was assumed. The notation $(u[d),s]$ represents the combination  $C^R_{ud,s}-C^R_{us,d}$ that is generated in Model 1, see Eqs.~\eqref{eq:Models} and \eqref{eq:FierzModel1}.}
\end{table}

\section{Hyperons in supernovae}
\label{sec:supernova}

Hyperons are predicted to coexist in equilibrium within the hot and dense proto neutron stars (PNS) forming in core-collapse supernovae~\cite{1960SvA,Oertel:2016bki}.
Therefore, dark sector particles produced in $\Lambda$ decays and leaving the star would drain energy from the PNS, leading to a new cooling mechanism that can be constrained by the observations of SN 1987A~\cite{MartinCamalich:2020dfe,Camalich:2020wac}.\footnote{Nonzero abundances are expected for all  hyperons, but the $\Lambda$ is the lightest and thus the most abundant at the thermodynamic conditions  inside of the PNS~\cite{Oertel:2016bki}.} In particular, the duration of the neutrino pulse would have been significantly shorter than the one observed if the dark luminosity $L_{\rm d}$ were of the same order as that for the neutrinos~\cite{Raffelt:1987yt,Turner:1987by,Mayle:1987as,Burrows:1988ah,Burrows:1990pk}. This implies a limit
\begin{align}
\label{eq:Raffelts_Criterion}
L_{\rm d}\lesssim3\times10^{52}~~\text{erg s$^{-1}$},    
\end{align}
imposed 1 second after the bounce~\cite{Raffelt:1996wa}. In the case of  dark baryon models, this limits the couplings to the $uds$ quarks via the emission of $\chi$'s from $\Lambda\to\chi\gamma$ or $\Lambda\to\chi\pi^0$ decays, or through the emission of $\xi$'s and $\phi$'s from $\Lambda\to\xi\phi$. One may also consider  processes affecting the $udd$ couplings via the neutron decays $n\to\chi\gamma$ and $n\to\xi\phi$. However, as discussed below, the constraints are weak compared to those obtained from neutron lifetime ($\tau_n$) measurements. 

Let us concentrate on the dark decay $\mathfrak B\to \mathfrak D\mathfrak b$ in the medium, where $\mathfrak B$ is an ordinary baryon,  $\mathfrak D$ is a dark baryon ($\chi$ or $\xi$), and $\mathfrak b$ is a photon, a pion or a dark boson $\phi$, depending on the decay mode under consideration. The spectrum of the energy-loss rate per unit volume with respect to the energy of $\mathfrak D$ in the star's frame, $E_{\mathfrak D}$, is given by~\cite{Camalich:2020wac} 
\begin{align}\label{eq:dQgen}
\frac{\diff Q_{\mathfrak B\to \mathfrak D\mathfrak b}}{\diff E_{\mathfrak D}}=&\frac{m_{\mathfrak B}^2\Gamma_{\mathfrak B\to \mathfrak D\mathfrak b}}{2\pi^2|\vec k|} \int^{E_+}_{E_-} \mathop{\diff E} E_{\rm d}\, f_{\mathfrak B}(E),
\end{align}
where $E$ is the energy of $\mathfrak B$ in the star's frame, $\Gamma_{\mathfrak B\to \mathfrak D\mathfrak b}$ is the decay rate in vacuum (with $\vec k$ being the corresponding three-momentum in the $\mathfrak B$ rest frame), and where $f_{\mathfrak B}(E)$ is the distribution function of the ${\mathfrak B}$ state inside the PNS. In Eq.~\eqref{eq:dQgen}, $E_{\rm d}$ refers to the energy carried by the dark particles, which is $E_{\rm d}= E_\chi$ in case of radiative and pionic decays, or $E_{\rm d}=E$ in case of purely dark decays. The limits of the integral are
\begin{align}
\label{eq:int_limits}
 E_{\pm}&=\frac{1}{2m_{\mathfrak D}^2}(E_{\mathfrak D}(m_{\mathfrak B}^2+m_{\mathfrak D}^2-m_{\mathfrak b}^2)\\
 &\pm q\sqrt{((m_{\mathfrak B}-m_{\mathfrak D})^2-m_{\mathfrak b}^2)((m_{\mathfrak B}+m_{\mathfrak D})^2-m_{\mathfrak b}^2)}),\nonumber
\end{align}
with $q=\sqrt{E_{\mathfrak D}^2-m_{\mathfrak D}^2}$, and the range of integration for $E_{\mathfrak D}$ is
\begin{align}\label{eq:rangeED}
E_{\mathfrak D}\in[m_{\mathfrak D},\infty].  \end{align}
The number density of ${\mathfrak B}$ follows a Fermi-Dirac distribution characterized by thermodynamical properties of the medium. We have also neglected a Bose-enhancement factor for the ${\mathfrak b}$ final state, which in case of photons and pions is negligible at energies relevant for the dark decays considered here. 

Equation~\eqref{eq:dQgen} can be employed directly to calculate the dark luminosity $L_{\rm d}$ for a given model of the PNS if the $\chi$ or the $\xi$ and $\phi$ escape freely from the star. However, this energy can be reabsorbed in the medium through inverse processes like $\chi \gamma\to \Lambda$ if the mean free path becomes much shorter than the typical distance the dark particles traverse in the star. If they are relativistic, this distance is roughly the radius of the PNS and they become trapped only for large interaction rates~\cite{Raffelt:1987yt,Turner:1987by}. 
On the other hand, this typical distance becomes effectively infinite if the emitted particles are nonrelativistic and get gravitationally trapped inside the star. 
These trapped particles can then be reabsorbed even if their interaction rates with the PNS constituents are small.
In this case, the SN cooling bound in Eq.~\eqref{eq:Raffelts_Criterion} can be conservatively applied to the luminosity from the dark particles that are energetic enough to escape the gravitational potential $V$ at the emission point.\footnote{The reabsorption of a significant number of gravitationally trapped particles could lead to anomalous heat transport within the PNS, which may affect its dynamics and lead to observable signatures. A careful analysis of this process, which may allow to place a bound even in the trapped regime, is a daunting task that lies beyond the scope of this paper.} This condition can be implemented by modifying Eq.~\eqref{eq:dQgen} with the replacements 
\begin{align}
\label{eq:Eesc1}
E_{\rm d}\longrightarrow E_{\chi}\Theta(E_\chi-E_{\chi}^{\rm esc}),
\end{align}
for $\Lambda\to\chi\gamma$ and $\Lambda\to\chi\pi^0$ and
\begin{align}
\label{eq:Eesc2}
E_{\rm d}\longrightarrow E_{\xi}\Theta(E_\xi-E_\xi^{\rm esc})+E_{\phi}\Theta(E_\phi-E_\phi^{\rm esc}),
\end{align}
for $\Lambda\to\xi\phi$. In this equation $\Theta(x)$ is the  Heaviside step function and $E_{\mathfrak D}^{\rm esc}$ is the escape energy defined as 
\begin{align}
E^{\rm esc}_{i}=m_i(1-V).
\end{align}
The gravitational potential can be approximated by $V\approx -G_N M/R\sim-0.22$ (in natural units) for $M=1.5M_{\odot}$ and $R=10$ km, which are typical estimates for the SN 1987A's PNS~\cite{Bollig:2020xdr}. 

\begin{figure}[t]
		\includegraphics[width=0.48\textwidth]{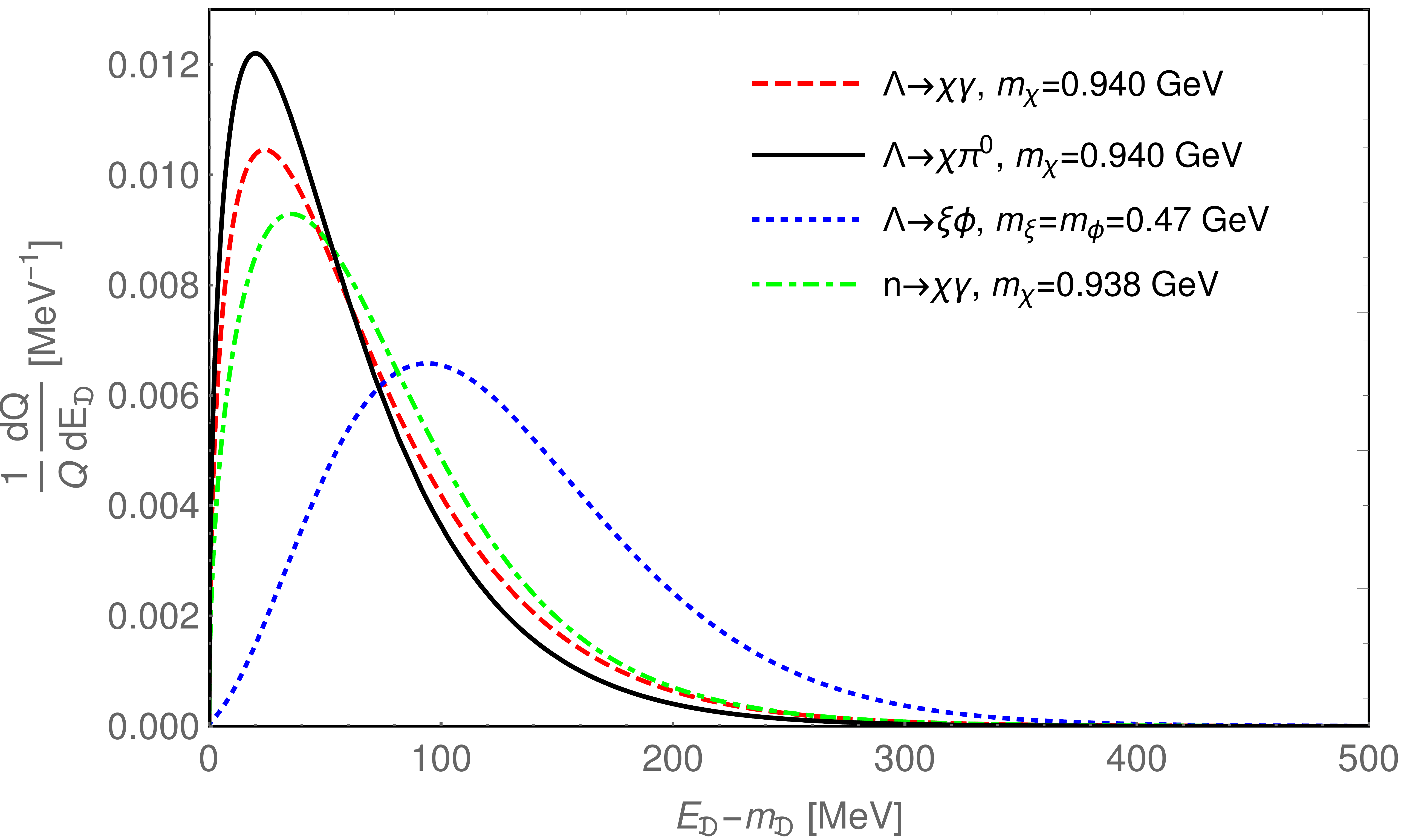}
\caption{Normalized spectra for the volume emission of the $\chi$ particles via $\Lambda\to\chi\gamma$ and $\Lambda\to\chi\pi$ or of $\xi$ in $\Lambda\to\xi\phi$ obtained using Eq.~\eqref{eq:dQgen}. For the baryon distribution $f_{\Lambda}(E)$, we use the standard Fermi-Dirac distribution for  typical conditions reached in the PNS: temperature $T=40$ MeV and chemical potential $\mu_\Lambda=957$ MeV. Also shown is the normalized spectrum for the $n\to\chi \gamma$ decay. \label{fig:SpectrumSN}
}
\end{figure}

In Fig.~\ref{fig:SpectrumSN}, we show the normalized spectrum for various $\Lambda$ decays assuming the extreme thermodynamical conditions reached in the PNS. For dark particles with $m_{\mathfrak D}\sim 1$ GeV, one has $E_{\mathfrak D}-m_{\mathfrak D}\sim0.22$ GeV and only the particles in the high-energy tail of the distribution can escape from the star. This represents a contribution to $L_{\rm d}$ of only a few percent of the total energy produced in the $\Lambda\to \chi\gamma$ and $\Lambda\to \chi\pi^0$ decays, the rest of which is then returned to the system. In the case of  pure dark decays the situation is different. If one of the particles is very light (e.g., $m_\phi\simeq0$), it will stream freely and the bound obtained would be very similar to the one derived for axions and dark photons in~\cite{Camalich:2020wac}. If the two particles are massive, with $m_{\xi}\sim m_{\phi}\simeq 0.5$ GeV, then about half of them would have enough energy to escape from the PNS. 

\subsection{Supernova 1987A}

\begin{table*}[t!]
\renewcommand{\arraystretch}{1.5}
  \setlength{\arrayrulewidth}{.2mm}
\centering
\small
\setlength{\tabcolsep}{0.5 em}
\begin{tabular}{ c  c  c  c  c c  }
    \hline\hline
\multirow{3}{*}{Simulation} & \multicolumn{2}{c}{$\Lambda\to \chi\gamma$} & $\Lambda\to \chi\pi^0$ & \multicolumn{2}{c}{$\Lambda\to \xi\phi$}\\
& \multicolumn{2}{c}{$m_\chi$ [GeV]} & $m_\chi$ [GeV] &  \multicolumn{2}{c}{($m_\xi$,~$m_\phi$) [GeV]}\\
\cline{2-6}
 & 0.94 & 1.05& 0.94 & (0.94,\,0) &(0.5,\,0.5) \\  
 \hline
 SFHo 18.60 & $2.2\times10^{-8}$ & $5.7\times10^{-8}$ & $4.2\times10^{-8}$ & $2.2\times10^{-9}$ & $1.2\times10^{-9}$\\
 SFHo 18.80 & $5.4\times10^{-8}$ & $\mathbf{1.6\times10^{-7}}$ & $\mathbf{1.1\times10^{-7}}$ & $\mathbf{5.0\times10^{-9}}$ & $\mathbf{2.4\times10^{-9}}$ \\
 SFHo 20.00 & $7.5\times10^{-9}$ & $1.7\times10^{-8}$ & $1.3\times10^{-8}$ &  $6.7\times10^{-10}$ & $4.6\times10^{-10}$\\
 
    \hline\hline
\end{tabular}
\caption{Upper limits on the branching ratios for $\Lambda$ dark decays based on the duration of the neutrino burst in SN 1987a. Bounds are shown for different masses of the dark sector particles and different PNS models (SFHo 18.60, 18.80, and 20.00). We use the results of the simulations in Ref.~\cite{Bollig:2020xdr}, and assume that each given channel provides the dominant cooling mechanism. We mark in bold face the limits that are taken as a reference for the SN 1987A bound on dark baryon emission in the remainder of this paper.}
\label{tab:SNbounds}
\end{table*}

In order to obtain a robust estimate of the dark luminosity emitted via different $\Lambda$ decay modes, we use recent simulations of SN 1987A~\cite{Bollig:2020xdr}, which are spherically symmetric (one-dimensional) and, hence, artificially triggered~\cite{Rampp:2002bq,Janka:2012wk}. We use only those simulations which implement the SFHo  (for Steiner, Hempel and Fischer) nuclear equation of state (EoS)~\cite{Steiner:2012rk,Oertel:2016bki} and  are labelled by SFHo-18.8, SFHo-18.6 and SFHo-20.0, depending on the mass of the progenitor star. The masses of the neutron-star remnant are $1.351M_{\odot}$, $1.553M_{\odot}$ and $1.947M_{\odot}$, respectively~\cite{Bollig:2020xdr}. The EoS employed in these simulations does not include hyperons. However, SFHo has been upgraded by adding the hyperons as explicit degrees of freedom in the system~\cite{Fortin:2017dsj}. Importantly, the hyperonic and non-hyperonic SFHo EoS lead to almost identical predictions of the system's thermodynamical properties for all the conditions reached in the SN simulations~\cite{Fortin:2017dsj} (they are also consistent with all known nuclear and astrophysical constraints~\cite{Oertel:2016bki}). 

The results of the simulations reported in Ref.~\cite{Bollig:2020xdr} can be found in~\cite{Garching}. We use the radial profiles of density, temperature and proton fraction as inputs in our calculation. The latter is relevant because we are describing the PNS during the phase in which neutrinos are partially trapped and have built up a chemical potential. These inputs are then used in the relevant hyperonic EoS (SFHoY), which is obtained by interpolating the tables provided in the CompOSE database~\cite{compose}, to predict all the other relevant thermodynamical quantities. Of particular importance are the chemical potentials and the parameters describing the nuclear-medium effects in the dispersion relation of the baryons~\cite{Fortin:2017dsj}. These enter the calculation through the energy dependence of the baryon's distribution $f_{\mathfrak B}$ in Eq.~\eqref{eq:dQgen}. We neglect the effect of the modified dispersion relations in the calculation of the rates in the medium, since this would lead only to small changes~\cite{Camalich:2020wac}. 

The total dark luminosity induced by a particular $\Lambda$ decay mode is then calculated as
\begin{align}
\label{eq:dark_lumi}
L_{\rm d}=\int \mathop{\diff^3\vec{r}}\int^\infty_{m_{\mathfrak D}} \mathop{\diff E_{\mathfrak D}} \frac{\diff Q(r)}{\diff E_{\mathfrak D}},  
\end{align}
where $\diff Q(r)/\diff E_{\mathfrak D}$ is the spectrum in Eq.~\eqref{eq:dQgen} produced assuming the thermodynamical conditions at the radius $r$ and with the replacements in Eqs.~\eqref{eq:Eesc1} and~\eqref{eq:Eesc2}. The escape energies at $r$ are calculated with the gravitational potential produced by a spherically-symmetric distribution of mass parametrized by the density $\rho(r^\prime)$~\cite{2008gady.book.....B},
\begin{align}
\!\!\!\!\!\! V(r)=-4\pi G_N\!\left(\frac{1}{r}\int^r_0dr^\prime r^{\prime 2}\rho(r^\prime)\!+\!\! \int^\infty_rdr^\prime r^{\prime}\rho(r^\prime)\right)\!.\!\!\!
\end{align}
In Eq.~\eqref{eq:dark_lumi} we neglected the reabsorbtion of the dark particles which have enough energy to escape the PNS in the large-coupling regime~\cite{Raffelt:1987yt,Turner:1987by}. For the processes involving $\Lambda$ hyperons, the maximal surface from where the dark particles can be radiated corresponds to a very hot region of the star and $L_{\rm d}$ is larger than in Eq.~\eqref{eq:Raffelts_Criterion} even in the strong coupling regime~\cite{Camalich:2020wac}. Therefore, we do not consider this type of effect in our calculation of $L_{\rm d}$.

In Table~\ref{tab:SNbounds} we show the results of imposing Eq.~\eqref{eq:Raffelts_Criterion} in our calculations, expressed as upper limits on the branching ratios and assuming that a given decay mode is the dominant dark-cooling mechanism. The results can vary by almost one order of magnitude when comparing different simulations, with the heavier, denser and hotter PNS leading to stronger limits. In case of the $\Lambda \to \chi\gamma$ and $\Lambda\to \chi\pi^0$ decays, we observe a weakening of the limits compared to the $\Lambda \to \xi \phi$ decays. As discussed above, this is due to the fact that only a small fraction of the $\chi$ particles emitted in the former case are energetic enough to escape the gravitational well of the PNS.  In Table~\ref{tab:SNbounds} we denote in boldface the final limits, which correspond to the most conservative ones stemming from the SFHo-18.80 simulations.

\begin{figure*}[t!]
\centering
\begin{tabular}{cc}
		\includegraphics[width=0.48\textwidth]{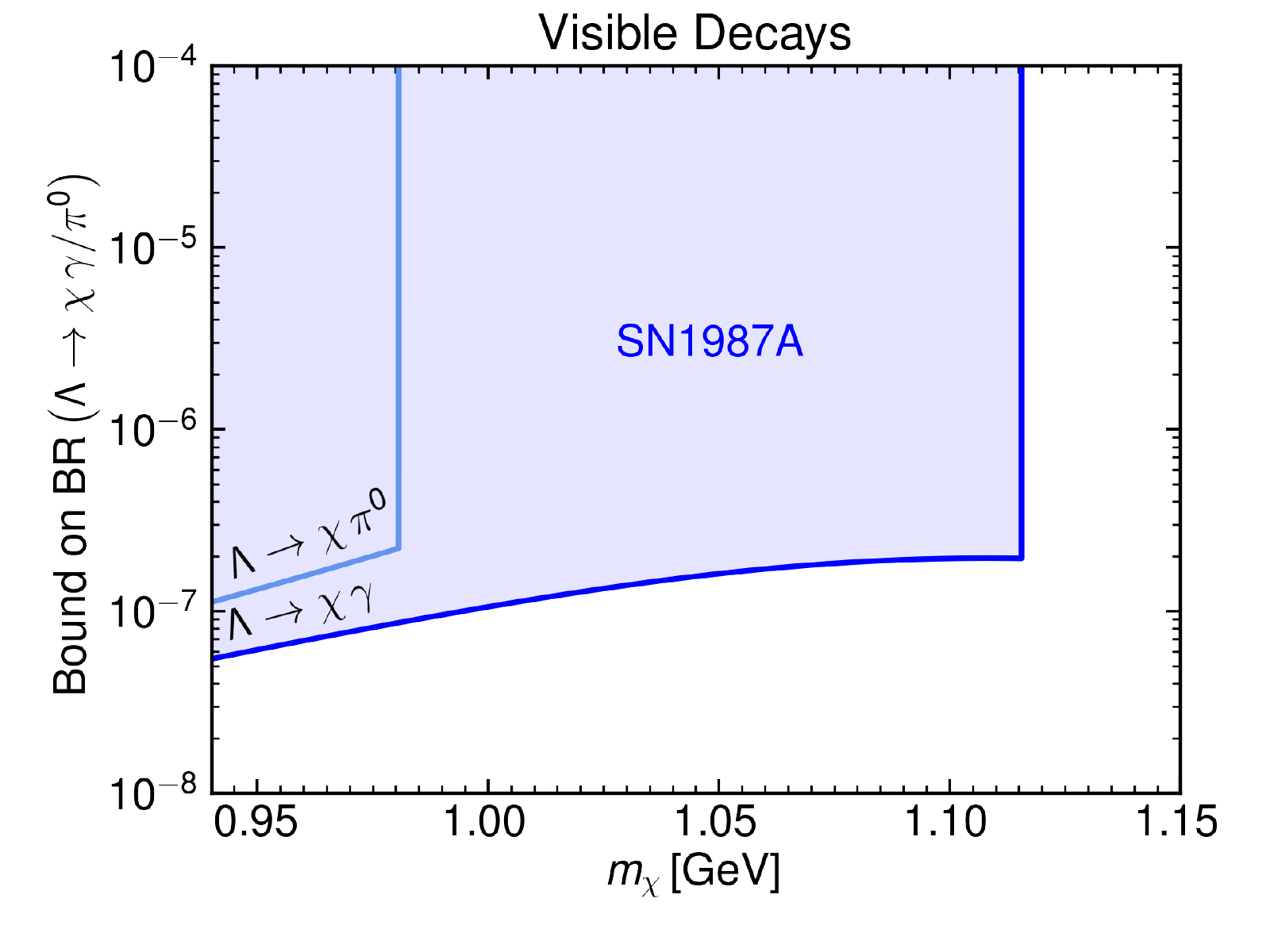}
&
		\includegraphics[width=0.48\textwidth]{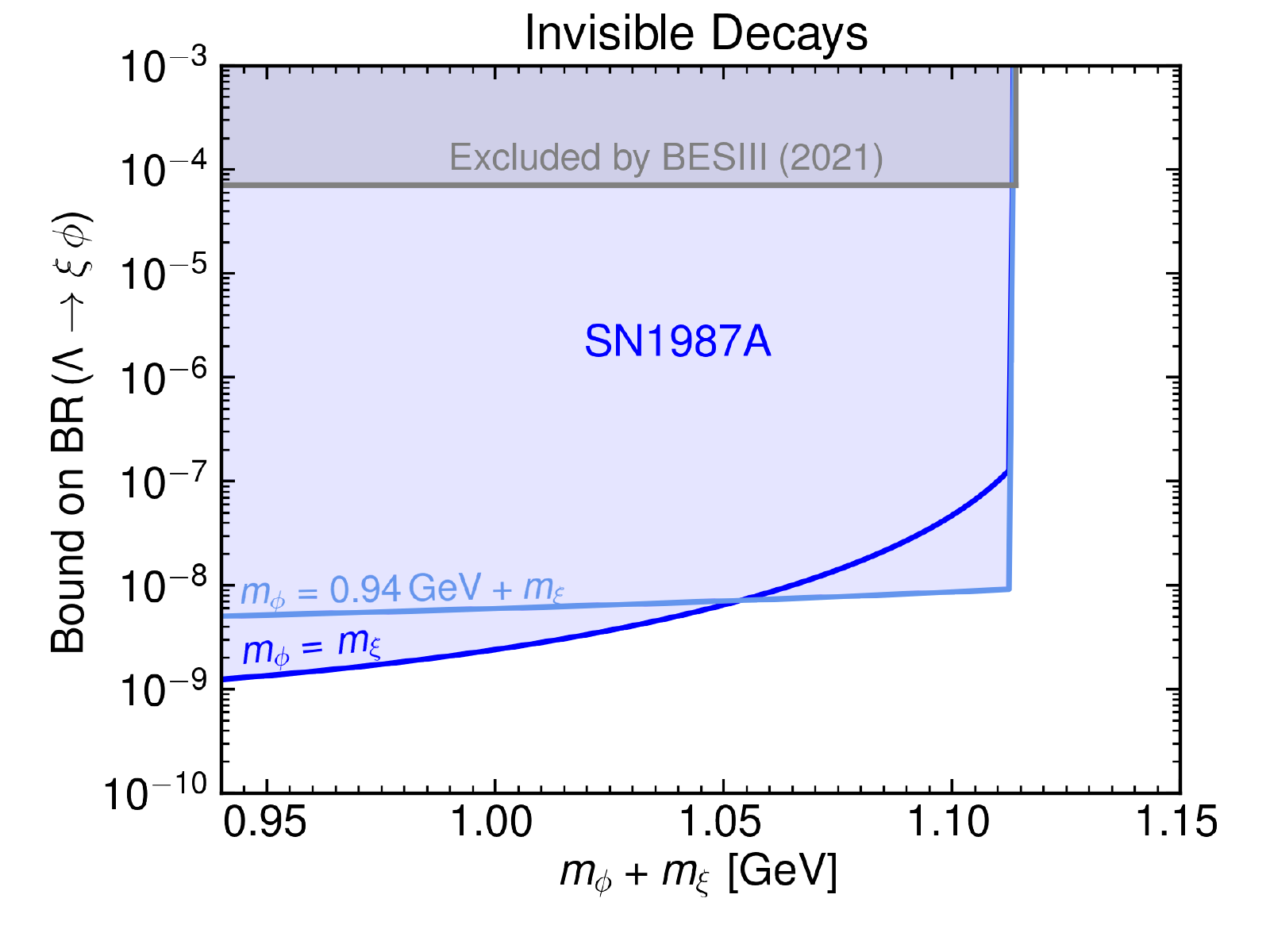}\\
\end{tabular}
\vspace{-0.2cm}
\caption{Supernova upper bounds on the branching ratios of $\Lambda\to\pi^0\chi$ and $\Lambda\to\gamma\chi$ (left panel) or $\Lambda\to\xi\phi$ (right panel). They are obtained from the analysis of the simulations labeled as SFHo18.80 in~\cite{Bollig:2020xdr} (see main text).
For invisible decays, we study two cases as a function of $m_\phi + m_\xi$: one in which $\xi$ and $\phi$ have the same mass ($m_\phi = m_\xi$, dark blue) and the other where $\phi$ is much heavier than $\xi$ ($m_\phi = 0.94\,{\rm GeV}+ m_\xi$, light blue). We also display the experimental upper limit on $\Lambda\to\text{invisible}$ as recently constrained by BESIII~\cite{BESIII:2021slv}.}
\label{fig:SN1987A-I}
\end{figure*}

The limit on the $\Lambda \to \chi\gamma$ branching ratio is slightly stronger than the one for $\Lambda\to \chi\pi^0$ (assuming $m_\chi=0.94$ GeV). However, the pionic decay implies a much stronger constraint on the size of the Wilson coefficients underpinning the  decays (see Sec~\ref{sec:DarkDecays}). On the other hand, for $m_\chi\gtrsim0.98$ GeV the pionic decay is blocked by phase space and the radiative decay provides the dominant dark cooling mechanism if $\chi$ cannot decay to other lighter dark sector particles. If it does decay, then the $\Lambda\to \xi\phi$ channel leads to the strongest supernova limits on the parameters of the dark baryon models. These conclusions also follow from the limits depicted in Fig.~\ref{fig:SN1987A-I} as a function  of the dark sector particle masses. For  invisible decays, we study two cases: \textit{(i)} $\xi$ and $\phi$ have the same mass, $m_\phi = m_\xi$; and \textit{(ii)} $\phi$ is substantially heavier than $\xi$, $m_\phi \simeq 0.94\,{\rm GeV}+m_\chi$. For comparison purposes, we also display in these figures the current experimental upper limits and estimated projections from the LHC (see Sec.~\ref{sec:LHC}) and the BESIII collaborations.

We have also performed a similar analysis for the neutron dark decay channels $n\to\chi\gamma$ and $n\to\xi\phi$.
We conclude that the SN upper limit on $\Gamma_{n\to\mathfrak{D}\mathfrak{b}}$ is too weak to be phenomenologically relevant, as it is several orders of magnitude smaller than the experimental value for $1/\tau_{n}$.

It is worth mentioning that the previous discussion implicitly assumes the dark baryons produced in the supernova to decay invisibly ($\chi \to \xi \phi$) or be stable. 
This is indeed predicted to be the case in $B$-Mesogenesis~\cite{Elor:2018twp} and in dark matter explanations of the neutron decay lifetime anomaly~\cite{Fornal:2018eol}. In other scenarios in which the $\chi$ particles can decay back into the SM, the constraints from SN~1987A could be even more stringent than those shown in Figure~\ref{fig:SN1987A-I}. As an example and given the lack of observed gamma-rays after the SN~1987A event~\cite{DeRocco:2019njg}, $\text{BR}(\Lambda \to \chi + \pi^0/\gamma)$ would be subject to strong constraints arising from $\chi\to n\gamma$ decays occurring in the outskirts of the supernova.
That said, this possibility involves the couplings of the dark baryon to $d$ and $s$ quarks to be simultaneously large, a situation that as we will see is strongly disfavored by neutral kaon mixing (see Sec.~\ref{sec:meson_mixing}).

\section{Phenomenological analysis}
\label{sec:phenomenology}

Having developed a formalism for computing branching ratios for various decay modes of strange hadrons into dark particles,  and having derived the supernova bounds on such processes, we now proceed to study the phenomenology of dark baryon sectors at terrestrial experiments. In particular, we present the LHC constraints on the colored scalar mediators needed to trigger the new strange hadron decays: $\Phi$, $\Psi$, and $X$ (see Table~\ref{tab:WCs}). These constraints can be recast into bounds on the Wilson coefficients for the operators under consideration. A similar procedure can be carried out using the SN~1987A bounds derived in Section~\ref{sec:supernova}. Finally, we derive meson-mixing constraints on various combinations of couplings, which helps to understand the possible flavor structures of the UV models.

\subsection{LHC limits}\label{sec:LHC}

The colored bosons in models of baryon decays to dark sector particles are constrained by various direct searches for colored states at the LHC. 

Firstly, ATLAS~\cite{ATLAS:2017jnp} and CMS~\cite{CMS:2018mts} have performed searches for pair-produced colored particles that decay into jets, leading to a 4-jet signature. These searches have been performed with 37 $\text{fb}^{-1}$ of data at $\sqrt{s} = 13\,\text{TeV}$ and have comparable sensitivity. The most constraining search, done by the  CMS, at 95\% CL rules out colored bosons with $M < 0.52\,\text{TeV}$~\cite{CMS:2018mts}. 

Secondly, ATLAS~\cite{Aad:2020aze} and CMS~\cite{Sirunyan:2019ctn} SUSY searches for pair-produced squarks decaying into a neutralino and a quark rule out the existence of this kind of strongly interacting bosons in the mass region below $1.2\,\text{TeV}$ with the current 139 $\text{fb}^{-1}$ of data, provided that the decay rate is $100\%$ to a jet and missing energy. 

Thirdly, at larger masses, searches for resonantly produced $qq'\rightarrow \Phi/\Psi/X$ decaying into quark pairs (dijet) or a quark and a dark baryon (jet+MET) can be used to place constraints on the colored bosons couplings.
For the dijet final state, we perform a recast of the CMS analysis presented in~\cite{Sirunyan:2018xlo}, which uses $36\,\text{fb}^{-1}$ of data at $13$~TeV, while for the jet+MET final state we employ a publicly available recast~\cite{recast_jetMET} of the ATLAS search~\cite{Aaboud:2017phn}, again based on $36\,\text{fb}^{-1}$ of data at $13$~TeV.
The details of the implementation are given in Appendix~\ref{app:LHC}.

\begin{figure*}[t!]
\centering
\begin{tabular}{cc}
		\label{fig:Y_LHC}
		\includegraphics[width=0.48\textwidth]{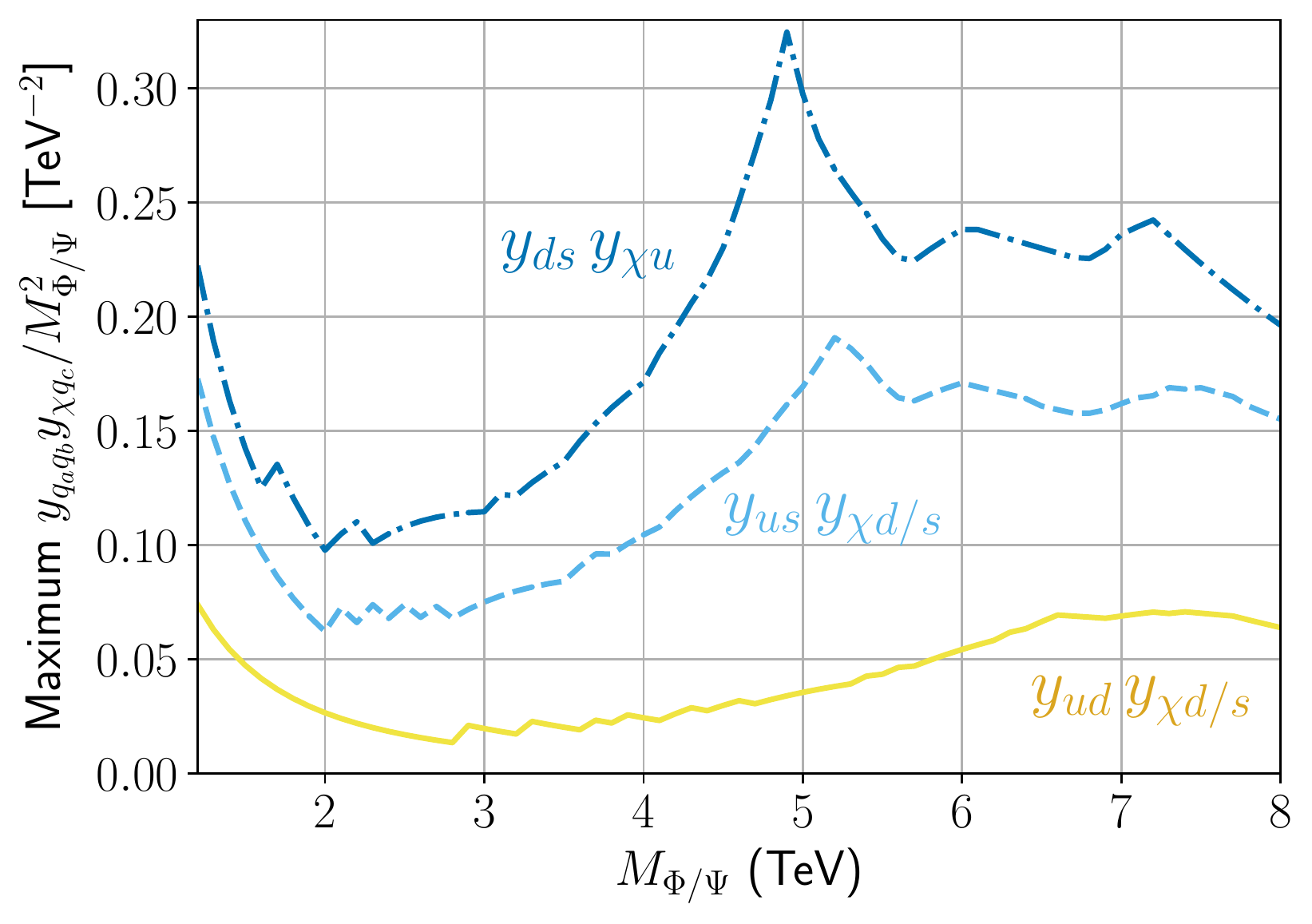}
&
		\label{fig:X_LHC}
		\includegraphics[width=0.48\textwidth]{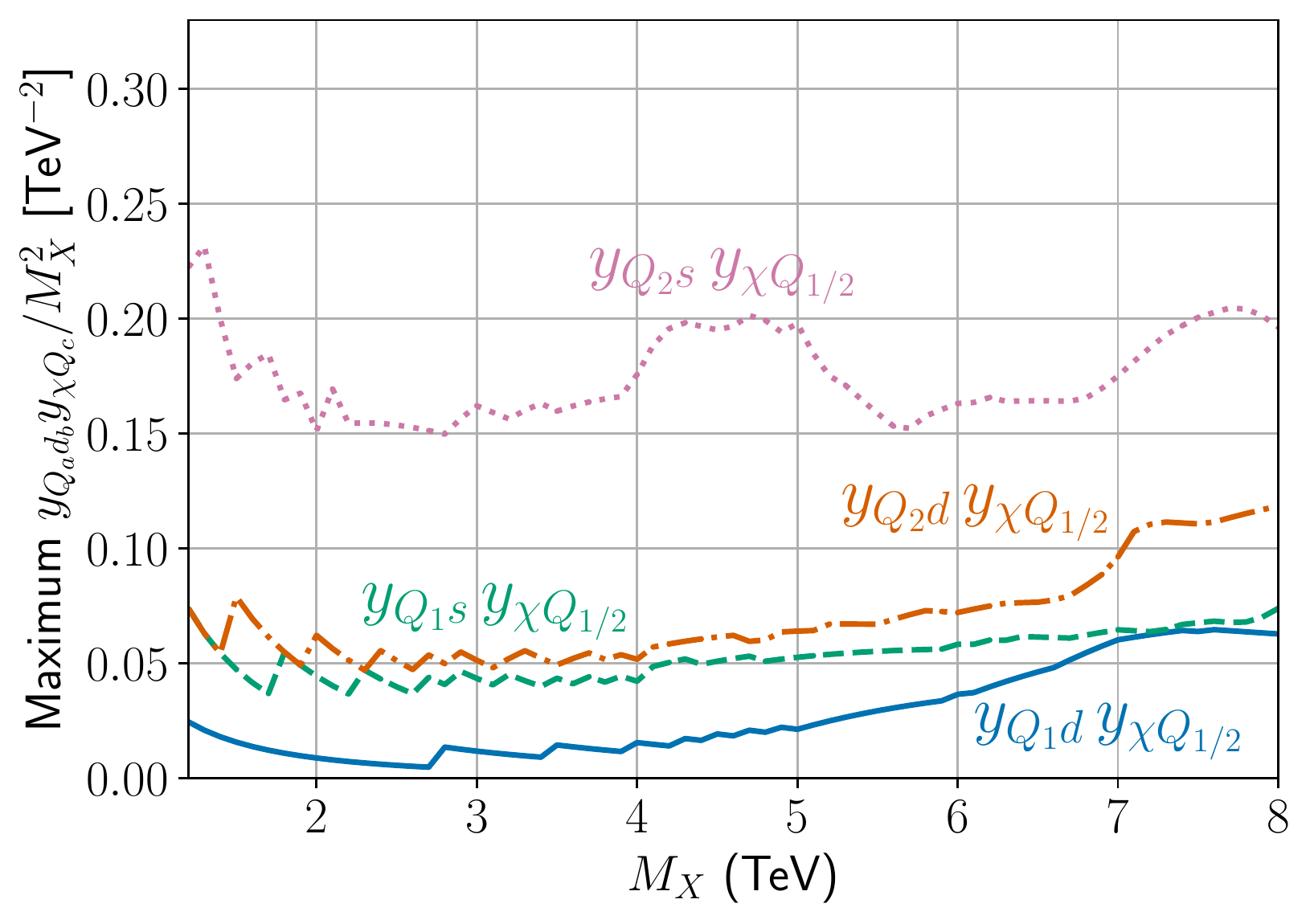}
\end{tabular}
\vspace{-0.2cm}
\caption{Collider constraints at 95\% CL on the flavor variations of the heavy colored mediator couplings: $\Phi$ and $\Psi$ (\emph{left}), and $X$ (\emph{right}). We show the bounds on the products $y_{qq'}y_{\chi q''}/M^2$, obtained from a combination of dijet~\cite{Sirunyan:2018xlo} and jet+MET~\cite{Aaboud:2017phn} searches at CMS and ATLAS, respectively.
The smallest mass considered corresponds to the limit $M\gtrsim 0.5$~TeV from 4-jet searches~\cite{CMS:2018mts}.
}
\label{fig:LHC}
\end{figure*}

The resulting bounds are shown in Figs.~\ref{fig:Y_LHC_BR} and~\ref{fig:X_LHC_BR} in the Appendix as a function of the branching ratios of the colored bosons into the respective channels. Those bounds can be combined to obtain a limit on the coupling products that enter the hyperon dark decay rates.
The procedure is described in Appendix~\ref{app:LHC}, and the resulting limits are shown in
Fig.~\ref{fig:LHC} as a function of $M_{\Phi/\Psi}$ and $M_X$ for all the relevant flavour combinations.
Given these, we obtain the following $95\%$ CL limits on the Wilson coefficients of the chiral EFT used to describe the hyperon dark decays:
\begin{align}
    \! \! \! C^L_{ud,d} &< 0.24\,\mathrm{TeV}^{-2}\,(X),\,\,\, &C^R_{ud,d} &< 0.07\,\mathrm{TeV}^{-2}\,(\Phi), \nonumber\\
    \! \! \! C^L_{ud,s} &< 0.28\,\mathrm{TeV}^{-2}\,(X),\,\,\, &C^R_{ud,s} &< 0.64\,\mathrm{TeV}^{-2}\,(\Psi), \nonumber\\ 
    \! \! \! C^L_{us,d} &< 0.48\,\mathrm{TeV}^{-2}\,(X),\,\,\, &C^R_{us,d} &< 0.64\,\mathrm{TeV}^{-2}\,(\Psi), \nonumber\\ 
    \! \! \! C^L_{us,s} &< 0.84\,\mathrm{TeV}^{-2}\,(X),\,\,\, &C^R_{us,s} &< 0.19\,\mathrm{TeV}^{-2}\,(\Phi).
    \label{eq:WC_LHC}
\end{align}
These limits have been obtained from the information in Fig.~\ref{fig:LHC}, and using Table~\ref{tab:WCs} to translate the bounds on the mediator mass and couplings to the Wilson coefficients in the chiral EFT.
In Eq.~\eqref{eq:WC_LHC} below we quote only the least stringent constraint among the ones obtained for the  mediators $\Phi$, $\Psi$, and $X$ (the parenthesis indicates to which one it corresponds).
In Table~\ref{tab:decays} and in Figs.~\ref{fig:ChPT_xi_phi},~\ref{fig:Baryontopi}, and~\ref{fig:Baryontogamma}, we normalize the branching ratios for the exotic hyperon decays to $C^{L/R}_{ab,c} = 0.5\,\mathrm{TeV}^{-2}$ as a representative figure for the LHC constraints in Eq.~\eqref{eq:WC_LHC}.

\subsection{Supernova bounds}
\label{subsec:SNpheno}
The SN 1987A cooling bounds on the hyperon dark decay rates shown in Fig.~\ref{fig:SN1987A-I} can be translated into bounds on the Wilson coefficients of the effective theory in Eq.~\eqref{eq:LEEFTchi}. 
This can be done straightforwardly by combining the upper limits on the branching fractions derived in Sec.~\ref{sec:supernova} with the theoretical predictions of the rates in Sec.~\ref{sec:DarkDecays}.

\begin{figure*}[t]
\centering
\begin{tabular}{cc}
		\includegraphics[width=0.48\textwidth]{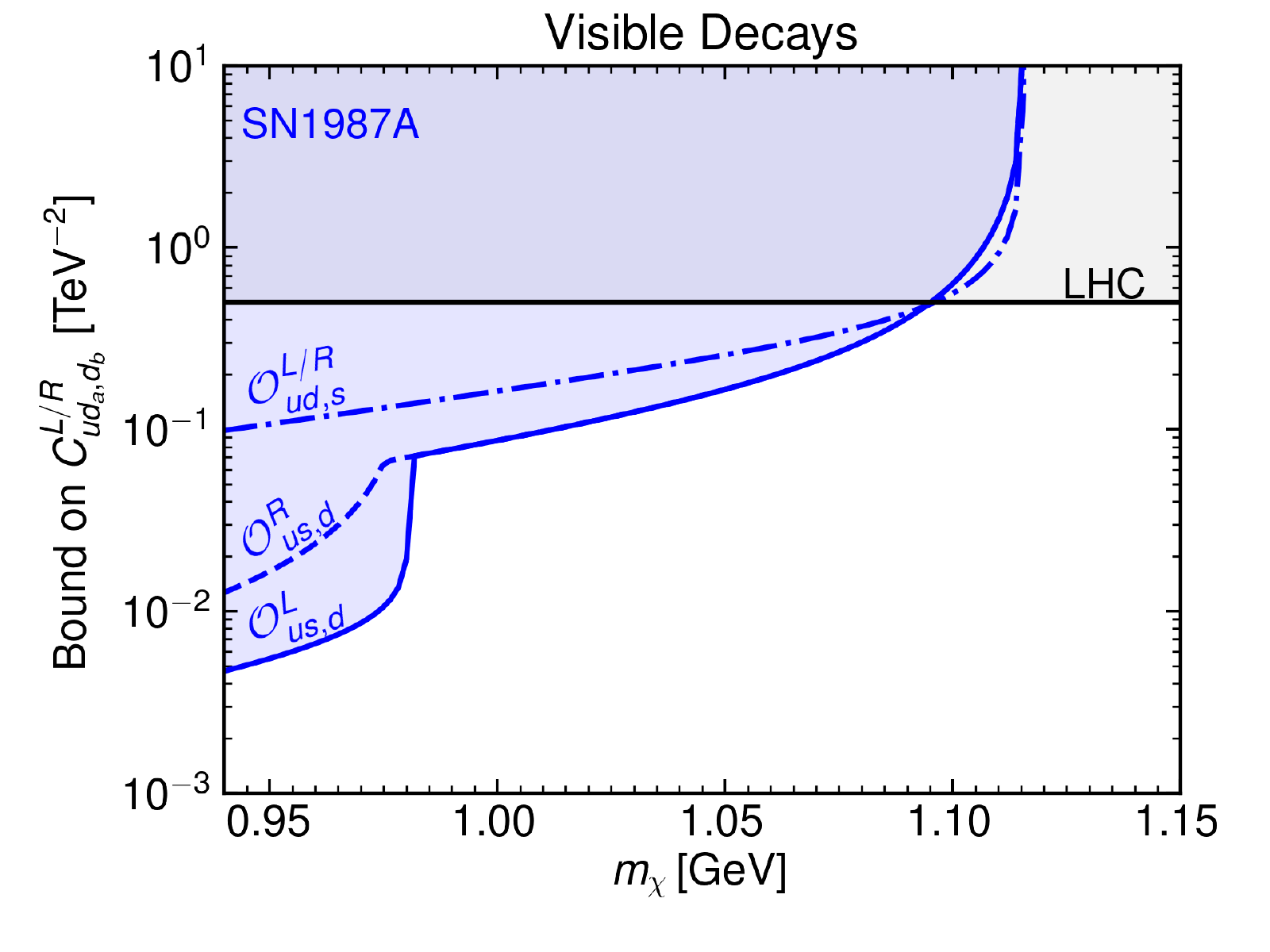} 
&
		\includegraphics[width=0.48\textwidth]{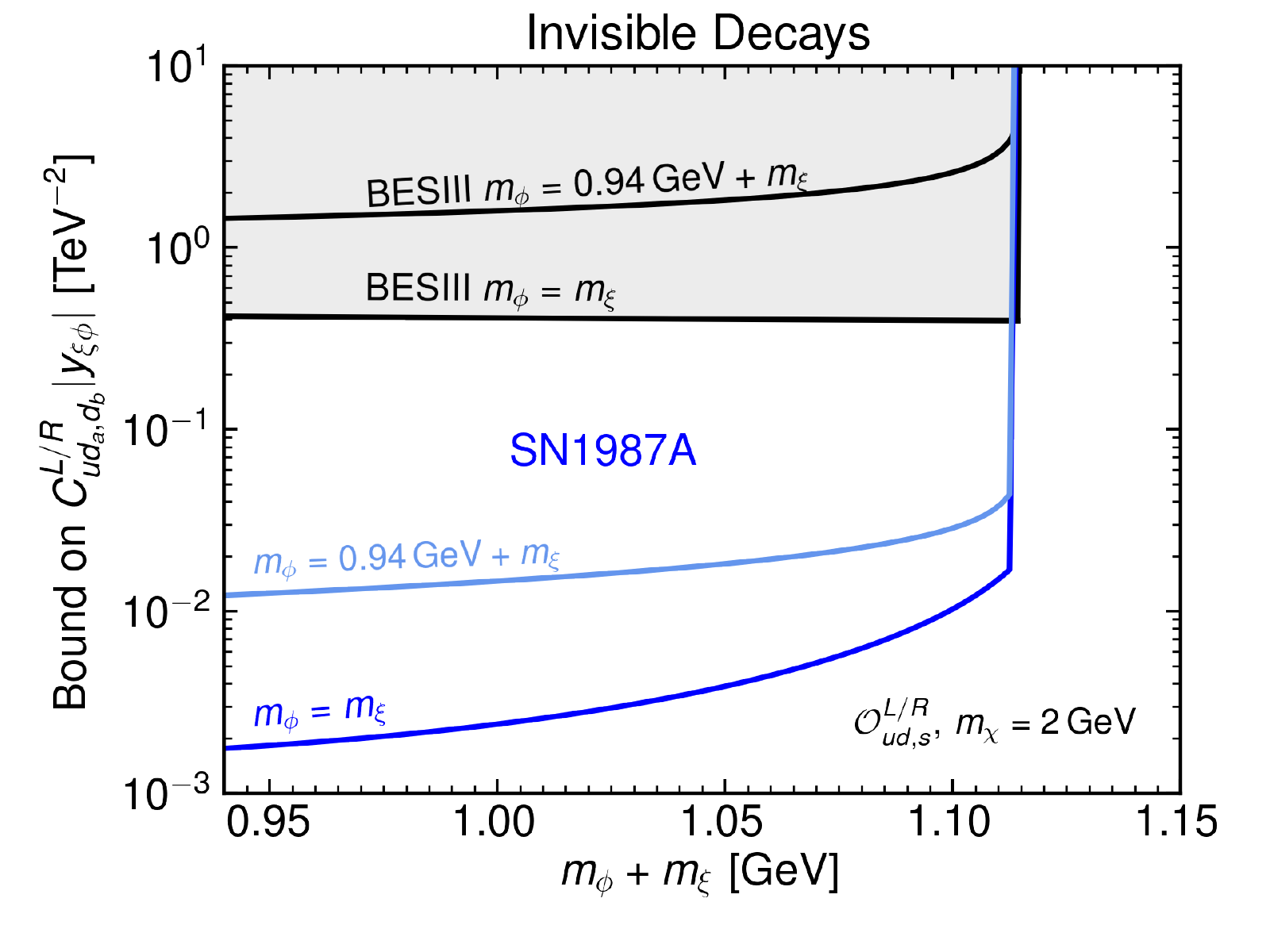} \\
\end{tabular}
\vspace{-0.2cm}
\caption{Constraints on the Wilson coefficients from SN~1987A. \textit{Left panel:} Constraints from visible decays. \textit{Right panel:} Constraints from invisible decays. For invisible decays, we show the bounds for the $|C_{ud,s}^{L/R}|$ coefficient and we have chosen $m_\chi = 2\,{\rm GeV}$ as an example. The bound for $|C_{us,d}^{L/R}|$ is simply a factor of two less stringent, see Table~\ref{tab:Btovac}. Note that the bound on the Wilson coefficient scales with $m_\chi$ according to Eq.~\eqref{eq:DecayRate}, and for $m_\chi > 2 \,{\rm GeV} $ roughly scales as $|C_{ud_a,d_b}^{L/R}| \propto m_\chi^{3/2}$. We also highlight regions of parameter space constrained by LHC searches, see Section~\ref{sec:LHC}, and by a recent search at BESIII~\cite{BESIII:2021slv}.}
\label{fig:SN1987A-II}
\end{figure*}

In Fig.~\ref{fig:SN1987A-II}, we show the results obtained for the different operators as a function of the mass of the dark sector particles in the case that the $\chi$ is taken to be a stable or very long-lived (left panel), or of the sum of dark sector particles $\phi$ and $\xi$. 
In the latter case, we consider two cases that were introduced already for Fig.~\ref{fig:SN1987A-I}: one in which $\xi$ and $\phi$ have same mass and another where $\phi$ is much heavier than $\xi$, $m_\phi = 0.94\,{\rm GeV}+ m_\xi$. 

The SN 1987A bounds are compared to those derived from direct searches of the colored mediators at the LHC in the previous section, and from the BESIII search of $\Lambda\to\xi\phi$~\cite{BESIII:2021slv}.
As illustrated by the right panel of Fig.~\ref{fig:SN1987A-II}, the SN~1987A bounds on the Wilson coefficients can be up to $\sim3$ orders of magnitude stronger than the collider ones. 

In case of the pionic and radiative decays, shown on the left panel of Fig.~\ref{fig:SN1987A-II}, the strongest bound for $m_\chi\lesssim970$ MeV is due to the former ones. This is especially true for $\mathcal O_{us,d}^L$, the operator to which the $\Lambda\to \pi^0\chi$ decays are most sensitive (see Tab.~\ref{tab:Ffunctions} and Fig.~\ref{fig:DecaysBNV1}).
These SN~1987A bounds are roughly 2 orders of magnitude stronger than the LHC constraints derived in the previous section.
For $\chi$ masses above 980 MeV, the $\Lambda\to \pi^0\chi$ process is kinematically forbidden and radiative decays take over as the main source of SN energy loss. The resulting limit is noticeably weaker, not more than $\sim1$ order of magnitude stronger than the LHC constraint and even subleading to the latter in the range $m_\chi\gtrsim 1.1$ GeV. 

Fig.~\ref{fig:SN1987A-II} showcases that it is challenging for terrestrial experiments to compete with the supernova limits on dark decays of the $\Lambda$ hyperon.
That said, in case of pionic and radiative decays, the charged $\Sigma$ baryons can be used to probe the same operators, potentially with much higher precision. 
Not only are $\Sigma^\pm$ decays much more sensitive to the corresponding Wilson coefficients (as per Tab.~\ref{tab:Ffunctions} and Fig.~\ref{fig:DecaysBNV1}), but they also give access to a broader range of $\chi$ masses.
For instance, consider the $C_{ud,s}^{L/R}$ operator, which the SN 1987A constrains at the $0.1$ TeV$^{-2}$ level for $m_\chi\simeq1$ GeV. This translates into a maximum rate
\begin{align}
\text{BR}(\Sigma^-\to\pi^-\chi)^{\rm SN}\lesssim5\times10^{-5}\;\;\;\text{for}\;\;\;\mathcal O_{ud,s}^{L/R}.
\end{align}
Another interesting example is the bound $C_{us,d}^{R}\lesssim 0.01$ TeV$^{-2}$ for $m_\chi\simeq0.94$ GeV, which implies that 
\begin{align}
\text{BR}(\Sigma^-\to\pi^-\chi)^{\rm SN}\lesssim2\times10^{-6}\;\;\;\text{for}\;\;\;\mathcal O_{us,d}^{R}.
\end{align}
Although these branching ratios are fairly small, they are not completely out of reach for hyperon facilities, as we discuss in Sec.~\ref{sec:decaysummary}.
The case of the purely invisible decays is less favorable as the only other hyperon that probes the same operators as $\Lambda\to\xi\phi$ is the neutral $\Sigma^0$, for which searches are highly penalized by its large electromagnetic width.

\subsection{Flavor constraints}\label{sec:meson_mixing}

Exotic colored bosons at the TeV scale coupling to quarks can mediate flavor mixing in neutral meson systems~\cite{Davidson:1993qk,Giudice:2011ak,Agrawal:2014aoa,Fajfer:2020tqf}.
The most sensitive observables in this regard are the mass differences in the $B$ mesons systems, $\Delta M_{B_d}$ and $\Delta M_{B_s}$, together with the CP-violating parameter $\epsilon_K$ in kaon mixing.
The SM predictions for these quantities match the experimental measurements very well, which imposes strong constraints on new physics contributions.
For other observables like the mass differences in the $K$ and $D$ meson systems, the SM prediction cannot be reliably calculated, and new effects can therefore only be constrained  to be smaller than the experimentally measured values.
Finally, CP-violating observables in the $B$ and $D$ meson systems are generally less constraining than $\Delta M$ and we do not consider them here. 

\begin{figure*}[t]
\centering
\begin{tabular}{ccc}
        \includegraphics[width=0.26\textwidth]{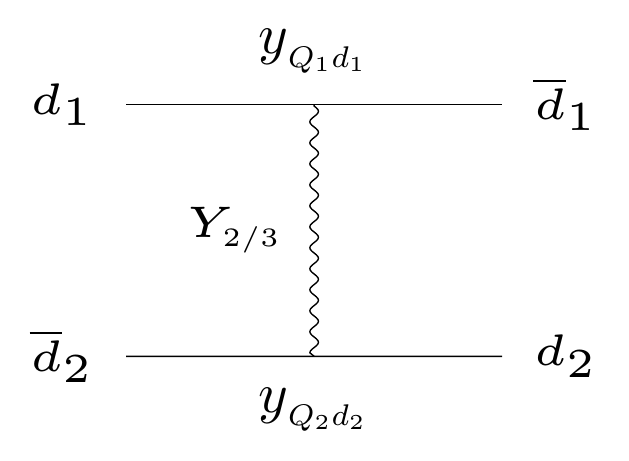}
		&
		\includegraphics[width=0.35\textwidth]{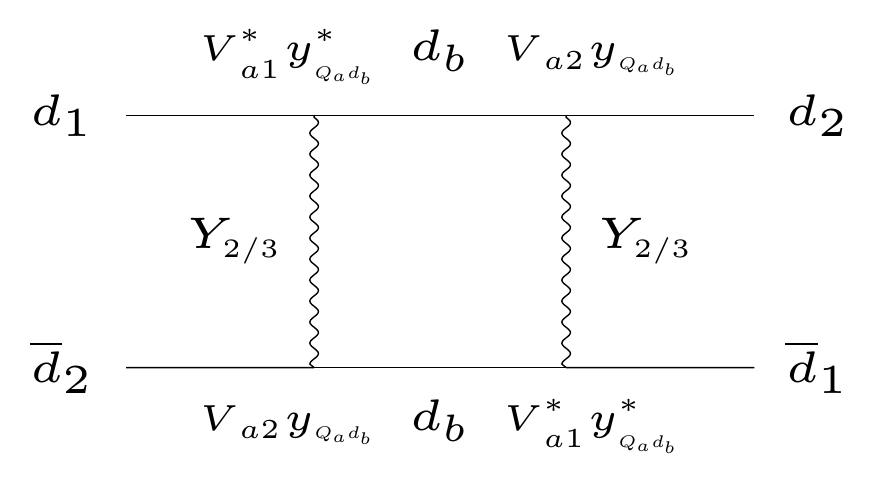}
		&
		\includegraphics[width=0.35\textwidth]{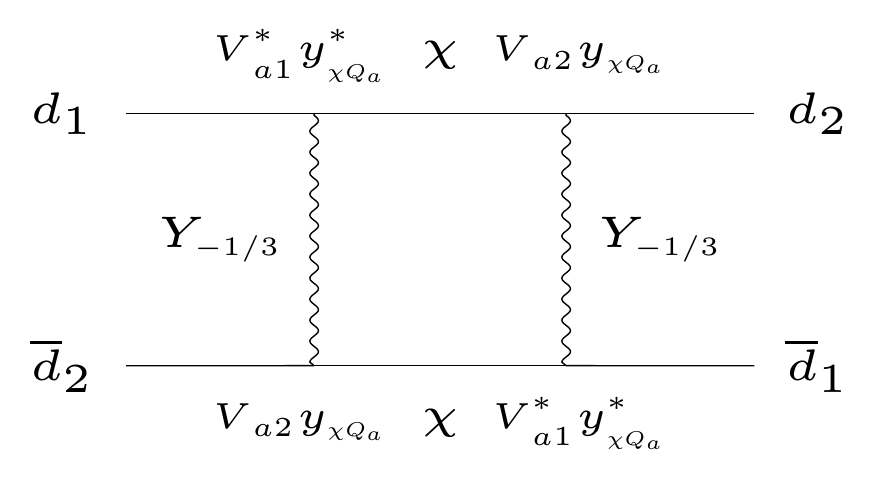}
		
\end{tabular}
\vspace{-0.2cm}
\caption{Some of the Feynman diagrams for the $\Delta F=2$ transitions induced by the vector colored mediator in Model 3.
The couplings are taken to be defined in the quark basis implied by Eq.~\eqref{eq:vector_decomposition}.
The left diagram induces flavor mixing at tree level.
The one-loop processes (center and left) require only one new physics coupling ($y_{Q_a d_b}$ and $y_{\chi Q_a}$, respectively) to be nonzero.}
\label{fig:Meson_mixing_V}
\end{figure*}

The $\Delta F=2$ phenomenology of scalar mediators with Lagrangians $\mathcal{L}_1$ and $\mathcal{L}_2$ in Eq.~\eqref{eq:Models} was recently studied in~\cite{Alonso-Alvarez:2021qfd}, and we summarize the main findings here.
We compare the most recent experimental measurements~\cite{pdg} and SM predictions~\cite{Lenz:2019lvd,Brod:2019rzc} to obtain the $95\%$ CL constraints shown in Table~\ref{tab:decays} for the color triplet scalars $\Phi$ and $\Psi$.
We refer the reader to Section V in~\cite{Alonso-Alvarez:2021qfd} for details regarding the calculation of the limits.\footnote{All of our calculations are made within the assumption of quark-hadron duality. They therefore neglect long-distance effects in intermediate states which may be relevant when internal light quarks are present in loop diagrams.}
Importantly, meson-antimeson mixing is only possible in the presence of at least two different nonzero flavor variations of the $y_{q_aq_b}$ or $y_{\chi q_a}$ couplings\footnote{In addition to the ones listed in Table~\ref{tab:decays}, bounds can also be placed on products of four different couplings. They are however less important for understanding the flavor structure of the models and decide to omit them here for the sake of brevity.}.
This means that $\Delta F=2$ processes cannot directly test individual Wilson coefficients leading to hyperon decays, but only products among them.
That said, these limits can play a crucial role in understanding the flavor structure of the couplings of the colored mediators.
As an example, assuming flavor universality in the $\Phi$ and $\Psi$ interactions would lead to limits on the Wilson coefficients of the chiral EFT at the level of $C^{L/R}_{ud_a,d_b}\lesssim 3\times 10^{-4}\,\mathrm{TeV}^{-2}$.
This would be stronger than the LHC and even the SN1987A limits, and highlights the fact that a nontrivial flavor structure for the colored mediator couplings is necessary to induce hyperon decays at rates detectable in current and upcoming experiments.

The $\Delta F=2$ phenomenology of the vector mediator is even more interesting.
The $X_\mu$ field with quantum numbers $(3, 2, 1/6)$ can be decomposed as
\begin{equation}
    X^\mu = \begin{pmatrix} 
    Y^\mu_{2/3} \\
    Y^\mu_{-1/3}
    \end{pmatrix},
\end{equation}
where the subscripts denote the electric charge of each component (which have the same mass but potentially different decay widths).
Using this to explicitly expand the $SU(2)$ indices, the Lagrangian in Eq.~\eqref{eq:Lag_dark} can be written in the quark mass basis as
\begin{align}
    \mathcal{L}_3 \supset &-y_{Q_a d_b} \epsilon_{ijk} \Big( Y^{\mu\,i}_{2/3}\, V_{a a'}\,\bar{d}_{a'}^{c\,j}\,\gamma_\mu\,P_R d^k_b \nonumber \\ 
     &\qquad\qquad\quad\,\,\,- Y^{\mu\,i}_{-1/3} \,\bar{u}_a^{c\,j}\, \gamma_\mu\, P_R d^k_b \Big) \nonumber\\
    &-y_{\chi Q_a} \left( Y^{\dagger\mu\,i}_{2/3}\,\bar{u}_{a}^{c\,i}\,\gamma_\mu\,P_R \chi + Y^{\dagger\mu\,i}_{-1/3}\, V_{aa'} \,\bar{d}_{a'}^{c\,i}\, \gamma_\mu\, P_R \chi \right)\,. \label{eq:vector_decomposition}
\end{align}
Here, we use $4$-component notation and have chosen to diaganolize the quark mass matrix by rotating the down-type components of the SU$_2$(L) doublets using the CKM matrix $V\equiv V^{\rm CKM}$.
The first thing to note is that in this model meson mixing can occur at tree level in the presence of two nonzero $y_{Q_ad_a}$ couplings via the left diagram in Fig.~\ref{fig:Meson_mixing_V}.
Using a generalization of the notation in~\cite{Bagger:1997gg} for the effective operators describing mixing in the $\bar{d_2}d_1$ meson system, we find that the left diagram in Fig.~\ref{fig:Meson_mixing_V} contributes to the operators $\mathcal{O}_4^{d_1 d_2} = \bar{d}_{1R}^i\, d_{2L}^i\, \bar{d}_{1L}^j\, d_{2R}^j$ and $\mathcal{O}_5^{d_1 d_2} = \bar{d}_{1R}^i\, d_{2L}^j\, \bar{d}_{1L}^j\, d_{2R}^i$ with a partonic amplitude
\begin{align}
    \delta C_4^{d_1d_2} & \simeq 2\, \frac{y_{Q_1d_1} y_{Q_2d_2}^*}{M_X^2}, \\
    \delta C_5^{d_1d_2} & = - \delta C_4^{d_1d_2},
\end{align}
where we have neglected external momenta and approximated the diagonal elements of the CKM matrix to unity.
We direct the reader to Appendix~E of~\cite{Alonso-Alvarez:2021qfd} for details on the conventions and the hadronic matrix elements needed to calculate the mixing parameters from the effective operators.
We compare these new physics contributions to the observations and SM predictions for the mixing parameters $\Delta M$ in the $K$, $B_d$, and $B_s$ systems, together with the CPV observable $\epsilon_K$.
To this end, we use the same procedure as described in~\cite{Alonso-Alvarez:2021qfd} for the scalar-mediated case.
The resulting $95\%$ CL bounds are summarized in Table~\ref{tab:meson_mixing_V_tree_level} for each flavor combination of the operators.
Note the steeper scaling of the bounds with the vector mass as compared with the loop-level limits.

The second important fact is that a single nonzero $y_{Q_a d_b}$ or $y_{\chi Q_a}$ in Eq.~\eqref{eq:vector_decomposition} induces couplings of $X$ to quarks of all generations due to CKM mixing.
In consequence, meson mixing can be induced by any of the individual couplings of the vector mediator via the loop diagrams shown in Fig.~\ref{fig:Meson_mixing_V}.
Note that the diagrams allowed are highly dependent on the original basis in which the couplings of the vector mediator are defined.
For concreteness, we work with the basis implied by Eq.~\eqref{eq:vector_decomposition}, taking a single $y_{Q_ad_b}$ or $y_{\chi Q_a}$ coupling to be nonzero at a time.
This is clearly an \emph{ad hoc} choice as other possibilities are equally valid, but it nonetheless allows us to obtain a rough understanding of the flavor-mixing phenomenology of the vector mediator.
The middle and left diagrams in Fig.~\ref{fig:Meson_mixing_V} contribute to the SM-like operator $\mathcal{O}_1^{d_1 d_2} = \bar{d}_{2L}^i\gamma_\mu d_{1L}^i \, \bar{d}_{2L}^j\gamma^\mu d_{1L}^j$ with partonic amplitudes\footnote{Given that we are dealing with a physical (as opposed to a gauged) vector boson, the diagrams in Fig.~\ref{fig:Meson_mixing_V} should be computed in unitary gauge. However, the $k^\mu k^\nu/M_X^2$ piece of the propagator results in a divergent loop integral.
This divergence can only be tamed in a full model that addresses the UV origin of the vector and is therefore necessarily model dependent.
In order to remain as general and conservative as possible, we choose to completely drop this term from the calculation, effectively using a Feynman gauge propagator for which no divergences arise.
}
\begin{align}
    \delta C_1^{d_1d_2} &= \frac{1}{8\pi^2} \frac{1}{M_X^2} \left( V_{u_ad_1}^* V_{u_a d_2} \right)^2 \left| y_{Q_a d_b} \right|^4\tilde{S}(x_{d_b}),\nonumber \\
    \delta C_1^{d_1d_2} &= \frac{1}{8\pi^2} \frac{1}{M_X^2} \left( V_{u_ad_1}^* V_{u_a d_2} \right)^2 \left| y_{\chi Q_a} \right|^4\tilde{S}(x_{\chi}),
    \label{eq:flavor_mixing_single_coupling}
\end{align}
for the center and right diagram, respectively.
Here $x_i = m_i^2 / M_X^2$, and
\begin{equation}
    \tilde{S}(x) = \int_{0}^{\infty} \frac{t^{2} \mathrm{d} t}{(t+1)^2(t+x)^2}\,.
\end{equation}
Due to the Lorentz structure of the interactions, the amplitudes of vector-mediated transitions are enhanced by a factor of $16$ compared to the case of the scalar mediator.
As before, these can be compared with the experimental measurements and SM predictions for meson-mixing observables.
The resulting $95\%$ CL bounds are summarized in Table~\ref{tab:meson_mixing_V_single_coupling}, where we only show the strongest constraint on each coupling.
They are weaker than the ones in Table~\ref{tab:decays} because of the CKM factors involved in the amplitudes in Eq.~\eqref{eq:flavor_mixing_single_coupling}.
As they apply to individual couplings rather than products among them, the limits in Table~\ref{tab:meson_mixing_V_single_coupling} can be translated into the following direct bounds on the Wilson coefficients in the chiral EFT:
\begin{align}
    C^L_{ud,d} &< 0.09\,\mathrm{TeV}^{-2}\,(X)\,, \nonumber\\
    C^L_{ud,s} &< 0.09\,\mathrm{TeV}^{-2}\,(X)\,, \nonumber\\ 
    C^L_{us,d} &< 0.08\,\mathrm{TeV}^{-2}\,(X)\,, \nonumber\\ 
    C^L_{us,s} &< 0.08\,\mathrm{TeV}^{-2}\,(X)\,.
    \label{eq:WC_meson_mixing_vector}
\end{align}
Importantly, these bounds  apply only for the vector mediator. The scalar mediators can lead to larger Wilson coefficients, since they are not directly constrained by flavor mixing, as explained above.

\begin{table}[t!]
\renewcommand{\arraystretch}{2.}
  \setlength{\arrayrulewidth}{.25mm}
\centering
\small
\setlength{\tabcolsep}{0.5 em}
\begin{tabular}{ | c | c | c | }
    \hline
    Coupling & Bound $\times\left(\frac{1.5 \ {\rm TeV}}{M}\right)$  & Origin   \\ 
    \hline \hline
    \multicolumn{3}{|c|}{$\Phi= (3,1)_{-\frac13}$}\\
    \hline
        $|y_{ud}\, y_{us}^\star|$ &  $ 10^{-3}$ & $\epsilon_K$  \\
    \hline
            $|y_{cd}\, y_{cs}^\star|$ &  $ 8 \times 10^{-4}$ & $\epsilon_K$  \\
    \hline
         $|y_{td}\, y_{ts}^\star|$ &  $ 3 \times 10^{-4}$ & $\epsilon_K$  \\
    \hline
            $|y_{cd_j}\, y_{ud_j}^\star|$ &  $ 2 \times 10^{-2}$ & $\Delta M_{D}$  \\
                \hline
            $|y_{u_ib}\, y_{u_id}^\star|$ &  $ (2-4) \times 10^{-2}$ & $\Delta M_{B_d}$  \\
    \hline
                $|y_{u_ib}\, y_{u_is}^\star|$ &  $ (1-2) \times 10^{-1}$ & $\Delta M_{B_s}$  \\
    \hline
        $|y_{\chi d}\, y_{\chi s}^\star|$ &  $  10^{-3}$ & $\epsilon_{K}$  \\
                   \hline
        $|y_{\chi b}\, y_{\chi d}^\star|$ & $(2-4) \times 10^{-2}$ & $\Delta M_{B_d}$   \\
    \hline
    $|y_{\chi b}\, y_{\chi s}^\star|$ & $(1-2) \times 10^{-1}$ & $\Delta M_{B_s}$   \\
    \hline
    \hline
        \multicolumn{3}{|c|}{$\Psi= (3,1)_{\frac23}$}\\
    \hline
                $|y_{db}\, y_{sb}^\star|$ & $ 10^{-3}$ & $\epsilon_{K}$   \\
                \hline
               $|y_{ds}\, y_{sb}^\star|$ & $(2-4)\times  10^{-2}$ & $\Delta M_{B_d}$   \\
                              \hline
               $|y_{db}\, y_{sd}^\star|$ & $(1-2)\times  10^{-1}$ & $\Delta M_{B_s}$   \\
           \hline
                          $|y_{\chi u}\, y_{\chi c}^\star|$ & $2\times  10^{-2}$ & $\Delta M_{D}$   \\
           \hline
\end{tabular}
\caption{$95\%$ CL flavor mixing constraints on products of couplings for the scalar colored mediators. The bounds are given for a benchmark mass of $1.5$~TeV to comply with the LHC bounds on strongly interacting bosons. The intervals in the bounds from $\Delta M_{B_{s/d}}$ cover all the possible complex phases of the couplings, ranging from aligned to anti-aligned with the SM contributions. The limits for the vector mediator $X$ are not listed, but they are a factor of $\sim4$ stronger due to the different Lorentz structure of the couplings.}
\label{tab:decays}
\end{table}

\begin{table}[t]
\renewcommand{\arraystretch}{2.}
  \setlength{\arrayrulewidth}{.25mm}
\centering
\small
\setlength{\tabcolsep}{0.5 em}
\begin{tabular}{ | c | c | c | }
    \hline
    Coupling & Bound $\times\left(\frac{1.5 \ {\rm TeV}}{M}\right)^2$  & Origin   \\ 
    \hline \hline
    \multicolumn{3}{|c|}{$X_\mu= (3,2)_{\frac16}$}\\
    \hline
    $|y_{Q_1 d}y_{Q_2 s}^*|$ &  $3.3\times10^{-8}$ & $\Delta M_K$  \\
    \hline
    $\mathrm{Im}(y_{Q_1 d}y_{Q_2 s}^*)$ &  $2.8\times10^{-11}$ & $\epsilon_K$  \\
    \hline
    $|y_{Q_3 b}y_{Q_1 d}^*|$ &  $(2.0-8.9)\times 10^{-7}$ & $\Delta M_{B_d}$  \\
    \hline
    $|y_{Q_3 b}y_{Q_2 s}^*|$ &  $(0.5-1.8)\times 10^{-5}$ & $\Delta M_{B_s}$  \\
    \hline
\end{tabular}
\caption{$95\%$ CL flavor mixing constraints on the vector couplings from tree-level meson mixing. The intervals in the bounds from $\Delta M_{B_{q}}$ cover all the possible complex phases of the couplings, ranging from aligned to anti-aligned with the SM contributions.}
\label{tab:meson_mixing_V_tree_level}
\end{table}

\begin{table}[h]
\renewcommand{\arraystretch}{2.}
  \setlength{\arrayrulewidth}{.25mm}
\centering
\small
\setlength{\tabcolsep}{0.5 em}
\begin{tabular}{ | c | c | c | }
    \hline
    Coupling & Bound $\times\left(\frac{1.5 \ {\rm TeV}}{M}\right)$  & Origin   \\ 
    \hline \hline
    \multicolumn{3}{|c|}{$X_\mu= (3,2)_{\frac16}$}\\
    \hline
    $|y_{Q_1 d_a}|^2$ and $|y_{\chi Q_1}|^2$ &  $5.0\times10^{-2}$ & $\Delta M_K$  \\
    \hline
    $|y_{Q_2 d_a}|^2$ and $|y_{\chi Q_2}|^2$ &  $4.3\times10^{-2}$ & $\epsilon_K$  \\
    \hline
    $|y_{Q_3 d_a}|^2$ and $|y_{\chi Q_3}|^2$ &  $0.5-1.0$ & $\Delta M_{B_d}$  \\
    \hline
\end{tabular}
\caption{$95\%$ CL flavor mixing constraints on the couplings of the vector mediator. The upper bound is normalized to the LHC reach. The experimental origin of each bound is also provided.}
\label{tab:meson_mixing_V_single_coupling}
\end{table}

In addition, flavor mixing bounds on the product of two (or more) different couplings can also be derived for the vector mediator.
The richer coupling structure and CKM mixing opens many possibilities for products of different $X$ couplings to be constrained using meson mixing generated at one loop.
It is however not the goal of this paper to perform an exhaustive analysis of these limits and we therefore omit them here.
A rough estimate of the order of magnitude of the constraints can be obtained based on the constraints on the scalar mediators.
The numerical values in Table~\ref{tab:decays}.
can be roughly translated to the $X$ couplings by strengthening them by a factor of $\sim 4$. This enhancement is caused by the different Lorentz structure in the vector-mediated box diagrams compared to the scalar case.
The main conclusion is that significant hierarchies in the flavor couplings of the vector mediator are necessary in order to obtain detectable hyperon decay rates while complying with flavor-mixing bounds.

In the derivation of the above bounds and for simplicity we have completely neglected any renormalization group (RG) running effects.
These can however be important as is highlighted, for instance, in~\cite{Aebischer:2020dsw}.
As an example, the Wilson coefficients for the (V-A)$\times$(V+A) operators ($\mathcal{O}_4$ and $\mathcal{O}_5$ in our notation) decrease by a factor of $\sim 10$ in running from the TeV to the GeV scale, as shown in Fig.~1 of~\cite{Aebischer:2020dsw}.
Taking this into account would weaken the bounds presented in Table~\ref{tab:meson_mixing_V_tree_level} by a factor of $\sim 10$.
In contrast, the running of the (V-A)$\times$(V-A) or (V+A)$\times$(V+A) operators that arise in the loop diagrams is much less significant.
All in all, a full RG-improved calculation would refine our quantitative understanding of the flavor mixing constraints, but it is not expected to alter our qualitative conclusions.
This task, which is outside the scope of the present paper, is deferred to future work.

The constraints derived in this section allow us to gain important insights about the flavor structure of the models.
Most importantly, the vector mediator ($\mathcal{L}_3$ in Eq.~\eqref{eq:Models}) is directly constrained as CKM mixing induces $\Delta F=2$ processes with a single nonzero coupling.
This results in the direct bounds in Eq.~\eqref{eq:WC_meson_mixing_vector} on the Wilson coefficients that control the exotic hyperon decay rates.
Furthermore and due to the existence of the tree-level diagram shown in Fig.~\ref{fig:Meson_mixing_V} (left), products of couplings are strongly constrained for the vector mediator.
Therefore, a huge hierarchy between different couplings would be necessary in order to yield a Wilson coefficient leading to an observable hyperon decay rate.

The situation is much more favorable for the scalar mediators corresponding to $\mathcal{L}_1$ and $\mathcal{L}_2$ in Eq.~\eqref{eq:Models}.
In this case, meson mixing is only induced at the loop level and as long as two different flavor combinations of couplings are simultaneously sizeable.
As a consequence, sizeable hyperon decay rates are possible with only mild ($10^{-2}-10^{-3}$) hierarchies in the coupling matrix in flavor space for the scalar mediators.

\section{Prospects for hyperon facilities}
\label{sec:theory}
In this section we summarize our results and  elaborate on the theoretical and experimental implications given the interplay of the various studies analyzed above. In particular, we comment on the  connections between searches for dark hyperon decays, the neutron lifetime puzzle, and the matter-antimatter asymmetry generation via Mesogenesis.

\subsection{Most promising channels}
\label{sec:decaysummary}

\begin{table*}[t]
\renewcommand{\arraystretch}{1.6}
  \setlength{\arrayrulewidth}{.25mm}
\centering
\small
\setlength{\tabcolsep}{0.5 em}
\begin{tabular}{  c | c | c | c | c  | c  | c }
    \hline
    \hline
    Initial State & Final State & ${\rm Max}[m_\chi]$ (MeV)  & Operator & $\text{Br} \times \Big[ \frac{C_{ud_a,d_b}^{L/R}}{0.5\,\text{TeV}^{-2}}\Big]^2$ (LHC) &  SN~1987A Limit &  BESIII \\
    \hline \hline
    \multirow{3}{*}{$\Lambda\, (uds)$}
    &  $\chi + \gamma$ & $1116$ & $\mathcal{O}^{L/R}_{us,d}$  & $3.5\times 10^{-6}$ & $\sim 10^{-7}$ &  -- \\
    &  $\xi + \phi$ & $1116$ & $\mathcal{O}^{L/R}_{ud,s}$  & $5.0 \times 10^{-4} \, |y_{\xi\phi}|^2$ & $\sim 10^{-8}$ & $7\times 10^{-5}$~\cite{BESIII:2021slv}\\
    &  $\chi + \pi^0$ & $981$ & $\mathcal{O}^{L}_{us,d}$  & $1.1 \times 10^{-3}$ & $\sim 10^{-7}$  & --\\
    \hline
    $\Sigma^+\, (uus)$ &  $\chi + \pi^+$ & $1050$ &  $\mathcal{O}^{L/R}_{us,d}$   & $1.6\times 10^{-3}$ & $\sim 10^{-4}$ & --\\
    \hline
    $\Sigma^-\, (dds)$ & $\chi + \pi^-$ & $1058$
    & $\mathcal{O}^{L/R}_{ud,s}$ 	  & $1.2\times 10^{-3}$ & $\sim 10^{-4}$ & --\\
    \hline
    \multirow{3}{*}{$\Xi^0\, (uss)$}
    & $\chi + \gamma$ & $1315$  &  $\mathcal{O}^{L/R}_{us,s}$ 	   & $ 1.5\times 10^{-5}$ & -- & --\\
    & $\xi + \phi$ & $1315$  &  $\mathcal{O}^{L/R}_{us,s}$	   & $3.7\times 10^{-4}\,|y_{\xi\phi}|^2$  & -- & --\\
    & $\chi + \pi^0$ & $1180$
    &  $\mathcal{O}^{R}_{us,s}$  & $6.2\times 10^{-3}$ & -- & --\\
    \hline
    $\Xi^-\, (dss)$ & $\chi + \pi^-$ & $1182$
    &  $\mathcal{O}^{R}_{us,s}$  & $7.1\times 10^{-3}$  & -- & --\\
    \hline
    \hline

\end{tabular}
\caption{The most relevant decay channels for hyperons into dark sector states. The columns provide: the initial and final state particles, the maximum  dark sector state mass that can be probed, a representative operator that contributes to the decay channel, and an estimate of the branching ratio assuming $m_\chi = 1.0\,\text{GeV}$, $m_\phi = 0.95\,\text{GeV}$ and $m_\xi = 0.04\,\text{GeV}$ (for $\Lambda\rightarrow\chi+\pi^0$ we use $m_\chi = 0.95\,\text{GeV}$). The branching ratios have been normalized to highlight their maximum possible values in light of the LHC constraints on color-triplet scalars (fifth column) and the SN~1987A bound (sixth column). 
\label{tab:decayChannels}}
\end{table*}

The most relevant hyperon decay channels involving dark baryons are identified in Table~\ref{tab:decayChannels}. 
For each channel, we specify the maximum possible mass of the dark sector particles and provide a representative operator that contributes to the corresponding  decay channel. We also show upper limits on the branching ratios that can be obtained indirectly by either LHC searches for the colored mediators (Sec.~\ref{sec:LHC}) or by cooling constraints from SN~1987A (Sec.~\ref{sec:supernova}). 
In light of a putative discovery of a hyperon dark decay mode, 
flavor-mixing observables (Sec.~\ref{sec:meson_mixing}) would assist in reconstructing the flavor structure of the possible models behind it.

Table~\ref{tab:decayChannels} helps showcasing the most important conclusions of our results:
\begin{enumerate}[label=\roman*]
\item[(a)] The branching ratios for dark decays of hyperons can be as large as $\sim10^{-3}$ in case of pionic decays of the $\Xi$ baryons. Note that these transitions are not constrained by SN~1987A and should thus be a priority target for laboratory searches. In particular, the decay $\Xi^-\to\pi^-\chi$ might be experimentally preferred because of the charged pion in the final state. Since the $\Xi$ hyperons are the heaviest among the baryon octet, they also probe a wider range of dark sector particle masses. 
\item[(b)] The decays of the $\Sigma$ and $\Lambda$ baryons can also lead to significant branching ratios when the LHC bounds are taken as reference. However, they probe the same operators (or even the same decay channels in case of the $\Lambda$) as those already constrained by the supernova  SN~1987A bound.
The best prospects in this case are for $\Sigma^\pm$ decays, which are more sensitive than $\Lambda$ ones to a given operator and therefore less constrained by SN~1987A.
\item[(c)] Radiative decay rates are suppressed roughly by $\alpha_{\rm em}$ as compared to, e.g., pionic decays. Nevertheless, they provide access to a wider range of dark baryon masses.  
\end{enumerate}

To contextualize the possibilities for testing these decays it would be ideal to have an estimate of the expected sensitivity at BESIII. Although a detailed quantitative analysis is beyond the scope of this work, given some properties of hyperons and BESIII features we can at least provide a qualitative discussion.

BESIII has already gathered data from $\sim 10^{10}\,J/\psi$ decays~\cite{BESIII:2021cxx}. The branching fraction of $J/\psi$ states into any flavor of hyperons pairs is $\text{BR}\sim (1-2) \times 10^{-3}$~\cite{pdg}. This in turn means that, at present, the maximum possible sensitivity to non-standard hyperon decays is $\text{BR}\sim 10^{-7}$.
In addition, since all of our decays involve missing energy in the final state, the kinematics of one the hyperons must be fully reconstructed so as to gain information about the missing energy in the event. This implies that one could expect a somewhat smaller sensitivity. In this context, $\Lambda$ hyperons have a large branching fraction into charged particles, $\text{BR}(\Lambda \to p\pi^-) \sim 64\%$~\cite{pdg}, which are easier to target. Following precisely this route, the BESIII collaboration has recently constrained $\Lambda$ decays into invisible particles and found $\text{BR}(\Lambda \to {\rm invisibles}) < 7.4\times10^{-5}$ at 90\% CL~\cite{BESIII:2021slv}. The difference between this number and the maximum sensitivity $\sim 10^{-7}$ stems from the fact that the tagging efficiency is not $100\%$ and that there are some small sources of background. 
 
Looking forward, one can consider the possibilities for other decay modes of hyperons and in other systems such as $\Sigma^+$, $\Sigma^-$, $\Xi^0$ and $\Xi^-$. By looking at the main decay modes of these states, we can readily see which ones could be more promising. $\Sigma^+$ baryons decay $\sim 50\%$ of the time into $p \pi^0$. Although the $\pi^0$ is a neutral particle, this decay could be used to target the missing energy in the final event. The $\Sigma^-$, on the other hand, decays $\sim 100\%$ to $n\pi^-$. Targeting the neutron may prove difficult, but one could use the antihyperon $\overline{\Sigma^-}$ since an antineutron would yield a clear signal in the detector. The $\Xi^0$ decays $\sim 100\%$ of the time into $\Lambda \pi^0$. These particles are neutral and light, but lead to final states that may be easily targeted. Finally, $\Xi^-$ decays mostly into $\Lambda \pi^-$. Given that the one of particles in the initial and final state are charged, this channel seems much easier to target.
 
So far, we have only discussed possibilities at BESIII, where the initial state energy is known as the decay happens at the $J/\psi$ peak. This is the ideal set up to search for decays with missing  energy in the final state, but in principle LHCb could search for them too. On the plus side, one expects a much larger number of hyperons to be produced at LHCb. On the negative one, the energy of the initial state is not known. This implies that in order to target these decays, one would need several particles in the final state to have a handle on missing energy. Experimentally, this approach has been studied in~\cite{Rodriguez:2021urv} to target $b$-hadron decays with large missing energy, showcasing very good sensitivities in these systems. From the theoretical perspective, we expect decays involving many pions in the final state to be possible only in $\Xi^0$ decays and only in a very small region of the parameter space.
Thanks to the generality of the ChPT framework, decays involving any number of pions can be easily computed employing the tools developed in Section~\ref{sec:DarkDecays}. 

\subsection{Connections to the neutron lifetime puzzle}

As discussed in Secs.\,\ref{intro} and \ref{sec:Models}, a neutron dark decay channel with a branching ratio around $1\%$ provides a possible explanation for the existing discrepancy in neutron lifetime measurements between the bottle and beam experiments \cite{Fornal:2018eol}. The possible dark decay channels include 
$n\to \chi\,\gamma$ and  $n\to \xi\,\phi$, as long as 
the final state mass lies in the range $937.993 \ {\rm MeV}<m< 939.565 \ {\rm MeV}$. Agreement with the observed neutron star masses requires dark sector self-interactions~\cite{McKeen:2018xwc,Baym:2018ljz,Motta:2018rxp} or additional neutron-dark matter interactions~\cite{Grinstein:2018ptl} to be present,  representative examples of which are described below~\cite{Cline:2018ami,Karananas:2018goc,Elahi:2020urr}. 

Given this extra ingredient, the decay channel $n\to \xi\,\phi$ with ${\rm BR}(n\to \xi\,\phi)= 1\%$ is consistent with all experimental data. On the other hand, the branching ratio  for  $n\to \chi\,\gamma$ is constrained by the Borexino data \cite{Borexino:2015qij} and requires $m_\chi \gtrsim 938.5 \ {\rm MeV}$ \cite{McKeen:2020zni}. In addition, the $n\to \chi\,\gamma$ channel has been searched for directly in the Los Alamos UCN experiment, and a ${\rm Br}(n\to \chi\,\gamma)= 1\%$ for a dark particle mass $937.901\,\text{MeV}<m_\chi<938.783\,\text{MeV}$ was excluded  with an overall significance of $2.2 \sigma$ \cite{PhysRevLett.121.022505}.

Within the framework of Model 2, described by the second Lagrangian in Eq.\,(\ref{eq:Models}), the neutron dark decay  $n\to \chi\,\gamma$ with a branching ratio 1\% would occur if 
\begin{align}
C_{ud,d}^{R} \approx \frac{2.7 \times 10^{-5}}{\rm TeV^2} \ .
\end{align}
Thus, the natural mass scale for $\Phi$ is $M_\Phi \sim \mathcal{O}(100 \ {\rm TeV})$.
However, if the combination of couplings entering the Wilson coefficient, $|y_{ud} \, y_{\chi d}|$, is  small, then  the  mass $M_\Phi$ required to explain the neutron lifetime puzzle can be  much smaller, in which case  the LHC constraints become relevant. As an example, for $M_\Phi = 7 \ {\rm TeV}$ the required combination of couplings is $|y_{ud} \, y_{\chi d}| \sim 10^{-3}$.  
This does not constrain the interactions of $\Phi$ with strange quarks -- a sizable rate for hyperon dark decays is still possible. Similar conclusions apply to Model 3. 

For the decay channel $n\to \xi\,\phi$, a branching ratio of $1\%$ is achieved when 
\begin{align}
C_{ud,d}^{R}\cdot|y_{\xi\phi}| \approx \frac{2 \times 10^{-8}}{\rm TeV^2} \ .
\end{align}
If $\xi$ is the dark matter particle, the coupling $y_{\xi\phi} \approx 0.04$ allows for the annihilation $\xi\bar\xi \to \phi \phi^*$ at a rate consistent with the observed dark matter relic density. Again, the natural scale  for $\Phi$ is high, $M_\Phi \sim \mathcal{O}(1000 \ {\rm TeV})$. In order to have $M_\Phi = 7 \ {\rm TeV}$, the couplings need to be small, $|y_{ud} \, y_{\chi d}| \sim 10^{-6}$. 
In the case of Model 1, it has  been  argued that the allowed branching ratio for $n\to \chi\,\gamma$ is smaller than $\sim 10^{-6}$ \cite{Fajfer:2020tqf}, due to the stringent flavor  constraints from kaon mixing.
This conclusion does however not apply to the decay channel $n\to \xi\,\phi$.

It should also be noted that in the minimal SM  the neutron lifetime can be extracted from $g_A/g_V$, i.e., the ratio of the axial-vector to vector current coefficient in the neutron beta decay matrix element  \cite{Czarnecki:2018okw}. The most recent precise measurement of  $g_A/g_V$  from the energy spectrum of electrons provided by the PERKEO III experiment agrees with the bottle result \cite{Dubbers:2018kgh}. However, the  value of $g_A/g_V$ derived from the energy spectrum of protons measured in the aSPECT experiment is consistent the beam result  \cite{Beck:2019xye}, favoring the neutron dark decay proposal as a solution to the neutron lifetime puzzle.

Finally, the existence of a neutron dark decay channel has  implications for neutron stars. Although not in conflict with their stability, the minimal neutron dark decay models predict neutron star masses that are lower than observed \cite{Baym:2018ljz,McKeen:2018xwc,Motta:2018rxp}. In particular, a dark particle leading to a neutron dark decay softens the EoS and implies a maximal neutron star mass of $\sim0.8 \,M_{\odot}$, well below  the observed value of $2\,M_{\odot}$. As mentioned above, this problem is solved  by introducing  extra interactions of the dark particle.

An example of a model for the neutron dark decay with self-interactions in the dark sector that is capable of producing neutron stars with masses $2\,M_{\odot}$ was proposed in  \cite{Cline:2018ami}. This involves a dark 
photon $A'$ and the minimal Lagrangian for the neutron dark decay $n\to \chi\,A'$ includes,
\begin{equation}
\label{LagKinMix}    
\mathcal{L} \ \supset \ -\frac14 F'_{\mu\nu}F^{\prime {\mu\nu}} -  \varepsilon \,F_{\mu\nu}F^{\prime {\mu\nu}} \ ,
\end{equation}

with $D_\mu \to D_\mu -i g'A'_\mu$. The coupling  between $\chi$ and $A'$  leads to repulsive interactions between the $\chi$ particles. There exists a range of parameters for which this model leads to ${\rm Br}(n\to \chi\,A') = 1\%$ while being consistent with neutron star constraints and other astrophysical bounds. The dark matter in the universe can consist of $\chi$ particles; however, if it was thermally produced, this model can only account for  $10\%$ of it. A further extension of this model \cite{Bringmann:2018sbs} can explain the matter-antimatter asymmetry  through low-scale baryogenesis. In a variation on this theme, the neutron decays into $\chi$ and a nonabelian light $W'$ that mixes kinetically with the photon, through a non-renormalizable generalization of \eqref{LagKinMix} \cite{Elahi:2020urr}.   

Instead of a dark photon, the gauge boson mediating self-interactions in the dark sector could be a dark $Z_D$ \cite{Karananas:2018goc}. Then the Lagrangian involves the terms,
\bea
\mathcal{L} \ \supset \ g' \bar\chi \,\gamma_\mu {Z}_{D}^\mu\chi - i\,g' Z^\mu_D\left(\phi^* \partial_\mu \phi-\phi\,\partial_\mu \phi^*\right) \ .
\eea
In this model, one can accommodate ${\rm Br}(n\to \chi\,\phi) = 1\%$, account for all of the dark matter in the universe, and solve the small-scale structure problems of the $\Lambda$CDM paradigm.

Apart from the dark sector self-interactions, neutron star constraints are overcome by  introducing additional repulsive neutron-dark matter interactions \cite{Grinstein:2018ptl}.  The  Lagrangian extension needed is,
\bea
\mathcal{L} \ \supset \ \mu \,H^\dagger H\,\phi + g_\chi\,\bar\chi \,\chi\,\phi \ ,
\eea
which induces, through the Higgs portal, an effective interaction $g_n \bar{n}n\phi$. This affects the  neutron star's EoS, allowing for $2\,M_{\odot}$ neutron stars and a simultaneous presence of the decay channel $n\to\chi\,\phi$.

\subsection{Connections to Mesogenesis}
Mesogenesis frameworks \cite{Elor:2018twp,Alonso-Alvarez:2021qfd,Elor:2020tkc,Elahi:2021jia} rely on the existence of baryonic dark sectors in order to generate the dark matter and the matter-antimatter asymmetry of the universe. 
In particular, the baryon asymmetry in $B$-Mesogenesis~\cite{Elor:2018twp,Alonso-Alvarez:2021qfd} is directly linked to the branching fraction of neutral $B$ meson decays into dark baryons via $b$-flavored operators among those in Eq.~\eqref{eq:Models}.
In turn, hyperon decays are excellent probes of $s$-flavored variations of these operators and therefore serve as indirect probes of this mechanism.
Other version of Mesogenesis like the ones proposed in~\cite{Elor:2020tkc,Elahi:2021jia} also invoke dark states carrying baryon and/or lepton number and may be tested using processes similar to the ones studied in this paper.
Overall, searches for hyperon decays into dark baryons could play a pivotal role in ultimately discovering Mesogenesis and furthermore discerning which variation is responsible for generating the baryon asymmetry and dark matter of the Universe. 

Let us focus on the neutral $B$-Mesogenesis framework \cite{Elor:2018twp,Alonso-Alvarez:2021qfd}, in which the matter anti-matter asymmetry of the Universe is directly related to two experimental observables:
\begin{eqnarray}\label{eq:YBMesogenesis}
\eta_{\mathfrak B} \,\, && \,\, \equiv \frac{n_{\mathfrak B}  - n_{\bar{{\mathfrak B} }}}{n_\gamma} 
\,\, \propto \,\,  \text{BR}(B^0_{q} \rightarrow \chi \, \mathfrak B \, \mathfrak{M}) \,A^q_{\rm SL}\,.
\end{eqnarray}
Here, $A_{\rm SL}^q$ is the CP charge asymmetry in $B_q^0$ decays, and $ \text{BR}(B^0_{s,d} \rightarrow \chi \, \mathfrak B \, \mathfrak{M})$ denotes the inclusive branching fraction of neutral $B$ mesons into a visible and a dark sector baryon---$ \mathfrak{M}$ stands for any number of accompanying light mesons. 

To successfully generate the observed baryon asymmetry, the product of the CP charge asymmetry and the branching fraction should be greater than $5\times 10^{-7}$~\cite{Alonso-Alvarez:2021qfd}. Given that the CP charge asymmetries are experimentally bounded to be $A_{\rm SL}^{s,d} \lesssim 5\times 10^{-3}$~\cite{pdg}, one would expect a branching fraction for $B$ meson decays into dark sector baryons larger than $10^{-4}$~\cite{Alonso-Alvarez:2021qfd} for successful $B$-Mesogenesis. At present, this branching fraction is constrained by an ALEPH search for $b$ decays with large missing energy resulting in $\text{BR}(B \rightarrow \chi \, \mathfrak B \, \mathfrak{M}) < 0.5\%$ at 95\% CL~\cite{Alonso-Alvarez:2021qfd,Barate:2000rc}, independent of the mass of the dark sector state $\chi$ and the flavor of the baryon $\mathfrak B$. 

Experimental searches can more easily target exclusive $B$ decays rather than inclusive quantities. Fortunately, the simplest exclusive final states are expected to have branching ratios not smaller than $10^{-6}$ within $B$-Mesogenesis~\cite{Alonso-Alvarez:2021qfd}. 
The exact flavor content of an exclusive final state depends on the operator at play among those in Eq.~\eqref{eq:generalOp} containing a $b$ quark: $\mathcal{O}_{ubd} =  \chi u b d$, $\mathcal{O}_{ubs} =  \chi u b s$, $\mathcal{O}_{cbd} =  \chi c b d$, and $\mathcal{O}_{cbs} =  \chi  c b s$.
The Belle collaboration has recently performed a search for $B_d^0 \to \Lambda \,\chi$ decays, leading to a constraint on the exclusive branching ratio at the $\mathcal{O}(10^{-5})$ level~\cite{Belle:2021gmc}. 
This has significantly constrained the possibility of $B$-Mesogenesis proceeding through the $\mathcal{O}_{ubs}$ operator, but the other three remain a perfectly viable possibility to explain the dark matter and baryon asymmetry of the Universe through this mechanism. Additionally, the parameter space for $B_c^+$-Mesogenesis~\cite{Elahi:2021jia} could accommodate smaller branching ratios arising from the $\mathcal{O}_{ubs}$ operator, thereby making this operator still viable even in light of the Belle result. 

In the full UV models of Eq.~\eqref{eq:Models}, the branching ratio controlling the generation of the baryon asymmetry in neutral $B$-Mesogenesis can be estimated to be~\cite{Elor:2018twp}
\begin{align}
& \text{BR} \left( B^0_{s,d} \rightarrow \chi \, \mathfrak B \, \mathfrak{M} \right) \label{eq:bdecayrate}  \\ \nonumber 
& \qquad \simeq 10^{-3} \left(\frac{\Delta m}{3~\rm GeV}\right)^4  \left(\frac{1.5~\rm TeV}{{M}} \frac{\sqrt{y^2}}{0.53} \right)^4 \,,
\end{align}
where $M$ denotes the colored mediator mass. The mass difference between initial and final states is given by $\Delta m = m_B - m_\chi - m_{\mathfrak B}- m_{\mathfrak{M}}$, and we write the coupling products schematically as $y^2$, where $y^2 \equiv y_{d_{i} d_{j}}\,y_{\chi u_k}$  for Model 1 and $y^2 \equiv y_{u_{i} d_{j}}\,y_{\chi d_k}$ for Model 2 in Eq.~\eqref{eq:Models}. 
Importantly, Model 3 cannot yield a large enough branching fraction because the couplings that enter Eq.~\eqref{eq:bdecayrate} are in this case directly constrained by meson mixing bounds as shown in Table~\ref{tab:meson_mixing_V_single_coupling}. 

The combination of UV model parameters $\sqrt{y^2}/M$ in Eq.~\eqref{eq:bdecayrate} are constrained by the same collider searches described in Section~\ref{sec:LHC} and which were explored systematically in~\cite{Alonso-Alvarez:2021qfd} for $b$ quark operators. Importantly, while only the couplings containing a $b$ quark in Eq.~\eqref{eq:Models} are required for $B$-Mesogenesis, the rest are expected to be present and can lead to hyperon decays into dark baryons. In that way, hyperon searches can indirectly constrain the parameter space of $B$-Mesogenesis.

For illustration, consider the following combination of theory parameters that leads to successful baryogenesis within the neutral $B$-Mesogenesis framework: $y_{db} = 0.3$, $y_{\chi u} = 2$, and $M_\Psi = 3\,\text{TeV}$, in Model 1 (see top left panel of figure 20 in~\cite{Alonso-Alvarez:2021qfd}).
Consistency with $B$-Meson oscillations requires $|y_{d b}\, y_{sd}^\star| \lesssim 0.2 \left(\frac{M_\Psi}{1.5\text{TeV}}\right) $. Consequently, the Wilson coefficients entering hyperon decays can be as large as 
\begin{align}
C_{ud,s}^R = -C_{us,d}^R \lesssim 0.02 \,\text{TeV}^{-2}.
\end{align}
This interaction strength could yield a branching fraction as large as $10^{-5}$ for the decay $\Sigma^- \to \chi + \pi^-$, as shown in Table~\ref{tab:decays}.
This branching fraction could well be tested at hyperon facilities. 
In this sense, searches at BESIII represent a relevant test of $B$-Mesogenesis models in a complementary way to other collider searches. 

\section{Summary and conclusions}
\label{sec:conclusions}

Several attractive solutions to some of the most outstanding problems in particle physics motivate the existence of dark sector particles at the GeV scale.
Some of these states may carry baryon number and couple to different flavors of SM quarks.
Surprisingly, although such particles would interact with hadrons, they remain relatively unconstrained by present experimental data. 
In this paper, we develop the theoretical framework needed to study the relevant aspects of strange baryon decays into the dark sector. 

We start by listing the UV-complete models that can lead to the decays of SM baryons into dark sector states carrying baryon number.
By integrating out the heavy mediators, we construct a low-energy effective theory that can be used to describe the interactions of hyperons with GeV-scale dark baryons.

The main technical development of our work involves the use of chiral perturbation theory methods to calculate the decay rates of strange baryons into dark sector states.
This approach allows us to account for the relevant strong dynamics taking place in these transitions.
Our predictions for the dark hyperon rates are shown to only depend on parameters that are experimentally measured or known from lattice calculations.
As a result, we identify the most promising dark hyperon decay channels and calculate their expected branching ratios.
These are summarized in Table~\ref{tab:decayChannels}.
We conclude that hyperon facilities like BESIII have great potential to discover baryonic dark sectors.

In addition to terrestrial experiments, we show that exotic hyperon decays have important implications for astrophysical setups.
In particular, the production of dark baryons from $\Lambda$ decays inside the proton-neutron star formed at the early stages of a core-collapse supernova acts as a source of anomalous cooling in the system.
Our calculation of the cooling rate includes the effect of gravitational trapping of the dark baryons inside the star, which is significant for nonrelativistic decay products.
Taking this into account, we derive stringent constraints on dark decays of $\Lambda$ baryons from observations of the supernova SN~1987A, which are displayed in Figs.~\ref{fig:SN1987A-I} and~\ref{fig:SN1987A-II}.

Although they are integrated out at hadronic energy scales, the heavy scalar bosons mediating the decays of baryons into dark sector states can have phenomenological consequences at other high-energy experiments.
If their masses are not far from the TeV scale, they could be directly produced at the LHC through their strong interactions or their couplings to quarks.
We use this to place bounds on the couplings and masses of these mediators, which translate into bounds on the Wilson coefficients of the low-energy effective theory.
Furthermore, we show that meson-mixing observables can also be very sensitive tests of these colored bosons.
In particular, the measured mixing parameters in the $K$, $B$, and $D$ systems allow us to gain insights on the flavor structure of their interactions. 

There are a few directions for future study that are motivated by this work.
The formalism developed in this paper to calculate the form factors for hyperon decays can also be used to compute higher-order processes, such as those involving any number of final-state light mesons. 

Follow-up work can also be done in understanding the role of dark baryons in neutron stars.
Hyperons are generically expected to be produced in the high-density regions of neutron stars~\cite{1960SvA}, and they act to soften the EoS. In fact, calculations that include hyperons have difficulties in stabilizing neutron stars with masses $\sim 2 M_{\odot}$ and they tend to run into conflict with observations~\cite{Vidana:2015rsa,Oertel:2016bki}. 
New physics solutions to this puzzle have been proposed~\cite{DelPopolo:2020pzh}, but a definite answer requires precise knowledge of the physics in the high-density cores of neutron stars, as well as of the hyperon-nucleon interaction strength, at a level that has not yet been achieved. In fact, the hyperonic EoS used in this work (labeled as SFHoY) is consistent with neutron star mass observations~\cite{Fortin:2017dsj}. 
Once a sufficient understanding of those topics is reached, it would be interesting to study what the role of dark baryons produced in hyperon decays could be in this situation, although, in general, it is expected that these bounds are weaker than those derived from SN~\cite{Raffelt:1996wa}. 

The scenario in which the dark baryon produced in $\Lambda$ decays is unstable and decays back to Standard Model particles is especially interesting.
If the decay products involve photons, as would be, e.g., for the $\chi\to n+\gamma$ channel, decays occurring in the atmosphere of a supernova or neutron star could give rise to distinct observational signatures in the form of $\gamma$ rays.

Our work highlights the rich interplay that exists between dark hyperon decays and theoretically motivated scenarios involving GeV-scale dark sector particles, like the neutron dark decay as an explanation of the neutron lifetime anomaly, and $B$-Mesogenesis as the origin of the DM and the baryon asymmetry of the universe.
Although these two models rely on dark baryons coupled to first- and third-generation quarks, respectively, in the lack of any symmetry argument couplings to second-generation quarks are also expected.
In this sense, studying exotic decays of strange baryons serves as an indirect probe of these scenarios.
In the event of a discovery of baryonic dark states in hyperon decays, a combination of all the probes discussed in this work would help to assess whether this particles could be responsible for dark matter, baryogenesis and/or the experimental discrepancies in measurements of the neutron lifetime. 

\section*{Acknowledgements}
We thank R.~Bollig and H.-Th.~Janka for providing us with the data of the core-collapse simulations used in this paper. We would also like to thank M.~Pospelov and J.~Zupan for useful discussions. 
G.A. is supported by NSERC (Natural Sciences
and Engineering Research Council, Canada) and the McGill Space Institute through a McGill Trottier Chair Astrophysics Postdoctoral Fellowship.
G.E. is supported by the Cluster of Excellence {\em Precision Physics, Fundamental Interactions and Structure of Matter\/} (PRISMA${}^+$ -- EXC~2118/1) within the German Excellence Strategy (project ID 39083149). M.E. is supported by a Fellowship of the Alexander von Humboldt Foundation and by the Collaborative Research Center SFB1258 and by the Deutsche Forschungsgemeinschaft (DFG, German Research Foundation) under Germany's Excellence Strategy - EXC-2094 - 39078331. The work of B.G. is supported in part by the Gordon and Betty Moore Foundation\footnote{The Gordon and Betty Moore Foundation fosters path-breaking scientific discovery, environmental conservation, patient care improvements and preservation of the special character of the Bay Area. Visit Moore.org and follow @MooreFound.} through Grant GBMF6210 and by the U.S. Department of Energy through Grant
No. DE-SC0009919. J.M.C. acknowledges support from the Spanish MINECO through the ``Ram\'on y Cajal'' program RYC-2016-20672 and the grant PGC2018-102016-A-I00.

\appendix
\newpage
\section{ChPT coefficients for hyperon decays} 
\label{sec:Appendix}

Here we list the coefficients in the analytical formulas for the matrix elements in Eq.~\eqref{eq:W01q2} (Tab.~\ref{tab:Btopipole}), Eq.~\eqref{eq:W0ct} (Tab.~\ref{tab:BtopiCT}) and Eq.~\eqref{eq:V01q2} (Tab.~\ref{tab:Btogamma}). 

\begin{table}[h!]
\renewcommand{\arraystretch}{1.4}
\begin{tabular}{cccc}
\hline\hline
Channel & Operator & $\mathfrak B^\prime$ &$b_{\mathfrak B}$\\
\hline \hline
$\Lambda\to\pi^0$&$(us)d$&$\Sigma^0$ & $\frac{D}{\sqrt{6}}$\\
\hline
\multirow{2}{*}{$\Sigma^{0}\to\pi^0$}& $(us)d$&$\Lambda$&$-\frac{D}{3\sqrt{2}}$\\
&$(ud)s$&$\Lambda$&$-\frac{\sqrt{2}D}{3}$\\
\hline
\multirow{2}{*}{$\Sigma^+\to\pi^+$}&$(us)d$ & $\Lambda$, $\Sigma^0$ & $-\frac{D}{3\sqrt{2}}$, $-\frac{F}{\sqrt{2}}$\\
&$(ud)s$&$\Lambda$&$-\frac{\sqrt{2}D}{3}$\\
\hline
\multirow{2}{*}{$\Sigma^-\to\pi^-$} &$(us)d$ & $\Lambda$, $\Sigma^0$ & $-\frac{D}{3\sqrt{2}}$, $+\frac{F}{\sqrt{2}}$\\
&$(ud)s$&$\Lambda$&$-\frac{\sqrt{2}D}{3}$\\
\hline
$\Xi^0\to\pi^0$&$(us)s$&$\Xi^0$&
$\frac{D-F}{2}$ \\
\hline
$\Xi^-\to\pi^-$&$(us)s$&$\Xi^0$&
$-\frac{D-F}{\sqrt{2}}$\\
\hline
\hline
\end{tabular}
\caption{\label{tab:Btopipole} Coefficients $b_{\mathfrak B}$ for the baryon-pole contributions to the form factors of the $\mathfrak B\to\pi$ transitions in Eq.~\eqref{eq:W01q2}. In our calculation we take $D=0.80$ and $F=0.46$~\cite{Ledwig:2014rfa}.}
\end{table}

\begin{table}[h]
\renewcommand{\arraystretch}{1.4}
\begin{tabular}{cccc}
\hline\hline
Channel & Operator & Chirality &$c_{\mathfrak B}$\\ 
\hline \hline
$\Lambda\to\pi^0$ &$(us)d$&$L$&$-\frac{1}{\sqrt{6}}$\\
\hline
$\Sigma^0\to\pi^0$ &$(us)d$&$L$&$\frac{1}{\sqrt{2}}$\\
\hline
$\Sigma^+\to\pi^+$&$(us)d$&$L/R$&$\frac{1}{\sqrt{2}}/\frac{1}{\sqrt{2}}$\\
\hline
$\Sigma^-\to\pi^-$&$(us)d$&$L/R$&$\frac{1}{\sqrt{2}}/-\frac{1}{\sqrt{2}}$\\
\hline
$\Xi^0\to\pi^0$&$(us)s$&$L/R$&$-\frac{1}{2}/\frac{1}{2}$\\
\hline
$\Xi^-\to\pi^-$&$(us)s$&$L/R$&$\frac{1}{\sqrt{2}}/-\frac{1}{\sqrt{2}}$\\
\hline\hline
\end{tabular}
\caption{\label{tab:BtopiCT}Contact-term contributions to the form factors of $\mathfrak B\to\pi$ in Eq.~\eqref{eq:W01q2} expressed in the form $W_{0\mathfrak B}^{L,\rm ct}= \alpha c^{L}_\mathfrak B / f$ and $ W_{0\mathfrak B}^{R,\rm ct}=\beta c^{R}_\mathfrak B / f$. 
}
\end{table}

\begin{table}[h!]
\renewcommand{\arraystretch}{1.4}
\begin{tabular}{cccc}
\hline\hline
Channel & Operator & $\mathfrak B^\prime$ & Coefficient\\
\hline
$n\to\gamma$&$(ud)d$&$n$& $\kappa_n$\\
\hline
\multirow{2}{*}{$\Lambda\to\gamma$}& $(ud)s$&$\Lambda$&$-\sqrt{\frac{2}{3}}\kappa_\Lambda$\\
&$(us)d$&$\Lambda$, $\Sigma^0$&$-\frac{\kappa_\Lambda}{\sqrt{6}}$,$\frac{\kappa_{\Lambda\Sigma^0}}{\sqrt{2}}$\\
\hline
\multirow{2}{*}{$\Sigma^0\to\gamma$}& $(ud)s$&$\Lambda$&$-\sqrt{\frac{2}{3}}\kappa_{\Lambda\Sigma^0}$\\
&$(us)d$&$\Lambda$, $\Sigma^0$&$-\frac{\kappa_{\Lambda\Sigma^0}}{\sqrt{6}}$,$\frac{\kappa_{\Sigma^0}}{\sqrt{2}}$\\
\hline
$\Xi^0\to\gamma$ & $(us)s$ & $\Xi^0$ &$\kappa_{\Xi^0}$\\
\hline
\hline
\end{tabular}
\caption{\label{tab:Btogamma}Coefficients for the baryon-pole contributions to the form factors of $\mathfrak B\to\gamma$ in Eq.~\eqref{eq:V01q2}, where we factored out the chiral parameters $\alpha$ and $\beta$. }
\end{table}

\begin{figure*}[t]
\centering
\begin{tabular}{cc}
		\label{fig:Y_BR1}
		\includegraphics[width=0.45\textwidth]{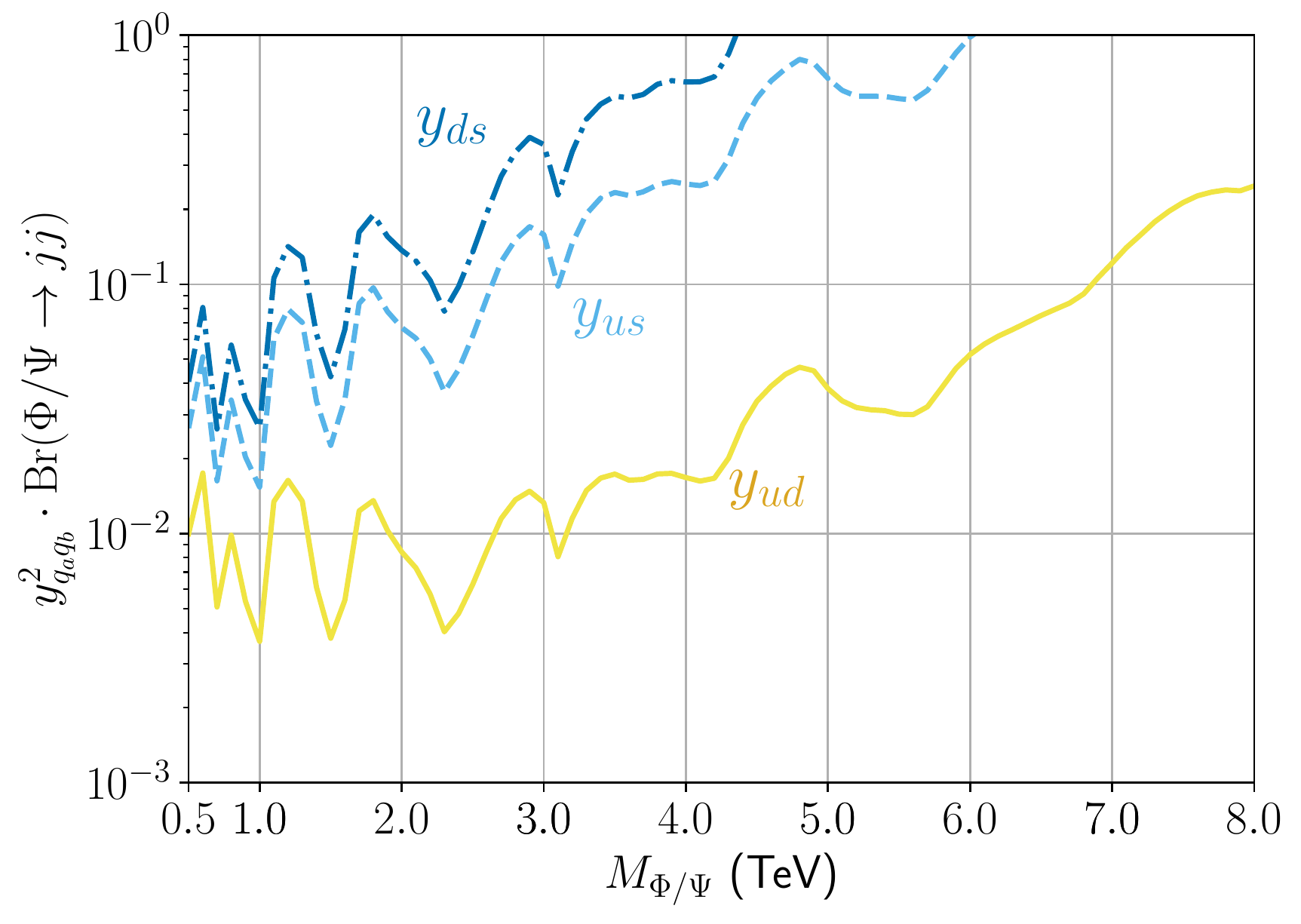}
&
		\label{fig:X_BR1}
		\includegraphics[width=0.45\textwidth]{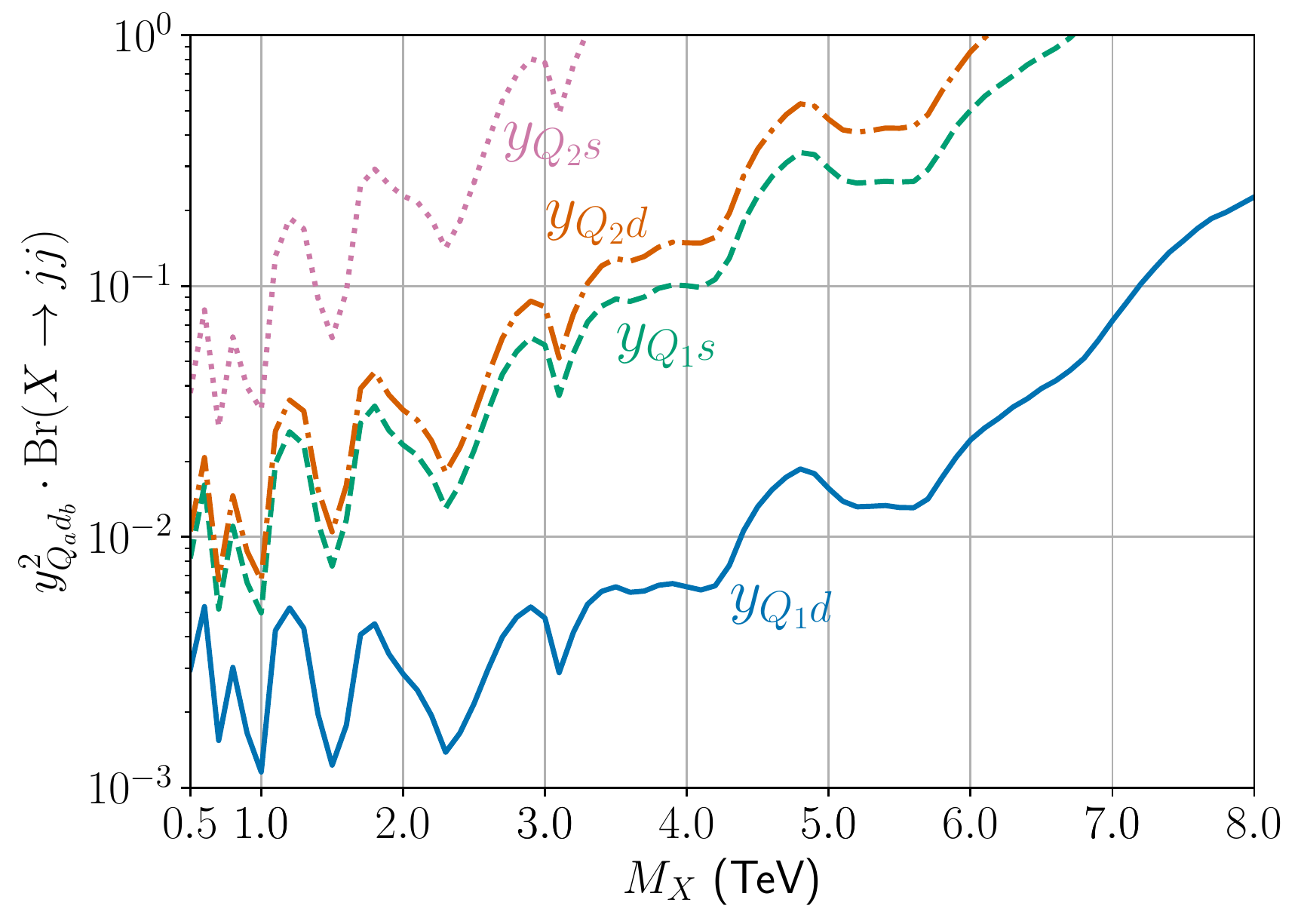}
\end{tabular}
\vspace{-0.2cm}
\caption{Constraints on $y_{q_iq_j}^2\, \text{BR}(\Phi/\Psi/X\to j \,j)$ from resonant dijet searches for the scalar (\emph{left}) and vector (\emph{right}) mediators. Bounds are based on a recast of the CMS search of~\cite{Sirunyan:2018xlo} that used $36\,\text{fb}^{-1}$ of data at 13 TeV. We assumed $\Gamma_\Phi/M_\Phi = \Gamma_\Psi/M_\Psi = 0.04$, $\Gamma_X/M_X = 0.05$ and, as discussed in~\cite{Sirunyan:2018xlo}, an acceptance of $A=0.56$.
The lowest mass shown in the plot corresponds to $0.5$~TeV, as smaller masses are excluded by 4-jet searches~\cite{ATLAS:2017jnp,CMS:2018mts}. 
}
\label{fig:Y_LHC_BR}
\end{figure*}

\begin{figure*}[t]
\centering
\begin{tabular}{cc}
        \label{fig:Y_BR2}
		\includegraphics[width=0.45\textwidth]{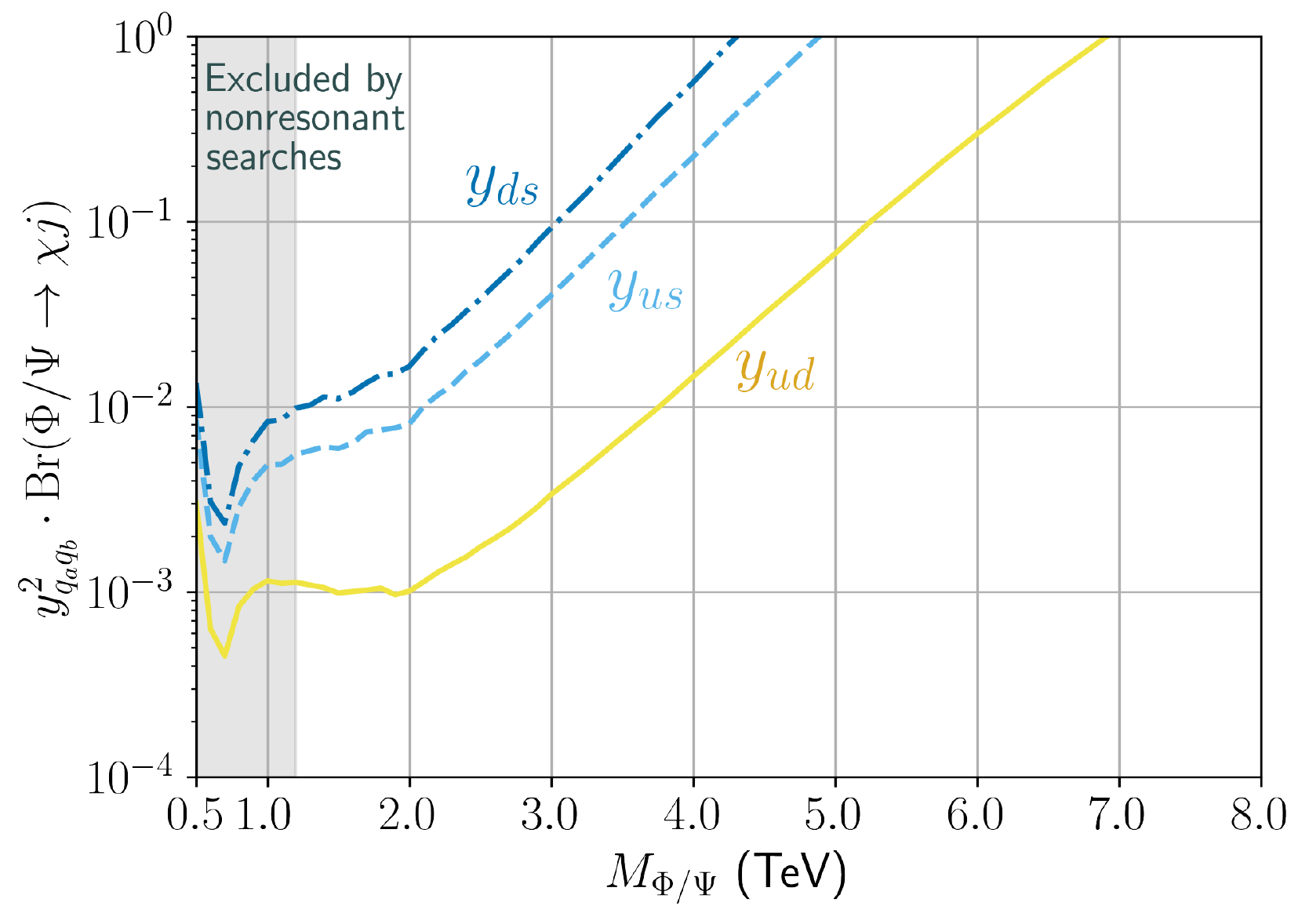}
&
        \label{fig:X_BR2}
		\includegraphics[width=0.45\textwidth]{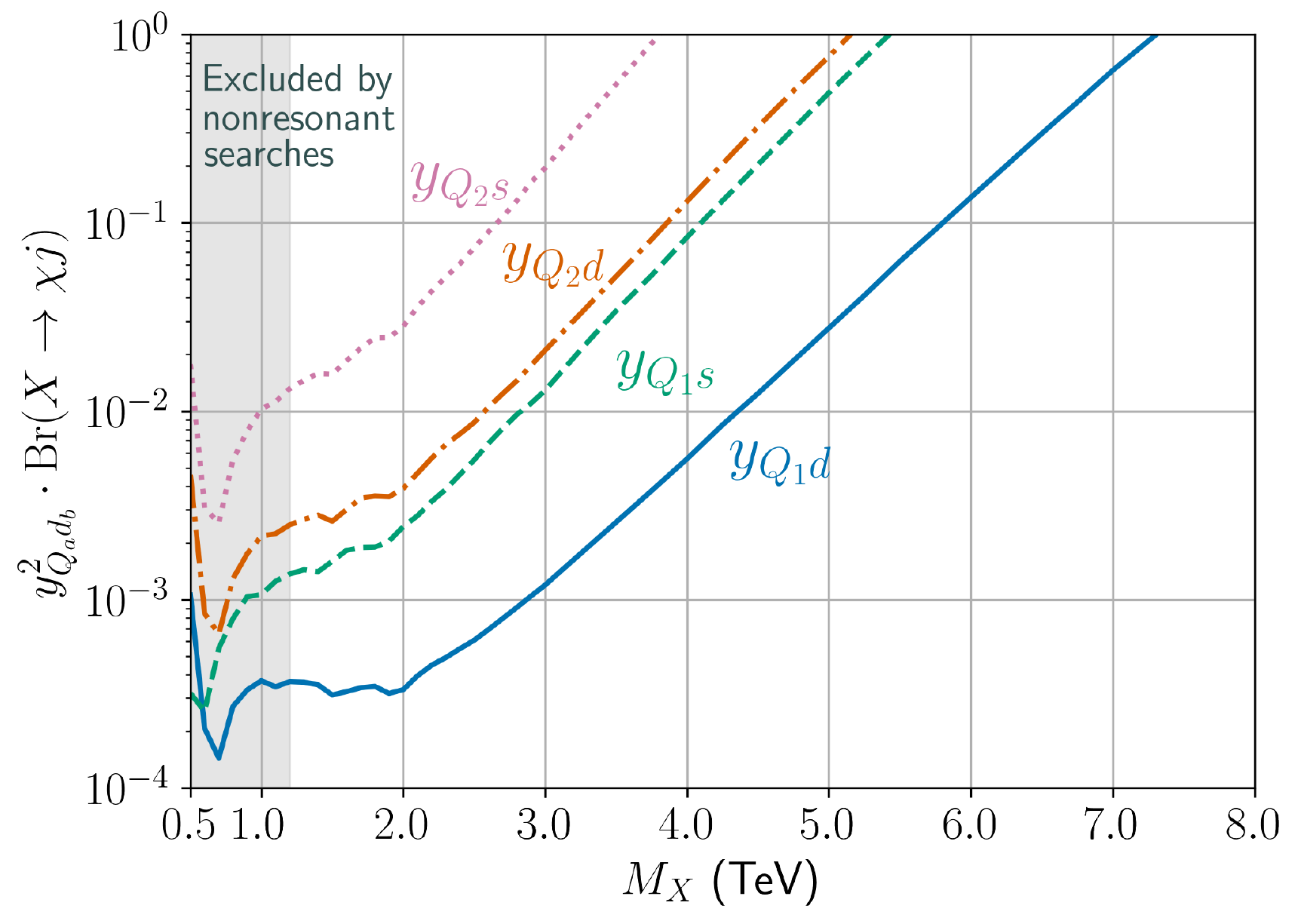}
\end{tabular}
\vspace{-0.2cm}
\caption{Constraints on $y_{q_iq_j}^2\, \text{BR}(\Phi/\Psi/X\to j \,\chi)$ from jet plus missing energy searches for the scalar (\emph{left}) and vector (\emph{right}) mediators. Bounds have been recasted from the analysis of ATLAS~\cite{Aaboud:2017phn}  using $36\,\text{fb}^{-1}$ of data at 13 TeV. We assumed $\Gamma_\Phi/M_\Phi = \Gamma_\Psi/M_\Psi = 0.02$, $\Gamma_X / M_X = 0.01$ and performed a fit without correlating errors to the exclusive signal regions of~\cite{Aaboud:2017phn}.
The gray-shaded area in both regions highlights the $M_{\Phi/\Psi/X}>1.2$~TeV exclusion from nonresonant SUSY searches~\cite{Aad:2020aze,Sirunyan:2019ctn}.
}
\label{fig:X_LHC_BR}
\end{figure*}

\section{Recast of dijet and jet+MET LHC searches}\label{app:LHC}

Our determination  of the bounds from the CMS and ATLAS searches for resonantly produced colored bosons are based on the procedure  described in Sec.\,V of~\cite{Alonso-Alvarez:2021qfd}, and we  only summarize the main points here.
Once produced, the colored boson decays into two quarks or a $\chi$-quark pair, with a partial decay rate:
\begin{align} \label{eq:PartialWidths}
\Gamma(\Phi\to \bar{u}_a \bar{d_b}) &= 2 \frac{y_{u_a d_b}^2}{16\pi} M_{\Phi} \,, &\!\!
\Gamma(\Phi\to \chi d_a) &=  \frac{y_{\chi d_a}^2}{16\pi} M_{\Phi} \,, \nonumber \\
\Gamma(\Psi\to \bar{d_a} \bar{d_b}) &= 2 \frac{y_{d_a d_b}^2}{16\pi} M_{\Psi} \,, &\!\!
\Gamma(\Psi\to \chi u_a) &=  \frac{y_{\chi u_a}^2}{16\pi} M_{\Psi} \,, \nonumber \\
\Gamma(X\to \bar{q}_a \bar{d_b}) &= \xi_{ab} \frac{y_{Q_a d_b}^2}{24\pi} M_{X} \,, &\!\!
\Gamma(X\to \chi q_a) &= \frac{y_{\chi Q_a}^2}{24\pi} M_{X} \,,
\end{align}
where $\xi_{ab}=1$ if $q_a=d_b$, and $\xi_{ab}=2$ otherwise.

The decays on the left of Eq.~(\ref{eq:PartialWidths}) can be tested via dijet searches like the one performed by CMS in~\cite{Sirunyan:2018xlo} based on $36\,\text{fb}^{-1}$ of data at $13$~TeV.
Our recast of this analysis has been implemented using \texttt{FeynRules}~\cite{Alloul:2013bka} and \texttt{MadGraph5\_aMC@NLO}~\cite{Alwall:2014hca} to calculate the leading order production cross section.
A comparison to the limits shown in Fig.~12 of~\cite{Sirunyan:2018xlo} allows to obtain a bound on the product $y_{q_aq_b}^2\, \text{BR}(\Phi/\Psi/X\to j j)$ using the narrow width approximation (NWA)\footnote{Our results are only valid as long as $y < 1$, which assures the NWA condition $\Gamma<M$. Larger couplings result in relaxed constraints due to the larger width of the colored boson~\cite{Sirunyan:2018xlo}.}.
The result is shown in Fig.~\ref{fig:Y_LHC_BR} for both scalar and vector mediators.

For the jet+MET decays on the right  of Eq.~\eqref{eq:PartialWidths} we employ the ATLAS search~\cite{Aaboud:2017phn}, once again based on the $36\,\text{fb}^{-1}$ of data collected at $13$~TeV\footnote{The more recent analysis using the full Run-2 dataset ($139\,\text{fb}^{-1}$) released in~\cite{ATLAS:2021kxv} is not included in our recast. We expect that considering the search of~\cite{ATLAS:2021kxv} could improve our constraints on Fig.~\ref{fig:X_LHC_BR} by as much as a factor of 2.}.
In this case, we also make use of the hadronization and detector simulation codes \texttt{PYTHIA8}~\cite{Sjostrand:2014zea} and \texttt{DELPHES3}~\cite{deFavereau:2013fsa}, in addition to a publicly available 
\texttt{MADANALYSIS5}~\cite{Conte:2012fm,Conte:2014zja,Conte:2018vmg} recast of the analysis~\cite{recast_jetMET}.
As before, the NWA allows us to obtain a limit on the product $y_{q_aq_b}^2\, \text{BR}(\Phi/\Psi/X\to j \,\chi)$ by combining the individual limits on the number of observed events in each exclusive signal region of~\cite{Aaboud:2017phn}.
The resulting limits are shown in Fig.~\ref{fig:X_LHC_BR}.

The limits on the couplings times the respective branching ratio for each analysis can be combined in order to obtain limits on the coupling products that control the hyperon dark decay rates.
This can be done using Eq.~\eqref{eq:PartialWidths}, which allows to compute the branching ratios in terms of the couplings within the NWA.
For each mediator mass, scanning over the possible values of $y_{q_aq_b}$ and $y_{\chi q_c}$, while keeping their product fixed, leads to the limits shown in Fig.~\ref{fig:LHC}. Those can be translated to bounds on the Wilson coefficients of the chiral EFT defined in Eq.~\eqref{eq:LEEFTchi}.


\bibliography{biblio.bib}


\end{document}